%% file: latsusy1_10.tex
\documentclass[a4paper,12pt]{article}

\usepackage{amsmath}
\usepackage[dvips]{graphicx}
\usepackage{epic}

\newcommand{\ii}{i}
\newcommand{\ee}{\mathrm{e}}

\def\be{\begin{equation}}
\def\beq{\begin{equation}}

\def\eeq{\end{equation}}
\newcommand{\en}{\end{equation}}
\def\ba{\begin{eqnarray}}
\def\bea{\begin{eqnarray}}
\def\ea{\end{eqnarray}}
\def\eea{\end{eqnarray}}
\newcommand{\eqa}{\begin{eqnarray}}
\newcommand{\ena}{\end{eqnarray}}

\def\Q{Q}
\def\D{D}
\def\U{U}
\def\uPhi{\underline{\Phi}}
\def\uvarphi{\underline{\varphi}}
\def\uFc{{\underline{\cal{F}}}}

\def\udelta{\underline{\delta}}
\def\uOc{\underline{\cal{O}}}

\def\Fc{{\cal{F}}}

\def\Oc{{\cal{O}}}

\def\tr{{\rm tr}}

\def\n{\mathbf{n}}

\def\m{\hat{n}_\mu}
\def\n{\hat{n}_{\nu}}
\def\a{\hat{a}}
\def\at{\hat{\tilde{a}}}

\def\th{\underline{\theta}}
\def\uQ{\underline{Q}}
\def\uD{\underline{D}}

\def\us{\underline{s}}
\def\uphi{\underline{\phi}}
\def\upsi{\underline{\psi}}
\def\uPsi{\underline{\Psi}}
\def\uchi{\underline{\chi}}
\def\uc{\underline{c}}
\def\uomega{\underline{\omega}}
\def\ub{\underline{b}}
\def\urho{\underline{\rho}}
\def\ulambda{\underline{\lambda}}
\def\upar{\underline{\partial}}
\def\auQ{\overrightarrow{\underline{Q}}}
\def\auD{\overrightarrow{\underline{D}}}

\def\arrowdel{\overrightarrow{\Delta}}

\newcommand{\phit}{\tilde{\phi}}
\newcommand{\chit}{\tilde{\chi}}
\newcommand{\xibar}{\bar{\xi}}

\def\varphit{\tilde{\varphi}}
\def\uvarphi{\underline{\varphi}}
\def\uvarphit{\underline{\varphit}}

\def\uphit{\underline{\phit}}
\def\uchit{\underline{\chit}}
\def\uxi{\underline{\xi}}
\def\uxibar{\underline{\xibar}}
\def\uF{\underline{F}}

\def\figlambda{\ulambda}
\def\figrho{\urho}
\def\figomega{\uomega}
\def\figc{\uc}
\def\figcb{\overline{\uc}}
\def\figphi{\uphi}
\def\figb{\ub}
\def\figvarphi{\uvarphi}
\def\figpsi{\upsi}
\def\figchi{\uchi}
\def\figphit{\uphit}
\def\figvarphit{\uvarphit}
\def\figchit{\uchit}

\begin{document}

\begin{titlepage}
\begin{flushright}
DFTT 14/2004 \\
EPHOU-04-004 \\
June, 2004
\end{flushright}

\vspace{5cm}

\begin{center}

{\Large Twisted Superspace on a Lattice }\\

\vspace{1cm}

{\sc Alessandro D'Adda}\footnote{dadda@to.infn.it}, 
{\sc Issaku Kanamori}\footnote{kanamori@particle.sci.hokudai.ac.jp}
{\sc Noboru Kawamoto}\footnote{kawamoto@particle.sci.hokudai.ac.jp} and 
{\sc Kazuhiro Nagata}\footnote{nagata@particle.sci.hokudai.ac.jp}\\

\vspace{0.5cm}
{\it{ INFN sezione di Torino, and Dipartimento di Fisica Teorica, Universita 
di Torino, I-10125 Torino, Italy }}\\
and \\
{\it{ Department of Physics, Hokkaido University }}\\
{\it{ Sapporo, 060-0810, Japan}}\\
\end{center}

\vspace{2cm}

\begin{abstract}
We propose a new formulation which realizes exact twisted supersymmetry 
for all the supercharges on a lattice by twisted superspace formalism. 
We show explicit examples of $N=2$ twisted supersymmetry invariant BF and 
Wess-Zumino models in two dimensions. We introduce mild lattice noncommutativity 
to preserve Leibniz rule on the lattice. The formulation is based on the 
twisted superspace formalism for $N=D=2$ supersymmetry which was proposed 
recently. From the consistency condition of the noncommutativity of 
superspace, we find an unexpected three-dimensional lattice structure which 
may reduce into two dimensional lattice where the superspace describes 
semilocally scattered fermions and bosons within a double size square lattice. 
\end{abstract}

\end{titlepage}

\newpage

\renewcommand{\theequation}{\arabic {section}.\arabic{equation}}
\makeatletter
\@addtoreset{equation}{section}
\makeatother
\setcounter{footnote}{0}
\def\thefootnote{\arabic{footnote}}

\section{Introduction} \label{intro}

We know by now that the lattice regularization is the existing most successful and 
systematic regularization scheme of field theory which covers not only perturbative 
but also non-perturbative regime. This is most successfully shown in lattice QCD. 
It was also shown numerically and analytically that dynamical lattice 
triangulation is the very powerful regularization scheme of two dimensional quantum 
gravity even with matter degrees of freedom for the observable like the fractal 
dimension of quantum surface\cite{2d-gravity1}. 

It is thus very natural to expect that the lattice regularization may play an 
important role even as a regularization scheme of unified theory of all interactions 
including gravity.  If we, however, try to consider to accommodate fermions and 
bosons on the lattice we face with the notorious chiral fermion difficulty. 
If we stick to obtain a chiral gauge theory, the difficulty may still remain 
or can be avoided by the new definition of chiral symmetry on the lattice 
{\it \`a la} Ginsparg-Wilson\cite{GW}. We claim that there is an alternative 
approach on this problem. 

The simplest version of the chiral fermion problem\cite{NN,KarstenSmit} could 
be phrased in the 
following: If we formulate a chiral invariant free Dirac Lagrangian on a lattice 
by simply replacing the derivative operator to a difference operator, we cannot 
avoid of having species doublers. This naive fermion formulation is equivalently 
transformed to the staggered fermion by the spin diagonalization\cite{KS}, 
which is equivalently transformed to the Kogut-Susskind 
fermion\cite{Kogut-Susskind} with an extra 
second derivative term\cite{G,KMN}. In fact all these formulations are 
equivalent each other and furthermore they are obtained from Dirac-K\"ahler 
fermion formalism which is formulated by differential form\cite{D-K}. 
In these frameworks the species doublers of the chiral fermion formulation on 
the lattice can turn into ``flavor'' degrees of freedom. 
It was then proposed that this ``flavor'' degrees of freedom is fundamentally related 
to the extended SUSY degrees of freedom, where the twisting procedure in 
the quantization of topological field theories is essentially equivalent 
to the Dirac-K\"ahler fermion mechanism\cite{KT}. It has then led to the proposal 
of a twisted superspace formalism \cite{KKU}.

It was shown that the quantized topological Yang-Mills action with instanton 
gauge fixing led twisted $N=2$ super Yang-Mills action where the BRST charge 
is equivalent to the scalar component of twisted supercharge\cite{W}. 
It was also shown that the two-dimensional version of quantized 
topological Yang-Mills action of the generalized gauge theory\cite{KW} with 
a two-dimensional instanton gauge fixing led to $N=2$ super Yang-Mills 
action\cite{KT}. In this investigation it was recognized that the twisting 
is nothing but the Dirac-K\"ahler fermion mechanism. Here is the explicit 
realization of the ``flavor'' suffixes of the Dirac-K\"ahler fermion as the 
$N=2$ extended SUSY suffixes. The $R$-symmetry of the $N=2$ twisted SUSY 
algebra is the ``flavor'' symmetry of Dirac-K\"ahler fermion.   
It turned out that this Dirac-K\"ahler twisting mechanism works universally 
in the quantization of topological field theory and an extended SUSY 
is generated. Furthermore the twisted superspace formulation is hidden behind 
the formulation. This was explicitly shown for $N=2$ twisted SUSY invariant 
BF action and Wess-Zumino action \cite{KKU}. 
In this context we can now recognize that 
the vector SUSY which was discovered in the quantization of 
topological fields theories is the vector counter part 
of the twisted SUSY\cite{vector-symm}. 
In this paper we formulate 
this $N=D=2$ twisted superspace on the lattice as 
a particular example of general formulation.    

Trials of formulating SUSY on the lattice has a long history which 
includes analyses of Wess-Zumino model with the ideas of Nicolai mapping, 
stochastic processes and others \cite{old-lattsusy}. 
In those early investigations the difficulty of lattice SUSY formulation 
was recognized since SUSY does not exist on the lattice due to the 
lack of Poincar\`e symmetry. 
The domain wall fermion idea triggered not only the investigations of chiral 
fermion problem on the lattice but also lattice SUSY investigations 
\cite{D-W-lattsusy}. 
There are deconstruction approaches of SUSY invariant lattices 
where the spatial lattice is created by ``orbifolding'' \cite{deconstruction}. 
The lattice Wess-Zumino model was investigated with the recent contexts by 
several authors \cite{other-lattsusy1}. 
There are some twisted SUSY approaches on the lattice where the scalar 
part of the nilpotent super charge is identified as the BRST 
charge\cite{twist-lattsusy}. A new type of fermionic symmetry was 
proposed for super Yang-Mills theory \cite{other-lattsusy2}.
There are many numerical lattice investigations to try to answer the dynamical 
questions of supersymmetric gauge theories, which are nicely reviewed in 
\cite{numerical-lattsusy} and references are therein.

Here we propose a new formulation of lattice SUSY which is different from any 
of past formulations in the following two points: Firstly we formulate a twisted 
superspace on the lattice. Secondly we introduce mild noncommutativity to preserve 
the lattice Leibniz rule. And then as a consequence we obtain exact lattice 
SUSY for all twisted supercharges.

There are two important well-known obstacles in formulating lattice SUSY: 
Firstly Poincar\`e invariance is lost on the lattice, thus SUSY is lost 
as well. Secondly Leibniz rule does not hold on the lattice.  
In the current approach superfields satisfy the algebraic relations of 
twisted SUSY with a difference operator. In other words SUSY holds exactly 
on the lattice with the help of superspace. 

The importance of the Leibniz rule for lattice SUSY is stressed 
in \cite{Leib-fujikawa}.
It has been recognized that the lattice Leibniz rule can be satisfied if 
we introduce mild noncommutativity between the difference operator and 
a function. This type of noncommutativity has been already investigated 
by several authors in the context of noncommutative differential form 
on a lattice\cite{non-comm-gauge}. 
Since differential form can be formulated on the lattice by the 
noncommutative geometry {\it \`a la} Connes\cite{Connes}, it is possible 
to formulate staggered fermion and Dirac-K\"ahler fermion formulation 
on the lattice in this framework \cite{non-comm-fermion}. 
In particular Clifford product was formulated on the lattice 
and thus Dirac-K\"ahler fermion can be successfully formulated 
with this noncommutative framework\cite{KK}. 
In this paper we use this noncommutative framework to establish 
twisted SUSY algebra. 

This paper is organized as follows: In Section \ref{Leibniz_rule} we first discuss the 
Leibniz rule and noncommutativity on the lattice. Then we summarize 
the continuum formulation of twisted $N=D=2$ SUSY and the 
corresponding superspace formulation in Section \ref{twist_cont}. We then reformulate 
the twisted superspace on the lattice in Section \ref{exact_lat}. We construct explicit 
examples of twisted SUSY invariant BF and Wess-Zumino actions in Section \ref{BF_WZ_lat}. 
We then summarize the results and discuss future problems in the last 
section.


\section{A new definition of the Leibniz rule on the lattice}
\label{Leibniz_rule}

One of the difficulties in formulating exact SUSY on a regular lattice
lies in the fact that the continuum Poincar\`e group is broken by the lattice
structure. The Lorentz group on a hypercubic lattice is reduced to the finite 
group of rotations by multiples of $\pi /2$ around the fundamental axis  and
translations are discretized to integer multiples of the lattice spacing.
As a consequence derivatives are replaced on the lattice by finite differences,
and these do not satisfy the Leibniz rule.  This is an important point, and
defining lattice derivatives that satisfy the Leibniz rule is 
a crucial step in establishing exact SUSY on a lattice. In view 
of that, we shall devote the present section to a detailed discussion of this
problem and show how  the Leibniz rule can be preserved on the lattice
at the price of introducing a mild noncommutativity in the definition of the
derivative operator.

The derivative $\partial_{\mu}$ is replaced on the lattice by the finite
difference $\upar_{+\mu}$ defined by:

\be
\upar_{+\mu} \Phi(x) = \Phi(x+2 \hat{n}_{\mu}) - \Phi(x),
\label{finite}
\eeq
where $\hat{n}_{\mu}$ corresponds to the shift of one lattice spacing in the $\mu$ 
direction. The difference is taken on two lattice spacings for reasons that will
be later clarified. The definition (\ref{finite}) does not satisfy the Leibniz
rule. In fact it is immediate to verify that 

\be
\upar_{+\mu}(\Phi_1(x) \Phi_2(x)) \not = \left( \upar_{+\mu} \Phi_1(x)\right) 
\Phi_2(x) + \Phi_1(x) \left( \upar_{+\mu} \Phi_2(x) \right),
\label{noliebnitz}
\eeq
and that we have instead the following relation:
\be
\upar_{+\mu}\left( \Phi_1(x) \Phi_2(x)\right) = \left( \upar_{+\mu} \Phi_1(x)
 \right) \Phi_2(x) + \Phi_1(x+2 \hat{n}_{\mu}) \left( \upar_{+\mu} \Phi_2(x)
 \right).
 \label{lrule}
 \eeq
 
 Eq. (\ref{lrule}) defines a modified Leibniz rule satisfied by the finite
difference $\upar_{+\mu}$ on the lattice. The new rule involves a certain degree of
 noncommutativity, in the sense that as $\upar_{+\mu}$ moves to the right of a
  function (for instance $\Phi_1(x)$  in (\ref{lrule})) it produces a 
 corresponding shift in the argument. 
We may then define the lattice difference operator which carries the shift $2\m$ 
as follows:
\be
\upar_{+\mu} \Phi(x) =  \arrowdel_{+\mu} \Phi(x) - \Phi(x+2\m) \arrowdel_{+\mu},
\label{arrowdel}
\end{equation}
which reproduces the lattice version of modified Leibniz rule by the successive 
operation of $\arrowdel_{+\mu}$ to the product of two functions. 
As we shall see in the following sections
 this is a general feature shared by all symmetry operators consistently defined
 on the lattice, including obviously the SUSY ones.
 
The violation of the Leibniz rule, and the modified rule(\ref{lrule}) can be
easily understood. In fact  the infinitesimal
generator $P_{\mu}= -i \partial_{\mu}$ is associated on the lattice with a finite
translation generated by the shift operator 
\be
T(\m) = \ee^{- \ii \hat{n}_\mu \cdot \vec{P}},
\label{ti}
\eeq 
where $\hat{n}_\mu$ is a unit vector in the $\mu$ direction:
$(\hat{n}_\mu)^{\nu}=\delta_{\mu}^{\nu}$.
The shift operator generates a corresponding shift in the argument of a field 
$\Phi(x)$ on the lattice:
\be
T(  \m) \Phi(x) = \Phi(x-  \hat{n}_{\mu}) T( \m).
\label{new}
\eeq
and can be used to define the difference operator. In fact if we define:
\be
\Delta_{+\mu} = - T(2 \m ),
\label{deltapiu}
\eeq
 we have:
\be
\left[\Delta_{+\mu}  , \Phi(x)\right] =  T(2\m)~\upar_{+\mu} \Phi(x). 
\label{newcomm}
\eeq
Eqs (\ref{arrowdel}) and (\ref{deltapiu}) are two distinct ways of introducing
the finite difference $\upar_{+\mu} \Phi(x)$ as a result of some type of
commutator. In that respect they are both the lattice equivalent of the
commutator defining the partial derivative in the continuum: 
\be
[P_{\mu}, \Phi(x) ] = -i \partial_{\mu} \Phi. 
\label{comm}
\eeq
The two operators $\arrowdel_{+\mu}$ and $\Delta_{+\mu}$ are obviously related.
In general, if an operator $\Oc$ satisfies the commutation relations
\be
\left[ \Oc , \Phi(x) \right] = T ( \a_{\Oc} )\udelta_{\Oc} \Phi(x) ,
\label{O}
\eeq
it follows that the operator $\overrightarrow{\uOc}$ defined by
\be
\Oc =  T ( \a_{\Oc} ) ~\overrightarrow{\uOc} ,
\label{Odef}
\eeq
obeys the following ``shifted'' commutator:
\be
\udelta_{\Oc} \Phi(x)
 = \overrightarrow{\uOc} \Phi(x) - \Phi(x + \a_{\Oc})  
\overrightarrow{\uOc}.
\label{arrowO}
\eeq
In the last equations $\a_{\Oc}$ is a shift of the two dimensional lattice
associated to the operator $\Oc$.
In this article we discriminate the operators and fields which carry the 
``shift'' from normal ones by underlining the corresponding operators and fields. 
All symmetry operators on the lattice that we shall consider in the rest of the
paper will satisfy commutation relations of the form (\ref{O}) or, in their 
``arrowed'' version, of the form (\ref{arrowO}). Needless to say, the two
formalism are entirely equivalent and the use of one or the other will be
dictated by convenience.
The modified Leibniz rule for a general operator on the lattice reads
\be
\udelta_{\Oc} \left( \Phi_1(x) \Phi_2(x) \right) 
= \left(\udelta_{\Oc} \Phi_1(x)
\right) \Phi_2(x) +  \Phi_1(x + \a_{\Oc}) 
\left(\udelta_{\Oc} \Phi_2(x) \right).
\label{lrulegen}
\eeq
This is a straightforward consequence of (\ref{arrowO}) or equivalently, if we 
take the definition (\ref{newcomm}), of the Leibniz rule for the commutator.

In addition to the finite difference operator $\Delta_{+\mu}$  that carries a
shift in the positive direction of the $\mu$-axis one can introduce a negative 
difference operator $\Delta_{-\mu}$ defined by
\be
\Delta_{-\mu} = T(-2 \m).
\label{deltaminus}
\eeq
Correspondingly we introduce a finite difference 
\be
\upar_{-\mu}\Phi(x)= \Phi(x) - \Phi(x-2\m),
\label{finmeno}
\eeq
which is obtained from the commutator
\be
\left[\Delta_{-\mu}  , \Phi(x)\right] =  T(-2\m)~\upar_{-\mu} \Phi(x),
\label{negcomm}
\eeq
or from the ``shifted commutator'' using the operator $\arrowdel_{-\mu}$:
\be
\upar_{-\mu} \Phi(x) =  \arrowdel_{-\mu} \Phi(x) - \Phi(x-2\m) \arrowdel_{-\mu}=
\Phi(x) - \Phi(x-2\m).
\label{arrowdelm}
\end{equation}
Notice that from the definition of the ``arrowed'' operators (\ref{Odef}) 
and Eqs
(\ref{deltaminus}) and (\ref{deltapiu}) we find:
\be
 \arrowdel_{\pm\mu} = \mp 1 , 
\label{arrowdelone}
\eeq
consistently with (\ref{arrowdel}) and (\ref{arrowdelm}).

In the following sections we shall also use instead of the variation 
$\udelta_{\Oc}$ of (\ref{O}) a symmetric variation 
$\udelta^{(s)}_{\Oc}$ defined
by:
\be
\left[ \Oc , \Phi(x) \right] = T ( \frac{\a_{\Oc}}{2} ) 
\udelta^{(s)}_{\Oc} 
\Phi(x) T ( \frac{\a_{\Oc}}{2} ) ,
\label{Os}
\eeq
or equivalently by 
\be
\udelta^{(s)}_{\Oc} \Phi(x) = \overrightarrow{\uOc} 
\Phi(x-\frac{\a_{\Oc}}{2}) -
 \Phi(x + \frac{\a_{\Oc}}{2}) \overrightarrow{\uOc} .
\label{arrowOs}
\eeq
Clearly $\udelta^{(s)}_{\Oc} \Phi(x)$ and $\udelta_{\Oc} \Phi(x)$ differ
simply by a shift in the argument:
\be
\udelta^{(s)}_{\Oc} \Phi(x)=\udelta_{\Oc} \Phi(x-\frac{\a_{\Oc}}{2}).
\label{s}
\eeq
The advantage of $\delta^{(s)}_{\Oc}$ is to make all formulas left-right
symmetric, for instance the modified Leibniz rule becomes:
\be
\udelta^{(s)}_{\Oc} \left( \Phi_1(x) \Phi_2(x) \right) = 
\left(\udelta^{(s)}_{\Oc} \Phi_1(x)
\right) \Phi_2(x-\frac{\a_{\Oc}}{2}) +  \Phi_1(x + \frac{\a_{\Oc}}{2}) 
\left(\udelta_{\Oc} \Phi_2(x) \right).
\label{lrulegens}
\eeq
Besides, with the definition (\ref{Os}) the positive and negative finite
difference operators $\Delta_{+\mu}$ and $\Delta_{-\mu}$ generate the same
symmetric finite difference $\upar_{\mu}^{(s)}$:
\be
\left[ \Delta_{\pm \mu} , \Phi(x) \right] = T ( \pm \m ) \upar_{\mu}^{(s)}
\Phi(x) T ( \pm \m ),
\label{Deltas}
\eeq
where
\be
\upar_{\mu}^{(s)} \Phi(x) = \Phi(x+ \m ) - \Phi(x - \m).
\label{sdiff}
\eeq


\section{$N=2$ Twisted SUSY in continuum two dimensions}
\label{twist_cont}


In this section, we summarize the continuum formulation of $N=2$ twisted 
superspace which was proposed recently\cite{KKU}.
Throughout this paper, we consider two-dimensional Euclidean space-time. 

We first introduce two-dimensional $N=D=2$ SUSY algebra:
\begin{equation}
\begin{split}
\{Q_{\alpha i},Q_{\beta j} \}
 &=2\delta_{ij} {\gamma^\mu}_{\alpha \beta}P_\mu,
\end{split}
\label{eq:N=2 algebra}
\end{equation}
where $Q_{\alpha i}$ is supercharge, where the left-indices $\alpha(=1,2)$ 
and the right-indices $i(=1,2)$ are Lorentz spinor and internal spinor 
suffixes labeling two different $N=2$ supercharges, respectively. We can take 
these operators to be Majorana in two dimensions. $P_\mu$ is generator of 
translation. 

Since the above supercharges have double spinor indices, we can decompose them 
into the following scalar, vector and pseudo-scalar components which 
we call twisted supercharges:
\begin{align}
 Q_{\alpha i}= \left(\mathbf{1} s + \gamma^\mu s_\mu + \gamma^5
\tilde{s}\right)_{\alpha i}.
\end{align}
%
%
Then the relations (\ref{eq:N=2 algebra}) can be rewritten by the twisted
generators:
\begin{equation}
\begin{split}
\{s,s_\mu\}&=P_\mu,\ \ \ 
\{\tilde{s},s_\mu\}=-\epsilon_{\mu\nu}P^\nu, \\ 
s^2=\tilde{s}^2&=\{s,\tilde{s}\}=\{s_\mu,s_\nu\}=0. \\
\end{split}
\label{eq:N=2 T-algebra}
\end{equation}
This is the twisted $N=D=2$ SUSY algebra. 
In this paper we do not consider $J$ and $R$ symmetry of the $N=2$ algebra. 
The notations of gamma matrices are given in Appendix \ref{app:notation}.

Similar to the supercharges we can introduce twisted super parameters as 
\begin{align}
 \theta_{\alpha i}= \frac{1}{2}\left(\mathbf{1} \theta + 
\gamma^\mu \theta_\mu + \gamma^5 \tilde{\theta}\right)_{\alpha i},
\end{align}
then we can define $N=2$ super symmetry transformation as
\begin{equation}
\delta_\theta \ = \ \theta_{\alpha i} Q_{\alpha i} \ = \  
\theta s+\theta^\mu s_\mu+\tilde{\theta}\tilde{s}.
\label{def_delta_theta}
\end{equation}

\subsection{Twisted superspace and superfield}
We consider the following super group element:
\begin{align}
G(x^\mu,\theta,\theta^\mu,\tilde{\theta})
 =\ee^{i(-x^\mu P_\mu + \theta s +\theta^\mu s_\mu +\tilde{\theta}\tilde{s})},
\end{align}
where $\theta$'s are anticommuting parameters. Twisted $N=D=2$ superspace
is defined in the parameter space of 
$(x^\mu,\theta,\theta^\mu,\tilde{\theta})$.

By using the relations (\ref{eq:N=2 T-algebra}), we can show the following 
relation:
\begin{align}
G(0,\xi,\xi^\mu,\tilde{\xi})
G(x^\mu,\theta,\theta^\mu,\tilde{\theta})
=G(x^\mu+a^\mu,\theta+\xi,\theta^\mu+\xi^\mu,\tilde{\theta}+\tilde{\xi}),
\end{align}
where $a^\mu=\frac{i}{2}\xi\theta^\mu+\tfrac{i}{2}\xi^\mu\theta
 +\tfrac{i}{2}\epsilon^\mu{_\nu}\xi^\nu\tilde{\theta}
 +\tfrac{i}{2}\epsilon^\mu{_\nu}\tilde{\xi}\theta^\nu$.
This multiplication induces a shift transformation in superspace
$(x^\mu,\theta,\theta^\mu,\tilde{\theta})$:
\begin{align}
(x^\mu,\theta,\theta^\mu,\tilde{\theta})
\rightarrow
(x^\mu+a^\mu,\theta+\xi,\theta^\mu+\xi^\mu,\tilde{\theta}+\tilde{\xi}),
\end{align}
which is generated by the following differential operators $Q$, 
$Q_\mu$, and $\tilde{Q}$:
\begin{equation}
\begin{split}
Q
 &=\frac{\partial}{\partial \theta}
  +\frac{i}{2}\theta^\mu\partial_\mu,\\
Q_\mu
 &=\frac{\partial}{\partial \theta^\mu}
  +\frac{i}{2}\theta\partial_\mu
  -\frac{i}{2}\tilde{\theta}\epsilon_{\mu\nu}\partial^\nu,\\
\tilde{Q}
 &=\frac{\partial}{\partial \tilde{\theta}}
  -\frac{i}{2}\theta^\mu\epsilon_{\mu\nu}\partial^\nu.\\
\end{split}
\label{eq:Q-operator}
\end{equation}
Indeed we find
\begin{align}
\delta_\xi
\begin{pmatrix}
x^\mu\\
\theta\\
\theta^\mu\\
\tilde{\theta}
\end{pmatrix}
=(\xi Q+\xi^\mu Q_\mu+\tilde{\xi}\tilde{Q})
\begin{pmatrix}
x^\mu\\
\theta\\
\theta^\mu\\
\tilde{\theta}
\end{pmatrix}
=
\begin{pmatrix}
a^\mu\\
\xi\\
\xi^\mu\\
\tilde{\xi}
\end{pmatrix}
.
\end{align}
These operators satisfy the following relations:
\begin{equation}
\begin{split}
& \{Q,Q_\mu\}=i\partial_\mu,\
 \{\tilde{Q},Q_\mu\}=-i\epsilon_{\mu\nu}\partial^\nu,\\
& Q^2=\tilde{Q}^2=\{Q,\tilde{Q}\}=\{Q_\mu,Q_\nu\}=0.
\end{split}
\label{N=2 left operator algebra}
\end{equation}

The general scalar superfields in twisted $N=D=2$ superspace are defined as
the functions of $(x^\mu,\theta,\theta^\mu,\tilde{\theta})$, and can be
expanded as follows:
\begin{equation}
\begin{split}
F(x^\mu,\theta,\theta^\mu,\tilde{\theta})
 =&\phi(x)+\theta^\mu\phi_\mu(x)+\theta^2 \tilde{\phi}(x)\\
 &+\theta\Bigl( \psi(x)+\theta^\mu\psi_\mu(x)
   +\theta^2 \tilde{\psi}(x)\Bigr)\\
 &+\tilde{\theta}\Big(\chi(x)+\theta^\mu\chi_\mu(x)
   +\theta^2 \tilde{\chi}(x)\Bigr)\\
 &+\theta\tilde{\theta}\Bigl(\lambda(x)+\theta^\mu\lambda_\mu(x)
   +\theta^2 \tilde{\lambda}(x)\Bigr),\\
\end{split}
\label{eq:F}
\end{equation}
where the leading component $\phi(x)$ can be taken to be not only bosonic
but also fermionic.

The transformation law of the superfield $F$ is defined as follows:
\begin{equation}
\begin{split}
\delta_\xi F(x^\mu,\theta,\theta^\mu,\tilde{\theta})
=&\delta_\xi\phi(x)+\theta^\mu \delta_\xi\phi_\mu(x)+\theta^2
\delta_\xi\tilde{\phi}(x)\\
&+\theta\Bigl(\delta_\xi\psi(x)+\theta^\mu \delta_\xi\psi_\mu(x)
   +\theta^2 \delta_\xi\tilde{\psi}(x)\Bigr)\\
&+\tilde{\theta}\Big(\delta_\xi\chi(x)+\theta^\mu \delta_\xi\chi_\mu(x)
  +\theta^2 \delta_\xi\tilde{\chi}(x)\Bigr)\\
&+\theta\tilde{\theta}\Bigl(\delta_\xi\lambda(x)+\theta^\mu
\delta_\xi\lambda_\mu(x)
  +\theta^2 \delta_\xi\tilde{\lambda}(x)\Bigr)\\
=&(\xi Q+\xi^\mu
Q_\mu+\tilde{\xi}\tilde{Q})F(x^\mu,\theta,\theta^\mu,\tilde{\theta}),
\end{split}
\label{eq:F-trans}
\end{equation}
where $Q$, $Q_\mu$ and $\tilde{Q}$ are the differential operators
(\ref{eq:Q-operator}). The transformation laws of the component fields
$\phi^A(x)=(\phi(x),\phi_\mu(x),\tilde{\phi}(x),\dots)$ are obtained 
by comparing coefficients of the same superspace parameters in 
(\ref{eq:F-trans}) (see Table \ref{T-laws_F}). Those
transformation laws lead to the following supercharge algebra:
\begin{equation}
\begin{split}
& \{s,s_\mu\}=-i\partial_\mu,\
 \{\tilde{s},s_\mu\}=i\epsilon_{\mu\nu}\partial^\nu,\\
& s^2=\tilde{s}^2=\{s,\tilde{s}\}=\{s_\mu,s_\nu\}=0,
\end{split}
\label{twisted_susy_algebra_1}
\end{equation}
which are the same of (\ref{eq:N=2 T-algebra}) with $P_\mu=-i\partial_\mu$.
It should be noted that the only difference between the supercharge 
algebra and the corresponding differential operator algebra is 
a sign difference for the derivative. 

\begin{table}[htbp]
\begin{center}
\renewcommand{\arraystretch}{1.3}
\begin{tabular}{c|c|c|c}
\hline
$\phi^A$ & $s\phi^A$ & $s_\mu\phi^A$ & $\tilde{s}\phi^A$ \\
\hline
$\phi$
 & $\psi$
 & $\phi_\mu$
 & $\chi$ \\
$\phi_\rho$
 & $-\psi_\rho-\frac{i}{2}\partial_\rho\phi$
 & $-\epsilon_{\mu\rho}\tilde{\phi}$
 & $-\chi_\rho+\frac{i}{2}\epsilon_{\rho\sigma}\partial^\sigma\phi$ \\
$\tilde{\phi}$
 & $\tilde{\psi}+\frac{i}{2}\epsilon^{\rho\sigma}\partial_\rho\phi_\sigma$
 & $0$
 & $\tilde{\chi}-\frac{i}{2}\partial^\rho\phi_\rho$ \\
\hline
$\psi$
 & $0$
 & $\psi_\mu-\frac{i}{2}\partial_\mu\phi$
 & $\lambda$ \\
$\psi_\rho$
 & $-\frac{i}{2}\partial_\rho\psi$
 & $-\epsilon_{\mu\rho}\tilde{\psi}+\frac{i}{2}\partial_\mu\phi_\rho$
 & $-\lambda_\rho+\frac{i}{2}\epsilon_{\rho\sigma}\partial^\sigma\psi$ \\
$\tilde{\psi}$
 & $\frac{i}{2}\epsilon^{\rho\sigma}\partial_\rho\psi_\sigma$
 & $-\frac{i}{2}\partial_\mu\tilde{\phi}$
 & $\tilde{\lambda}-\frac{i}{2}\partial^\rho\psi_\rho$ \\
\hline
$\chi$
 & $-\lambda$
 & $\chi_\mu+\frac{i}{2}\epsilon_{\mu\nu}\partial^\nu\phi$
 & $0$ \\
$\chi_\rho$
 & $\lambda_\rho-\frac{i}{2}\partial_\rho\chi$
 & $-\epsilon_{\mu\rho}\tilde{\chi}-\frac{i}{2}
\epsilon_{\mu\nu}\partial^\nu \phi_\rho$
 & $+\frac{i}{2}\epsilon_{\rho\sigma}\partial^\sigma\chi$ \\
$\tilde{\chi}$
 & $-\tilde{\lambda}+\frac{i}{2}\epsilon^{\rho\sigma}\partial_\rho\chi_\sigma$
 & $\frac{i}{2}\epsilon_{\mu\nu}\partial^\nu\tilde{\phi}$
 & $-\frac{i}{2}\partial^\rho\chi_\rho$ \\
\hline
$\lambda$
 & $0$
 & $\lambda_\mu+\frac{i}{2}\partial_\mu\chi+
 \frac{i}{2}\epsilon_{\mu\nu}\partial^\nu\psi$
 & $0$ \\
$\lambda_\rho$
 & $-\frac{i}{2}\partial_\rho\lambda$
 & $-\epsilon_{\mu\rho}\tilde{\lambda}
  -\frac{i}{2}\partial_\mu\chi_\rho
  -\frac{i}{2}\epsilon_{\mu\nu}\partial^\nu\psi_\rho$
 & $\frac{i}{2}\epsilon_{\rho\sigma}\partial^\sigma\lambda$ \\
$\tilde{\lambda}$
 & $\frac{i}{2}\epsilon^{\rho\sigma}\partial_\rho\lambda_\sigma$
 & $\frac{i}{2}\partial_\mu\tilde{\chi}
  +\frac{i}{2}\epsilon_{\mu\nu}\partial^\nu\tilde{\psi}$
 & $-\frac{i}{2}\partial^\rho\lambda_\rho$ \\
\hline
\end{tabular}
\caption{$N=2$ twisted SUSY transformation laws of the component fields of $F$.}
\label{T-laws_F}
\end{center}
\end{table}

Given the transformation laws of the component fields, we can expand the
superfield $F$ as follows:
\begin{equation}
\begin{split}
F(x^\mu,\theta,\theta^\mu,\tilde{\theta})
 &=e^{\delta_\theta}\phi(x)\\
 &=\phi(x)+\delta_\theta\phi(x)
  +\frac{1}{2}\delta_\theta{^2}\phi(x)
  +\frac{1}{3!}\delta_\theta{^3}\phi(x)
  +\frac{1}{4!}\delta_\theta{^4}\phi(x),
\end{split}
\end{equation}
where $\delta_\theta$ is defined in (\ref{def_delta_theta}).

As we have seen, the differential operators in (\ref{eq:Q-operator}) generate
the shift transformation of superspace induced by left multiplication
$G(0,\xi,\xi^\mu,\tilde{\xi})G(x^\mu,\theta,\theta^\mu,\tilde{\theta})$. On
the other hand, there exist the differential operators which generate the
shift transformation induced by right multiplication
$G(x^\mu,\theta,\theta^\mu,\tilde{\theta})G(0,\xi,\xi^\mu,\tilde{\xi})$:
\begin{equation}
\begin{split}
D
 &=\frac{\partial}{\partial \theta}
  -\frac{i}{2}\theta^\mu\partial_\mu,\\
D_\mu
 &=\frac{\partial}{\partial \theta^\mu}
  -\frac{i}{2}\theta\partial_\mu
  +\frac{i}{2}\tilde{\theta}\epsilon_{\mu\nu}\partial^\nu,\\
\tilde{D}
 &=\frac{\partial}{\partial \tilde{\theta}}
  +\frac{i}{2}\theta^\mu\epsilon_{\mu\nu}\partial^\nu ,
\end{split}
\label{eq:D-operator}
\end{equation}
which satisfy the relations:
\begin{equation}
\begin{split}
\{D,D_\mu\}&=-i\partial_\mu,\
\{\tilde{D},D_\mu\}=i\epsilon_{\mu\nu}\partial^\nu,\\
D^2&=\tilde{D}^2=\{D,\tilde{D}\}=\{D_\mu,D_\nu\}=0,\\
\end{split}
\label{N=2 right operator algebra}
\end{equation}
where only the sign of $\partial_\mu$ is changed from 
the left operator algebra (\ref{N=2 left operator algebra}).
$Q^A=(Q,Q_\mu,\tilde{Q})$ and $D^A=(D,D_\mu,\tilde{D})$ anticommute:
\begin{align}
\{Q^A,D^B\}=0.
\end{align}


\subsection{Chiral and anti-chiral superfields} 

The chiral conditions for the twisted chiral superfield can be given by
\begin{align}
D\Psi(x^\mu,\theta,\tilde{\theta},\theta^\mu) \ = \ 0 , \ \ 
\tilde{D}\Psi(x^\mu,\theta,\tilde{\theta},\theta^\mu) \ = \ 0.
\label{scalar_condition_1}
\end{align}
The details of the twisted chiral superfield formulation can be 
found in \cite{KKU}.
It is convenient to rewrite the chiral conditions by using the operator 
which satisfy the following relations for the differential operators:
\begin{align}
UDU^{-1} \ &= \ \frac{\partial}{\partial \theta} , \ \ 
U\tilde{D}U^{-1} \ = \ \frac{\partial}{\partial \tilde{\theta}}, \nonumber \\
U^{-1} D_\mu U \ &= \ \frac{\partial}{\partial \theta^\mu}, 
\label{diagonalization}
\end{align}
where 
\begin{equation}
U \ = \ \ee^{-\frac{i}{2}(\theta\theta^\mu\partial\mu - 
           \tilde{\theta} \epsilon^{\mu\nu}\theta_\mu\partial_\nu)}.
\label{def_of_U}
\end{equation} 
Then the chiral conditions (\ref{scalar_condition_1}) can be transformed 
into 
\begin{align}
\frac{\partial}{\partial \theta} \ 
U \Psi(x^\mu,\theta,\tilde{\theta},\theta^\mu) U^{-1} \ = \ 0, \ \ 
\frac{\partial}{\partial \tilde{\theta}} \ 
U \Psi(x^\mu,\theta,\tilde{\theta},\theta^\mu) U^{-1} \ = \ 0, 
\label{scalar_condition_2}
\end{align}
which leads 
\begin{align}
U \Psi(x^\mu,\theta,\tilde{\theta},\theta^\mu) U^{-1} \ = \ 
\Psi^{\prime} (x^\mu,\theta^\mu).
\label{solution_of_scalar_condition_2}
\end{align}
Then the solution for the original chiral condition 
(\ref{scalar_condition_1}) is obtained as 
\begin{align}
\Psi(x^\mu,\theta,\tilde{\theta},\theta^\mu)  \ = \ 
U^{-1} \Psi^{\prime} (x^\mu,\theta^\mu) U \ = \ 
\Psi^\prime (z^\mu,\theta^\mu), 
\label{solution_of_scalar_condition_3}
\end{align}
where 
\begin{equation}
z^\mu=x^\mu+\frac{i}{2}\theta\theta^\mu-\frac{i}{2}\epsilon^\mu{_\nu}\theta
^\nu\tilde{\theta}. 
\label{def_of_z}
\end{equation}
This solution can be expanded as follows:
\begin{equation}
\begin{split}
\Psi(x^\mu,\theta,\tilde{\theta},\theta^\mu)&=\Psi^\prime (z^\mu,\theta^\mu)\\
 &=\phi(z)+\theta^\mu \psi_\mu(z) +\theta^2 \tilde{\phi}(z)\\
 &=\phi(x)+\theta^\mu\psi_\mu(x)\\
 &\quad
  +\tfrac{i}{2}\theta\theta^\mu \partial_\mu \phi(x)
  +\theta^2\tilde{\phi}(x)
  -\tfrac{i}{2}\epsilon^\mu{_\nu}\theta^\nu\tilde{\theta}\partial_\mu
\phi(x)\\
 &\quad
  +\tfrac{i}{2}\theta\theta^2\epsilon^{\mu\nu}\partial_\mu\psi_\nu(x)
  -\tfrac{i}{2}\theta^2\tilde{\theta}\partial^\mu\psi_\mu(x)
  +\tfrac{1}{4}\theta^4\partial^2 \phi(x),
\label{psi_superfield}  
\end{split}
\end{equation}
where $\theta^2=\frac{1}{2}\epsilon_{\mu\nu}\theta^\mu\theta^\nu$ 
and $\theta^4=\theta\tilde{\theta}\theta^2$. 

Anti-chiral conditions for the anti-chiral superfield are given by 
\begin{equation}
D_\mu\overline{\Psi}(x^\mu,\theta,\tilde{\theta},\theta^\mu) \ = \ 0.
\label{anti-chiral-condition_1}
\end{equation}
Similar to the chiral condition (\ref{scalar_condition_2}), 
we can transform the original anti-chiral condition 
(\ref{anti-chiral-condition_1}) into the following form: 
\begin{equation}
U^{-1} D_\mu U 
U^{-1} \overline{\Psi}(x^\mu,\theta,\tilde{\theta},\theta^\mu)U \ = \ 
\frac{\partial}{\partial \theta^\mu} 
U^{-1} \overline{\Psi}(x^\mu,\theta,\tilde{\theta},\theta^\mu)U \ = \ 0,
\label{anti-chiral-condition_2}
\end{equation}
which leads 
\begin{align}
U^{-1} \overline{\Psi}(x^\mu,\theta,\tilde{\theta},\theta^\mu)U \ = \ 
\overline{\Psi}^\prime (x^\mu,\theta,\tilde{\theta}).
\label{solution_of_anti-chiral_condition_2}
\end{align}
Then the solution for the original anti-chiral condition 
(\ref{anti-chiral-condition_1}) is obtained as 
\begin{align}
\overline{\Psi}(x^\mu,\theta,\tilde{\theta},\theta^\mu) \ = \ 
U \overline{\Psi}^\prime (x^\mu,\theta,\tilde{\theta}) U^{-1} \ = \ 
\overline{\Psi}^\prime (\tilde{z}^\mu,\theta,\tilde{\theta}), 
\label{solution_of_anti-chiral_condition_3}
\end{align}
where 
\begin{equation}
\tilde{z}^\mu=x^\mu-\frac{i}{2}\theta\theta^\mu+
\frac{i}{2}\epsilon^\mu{_\nu}\theta^\nu\tilde{\theta}. 
\label{def_of_tilde_z}
\end{equation}
This solution can be expanded as follows:
\begin{equation}
\begin{split}
\overline{\Psi}(x^\mu,\theta,\tilde{\theta},\theta^\mu) \ 
&= \ \overline{\Psi}^\prime (\tilde{z}^\mu,\theta,\tilde{\theta}) \\
 &=\varphi(\tilde{z})+\theta \chi(\tilde{z})
  +\tilde{\theta}\tilde{\chi}(\tilde{z})
  +\theta \tilde{\theta}\tilde{\varphi}(\tilde{z})\\
 &=\varphi(x)+\theta\chi(x)+\tilde{\theta}\tilde{\chi}(x)\\
 &\quad
  +\theta\tilde{\theta}\tilde{\varphi}(x)
  -\tfrac{i}{2}\theta\theta^\mu \partial_\mu \varphi(x)
  +\tfrac{i}{2}\epsilon^\mu{_\nu}\theta^\nu\tilde{\theta}\partial_\mu
\varphi(x)\\
 &\quad
  -\tfrac{i}{2}\theta\theta^\mu\tilde{\theta}
   (\epsilon_{\mu\nu}\partial^\nu\chi(x)+\partial_\mu\tilde{\chi}(x))
  +\tfrac{1}{4}\theta^4\partial^2 \varphi(x).
\label{psibar_superfield}
\end{split}
\end{equation}

The SUSY transformations of the chiral and anti-chiral 
superfields are given by 
\begin{align}
s_A \Psi(x^\mu,\theta,\tilde{\theta},\theta^\mu) \ &= \ 
Q_A \Psi(x^\mu,\theta,\tilde{\theta},\theta^\mu), \nonumber \\  
s_A \overline{\Psi}(x^\mu,\theta,\tilde{\theta},\theta^\mu) \ &= \ 
Q_A \overline{\Psi}(x^\mu,\theta,\tilde{\theta},\theta^\mu), 
\label{susy_trasformation_1}
\end{align}
where $s_A = (s,\tilde{s},s_\mu )$ and $ Q_A = (Q,\tilde{Q},Q_\mu)$. 
These SUSY transformation can be transformed into the 
following form by using the operator (\ref{def_of_U}): 
\begin{align}
s_A U \Psi(x^\mu,\theta,\tilde{\theta},\theta^\mu)U^{-1} \ &= \ 
UQ_AU^{-1} U\Psi(x^\mu,\theta,\tilde{\theta},\theta^\mu)U^{-1},\nonumber \\  
s_A U^{-1}\overline{\Psi}(x^\mu,\theta,\tilde{\theta},\theta^\mu)U \ &= \ 
U^{-1}Q_AU U^{-1}\overline{\Psi}(x^\mu,\theta,\tilde{\theta},\theta^\mu)U, 
\label{susy_trasformation_2}
\end{align}
which can be equivalently written as 
\begin{align}
s_A \Psi^{\prime} (x^\mu,\theta^\mu) \ &= \ 
Q^{\prime}_A \Psi^{\prime} (x^\mu,\theta^\mu),\nonumber \\  
s_A \overline{\Psi}^\prime (x^\mu,\theta,\tilde{\theta}) \ &= \ 
Q^{\prime\prime}_A \overline{\Psi}^\prime (x^\mu,\theta,\tilde{\theta}), 
\label{susy_trasformation_3}
\end{align}
where $Q^{\prime}_A=UQ_AU^{-1}$ and $Q^{\prime \prime}_A=U^{-1}Q_AU$ 
are given by 
\begin{align}
Q^{\prime} \ &= \ \frac{\partial}{\partial \theta}
  + i \theta^\mu\partial_\mu, \ 
&\tilde{Q}^{\prime}& \ = \ \frac{\partial}{\partial \tilde{\theta}}
  - i \theta^\mu\epsilon_{\mu\nu}\partial^\nu, \ 
&Q^{\prime}_\mu& \ = \ \frac{\partial}{\partial \theta^\mu}, 
\label{susy_trasformation_4}\\
Q^{\prime\prime} \ &= \ \frac{\partial}{\partial \theta}, \ 
&\tilde{Q}^{\prime\prime}& \ = \ \frac{\partial}{\partial \tilde{\theta}}, \ 
&Q^{\prime\prime}_\mu& \ = \ \frac{\partial}{\partial \theta^\mu} 
 + i \theta\partial_\mu - i \tilde{\theta}\epsilon_{\mu\nu}\partial^\nu. 
\label{susy_trasformation_5}
\end{align}

The twisted $N=2$ SUSY transformations of the chiral set 
$(\phi,\psi_\mu,\tilde{\phi})$ and anti-chiral set 
$(\varphi,\chi,\tilde{\chi},\tilde{\varphi})$ are given in Table \ref{T-laws_chiral}.

\begin{table}[htbp]
\begin{center}
\begin{tabular}{c|c|c|c}
\hline
$\phi^A$ & $s\phi^A$ & $s_\mu\phi^A$ & $\tilde{s}\phi^A$\\
\hline
$\phi$ & $0$ & $\psi_\mu$ & $0$\\
$\psi_\rho$ & $-i\partial_\rho \phi$ & $-\epsilon_{\mu\rho}\tilde{\phi}$
  & $i\epsilon_{\rho\nu}\partial^\nu \phi$\\
$\tilde{\phi}$ & $i\epsilon^{\rho\sigma}\partial_\rho\psi_\sigma$
  & $0$ & $-i\partial^\rho \psi_\rho$\\
\hline
$\varphi$ & $\chi$ & $0$ & $\tilde{\chi}$\\
$\chi$ & $0$ & $-i\partial_\mu\varphi$ & $\tilde{\varphi}$\\
$\tilde{\chi}$ & $-\tilde{\varphi}$ 
& $i\epsilon_{\mu\nu}\partial^\nu \varphi$ & $0$\\
$\tilde{\varphi}$ & $0$
 & $i\epsilon_{\mu\nu}\partial^\nu \chi+i\partial_\mu\tilde{\chi}$ & $0$\\  
\hline
\end{tabular}
\caption{$N=2$ twisted SUSY transformation of chiral and anti-chiral 
twisted super multiplet $(\phi,\psi_\mu,\tilde{\phi})$ and 
$(\varphi,\chi,\tilde{\chi},\tilde{\varphi})$.}
\label{T-laws_chiral}
\end{center}
\end{table}
After obtaining the twisted $N=2$ SUSY transformation, we find 
the following natural relations:
\begin{equation}
\begin{split}
\Psi^\prime (x^\mu,\theta^\mu) & = \ee^{\theta^\mu s_\mu}\phi(x) \\
 &=\phi(x)+\theta^\mu s_\mu \phi(x) +\theta^2 s_2 s_1 \phi(x)\\
 &=\phi(x)+\theta^\mu \psi_\mu(x) +\theta^2 \tilde{\phi}(x),\\
 \overline{\Psi}^\prime (\tilde{x}^\mu,\theta,\tilde{\theta}) &=
 \ee^{\theta s + \tilde{\theta} \tilde{s}}\psi(x) \\
 &=\varphi(x)+\theta s \varphi(x)
  +\tilde{\theta}\tilde{s}\varphi(x)
  +\theta \tilde{\theta} \tilde{s} s \varphi(x) \\
 &=\varphi(x)+\theta \chi(x)
  +\tilde{\theta}\tilde{\chi}(x)
  +\theta \tilde{\theta}\tilde{\varphi}(x),\\
\label{chi_anti-chi_superfield}
\end{split}
\end{equation}
which shows that the components of the chiral and anti-chiral fields are 
defined by operating the twisted supercharges to the leading field.

\subsection{$N=2$ twisted supersymmetric BF and Wess-Zumino actions}  


We now introduce off-shell $N=2$ twisted supersymmetric action:
\begin{align}
S=\int d^2x \int d^4\theta
 \Bigl(i^{\epsilon_\Psi} \overline\Psi(x^\mu,\theta,\theta^\mu,\tilde{\theta})
 {\Psi}(x^\mu,\theta,\theta^\mu,\tilde{\theta})
 \Bigr).
\label{eq:WZ action}
\end{align}
We can take the chiral superfields to be not only bosonic but also
fermionic. $\epsilon_\Psi$ should be taken $0$ or $1$ for bosonic or
fermionic $(\Psi,\overline{\Psi})$, respectively.

For fermionic $(\Psi,\overline{\Psi})$, the fields in the expansion 
of the superfield (\ref{psi_superfield}) and (\ref{psibar_superfield}) 
can be renamed as:
\begin{equation}
\begin{split}
\Psi(x^\mu,\theta,\tilde{\theta},\theta^\mu)  \ &= \ 
\Psi^\prime (z^\mu,\theta^\mu)
 =i\ee^{\theta^\mu s_\mu}c(z) 
 =ic(z)+\theta^\mu \omega_\mu(z) +i\theta^2 \lambda(z) \\
\overline{\Psi}(x^\mu,\theta,\tilde{\theta},\theta^\mu) \ &= \ 
\overline{\Psi}^\prime (\tilde{z}^\mu,\theta,\tilde{\theta})
  =i\ee^{\theta s + \tilde{\theta} \tilde{s}}\overline{c}(\tilde{z})
  =i\overline{c}(\tilde{z})+\theta b(\tilde{z})
  +\tilde{\theta}\phi(\tilde{z})
  -i\theta \tilde{\theta}\rho(\tilde{z}),
\label{expansion_psi_psibar}  
\end{split}
\end{equation}
where we have the correspondence of the fields; 
$(\varphi,\chi,\tilde{\chi},\tilde{\varphi})\rightarrow 
(i\overline{c},b,\phi,-i\rho)$ and
$(\phi,\psi_\mu,\tilde{\phi})\rightarrow (ic,\omega_\mu,i\lambda)$.
$N=2$ twisted SUSY transformations of the renamed fields can be 
read off from Table \ref{T-laws_chiral}.
Then the action (\ref{eq:WZ action}) leads
\begin{align}
S_f&= \int d^2x \int d^4\theta \  
 ( i \overline{\Psi} \Psi )
 = \int d^2x \int d^4\theta \ \ee^{\delta_\theta}(-i\overline{c}c)
 \nonumber \\
 &=\int d^2x \ 
s\tilde{s}\frac{1}{2}\epsilon^{\mu\nu}s_\mu s_\nu (-i\bar{c}c)\nonumber\\
 &=\int d^2x
\Bigl(
\phi\epsilon^{\mu\nu}\partial_\mu\omega_\nu
+b \partial^\mu\omega_\mu
+i\partial_\mu\overline{c}\partial^\mu c
+i\rho\lambda
\Bigr),
\label{eq:WZ 3}
\end{align}
where $\delta_\theta$ is defined in (\ref{def_delta_theta}).
This action can be identified as the quantized BF action 
with auxiliary fields and has off-shell $N=2$ twisted SUSY 
up to the surface terms by construction. 

For bosonic $(\Psi,\overline{\Psi})$ the action (\ref{eq:WZ action}) can be
written as follows:
\begin{align}
S_b&= \int d^2x \int d^4\theta \  
 (  \overline{\Psi} \Psi )
 = \int d^2x \int d^4\theta \ e^{\delta_\theta}(\varphi\phi)
 \nonumber \\
 &=\int d^2x \ 
s\tilde{s}\frac{1}{2}\epsilon^{\mu\nu}s_\mu s_\nu (\varphi\phi)\nonumber\\
 &=\int d^2x
\Bigl(
i\tilde{\chi}\epsilon^{\mu\nu}\partial_\mu\psi_\nu
+i\chi\partial^\mu\psi_\mu
-\partial^\mu\varphi\partial_\mu\phi
+\tilde{\varphi}\tilde{\phi}
\Bigr),
\label{eq:WZ action 2}
\end{align}
where $(\phi,\varphi,\tilde{\phi},\tilde{\varphi})$ and
$(\psi_\mu,\chi,\tilde{\chi})$ are bosonic and fermionic fields,
respectively.
The fermionic terms in (\ref{eq:WZ action 2}) change into matter fermions via 
Dirac-K\"ahler fermion mechanism:
\begin{align}
\int d^2x\ \left(
  i\tilde{\chi} \epsilon^{\mu\nu}\partial_\mu\psi_\nu 
  +i\chi \partial^\mu\psi_\mu 
 \right)
=\int d^2x\
  \text{Tr} \left(
   i\overline{\xi} \gamma^\mu \partial_\mu \xi
  \right),
\end{align}
where the Dirac-K\"ahler fermion $\xi$ is defined as
\begin{align}
\xi_{\alpha\beta}
=\frac{1}{2}\left(
 \mathbf{1}\chi+\gamma^\mu\psi_\mu+\gamma^5\tilde{\chi}
\right)_{\alpha\beta},
\label{DKtr1}
\end{align}
with $\overline{\xi}=C\xi^TC^{-1}=\xi^T$. 
It is natural to identify the fermionic antisymmetric tensor fields, 
$\chi,\psi_\mu,\tilde{\chi}$, as Dirac-K\"ahler fields of 0-,1-,2-forms, 
respectively.  
We can recognize that each spinor suffix of this Dirac-K\"ahler fermion 
has the Majorana Weyl fermion nature. 
We now redefine the bosonic fields as
follows:
\begin{equation}
\begin{split}
\phi_0 &=\frac{1}{2}(\phi+\varphi),
 \phi_1 =\frac{1}{2}(\phi-\varphi),\\
F_0&=\frac{1}{2}(\tilde{\phi}+\tilde{\varphi}),
 F_1 =\frac{1}{2}(\tilde{\phi}-\tilde{\varphi}).
\end{split}
\end{equation}
Then the action (\ref{eq:WZ action 2}) can be rewritten by the new fields:
\begin{align}
S_b&=\int d^2x \sum_{i=1}^2
\Bigl(
 i{\xi_\alpha}^i {\gamma^\mu}_{\alpha\beta} \partial_\mu {\xi_\beta}^i
 +\partial_\mu\phi^i\partial^\mu\phi^i
 -F^iF^i
\Bigr),
\end{align}
where we further redefine 
$i\phi_0\rightarrow\phi_2$ and $iF_0\rightarrow F_2$. 
This is the 2-dimensional version of $N=2$ Wess-Zumino action which 
has off-shell $N=2$ SUSY invariance of standard $N=2$ SUSY 
algebra (\ref{eq:N=2 algebra}). 
It is important to recognize at this stage that the ``flavor'' suffix 
of the Dirac-K\"ahler matter fermion in the action is $N=2$ extended 
SUSY suffix. 

\subsection {Non-Abelian Extension}

The Abelian version of the chiral and anti-chiral conditions 
(\ref{scalar_condition_1}) and (\ref{anti-chiral-condition_1}) 
can be covariantly extended to the following non-Abelian conditions:
\begin{eqnarray}
D\Phi(x^\mu,\theta,\tilde{\theta},\theta^\mu) 
-i\Phi^2(x^\mu,\theta,\tilde{\theta},\theta^\mu) &=& 0, \ \ 
 \tilde{D}\Phi(x^\mu,\theta,\tilde{\theta},\theta^\mu)=0,
 \label{nonab_chiral_conditions1}\\
D_\mu\overline{\Psi}(x^\mu,\theta,\tilde{\theta},\theta^\mu)
&=&0,\label{nonab_anti-chiral_conditions}
\end{eqnarray}
where $\Phi(x^\mu,\theta,\tilde{\theta},\theta^\mu)$ and 
$\overline{\Psi}(x^\mu,\theta,\tilde{\theta},\theta^\mu)$ are, 
respectively, non-Abelian  chiral and anti-chiral superfields.
Notice that the above extension makes sense only in the 
case of $\Phi$ to be fermionic.
Similar to the Abelian case we can transform the non-Abelian chiral condition 
(\ref{nonab_chiral_conditions1}) into the following form: 
\begin{eqnarray}
\frac{\partial}{\partial\theta}\Phi'-i(\Phi')^2&=&0,\label{no5}\\
\frac{\partial}{\partial\tilde{\theta}}\Phi'&=&0,\label{no6}
\end{eqnarray}
where $\Phi'\equiv U\Phi(x^\mu,\theta,\tilde{\theta},\theta^\mu) U^{-1}$ 
and $U$ is given by (\ref{def_of_U}). 
Unlike the Abelian case, $\Phi'$ is still a function of 
$(x^\mu,\theta,\tilde{\theta},\theta^\mu)$ due to inhomogeneous 
component in the non-Abelian chiral condition 
(\ref{no5}). 
In order to find solutions for these chiral conditions, we 
expand $\Phi'$ as 
\begin{equation}
\Phi'(x^\mu,\theta,\tilde{\theta},\theta^\mu)=F_1(x^\mu,\theta^\mu)
+\theta B_1(x^\mu,\theta^\mu)
+\tilde{\theta} B_2(x^\mu,\theta^\mu)
+\theta\tilde{\theta} F_2(x^\mu,\theta^\mu).
\end{equation}
Then the condition (\ref{no5}) leads
\begin{eqnarray}
B_1-i(F_1)^2&=&0,\label{c1}\\
F_2+i \left[ F_1,B_2 \right]&=&0,\\
\left[ F_1,B_1 \right ] &=& 0,\\
\{F_1,F_2\}+\left[ B_1,B_2 \right]&=&0,
\end{eqnarray}
while the other condition (\ref{no6}) leads
\begin{equation}
B_2=0, \ \ F_2 = 0.\label{c5}
\end{equation}
These conditions (\ref{c1})$\sim$(\ref{c5})
are solved as
\begin{eqnarray}
B_1=i(F_1)^2,\ \ B_2=0,\ \ F_2=0.
\end{eqnarray}
Therefore the possible expression for $\Phi'$ turns out to be
\begin{equation}
\Phi'(x^\mu,\theta,\tilde{\theta},\theta^\mu)= 
\Psi'(x^\mu,\theta^\mu)+i\theta(\Psi'(x^\mu,\theta^\mu))^2,\label{no7}
\end{equation}
where we have renamed $F_1(x^\mu,\theta^\mu)$ as $\Psi'(x^\mu,\theta^\mu)$ 
which satisfies the same form of the Abelian chiral condition 
(\ref{scalar_condition_2}) by definition, 
\begin{align}
\frac{\partial}{\partial \theta} \ 
\Psi'(x^\mu,\theta^\mu)  \ = \ 0, \ \ 
\frac{\partial}{\partial \tilde{\theta}} \ 
\Psi'(x^\mu,\theta^\mu)  \ = \ 0. 
\label{nonab_scalar_condition_2}
\end{align}
We can then obtain the non-Abelian version of chiral 
superfield as: 
\begin{eqnarray}
\Phi(x^\mu,\theta,\tilde{\theta},\theta^\mu)=
\Psi(x^\mu,\theta,\tilde{\theta},\theta^\mu)+
i\theta(\Psi(x^\mu,\theta,\tilde{\theta},\theta^\mu))^2
\end{eqnarray}
where $\Psi(x^\mu,\theta,\tilde{\theta},\theta^\mu)\equiv 
U^{-1}\Psi'(x^\mu,\theta^\mu) U = \Psi'(z^\mu,\theta^\mu)$ 
satisfies the same form of the Abelian chiral conditions as 
(\ref{scalar_condition_1}), 
\begin{align}
D\Psi(x^\mu,\theta,\tilde{\theta},\theta^\mu) \ = \ 0 , \ \ 
\tilde{D}\Psi(x^\mu,\theta,\tilde{\theta},\theta^\mu) \ = \ 0.
\label{nonab_scalar_condition_1}
\end{align}

Since the non-Abelian version of the anti-chiral condition 
(\ref{nonab_anti-chiral_conditions}) has the 
same form as the Abelian condition (\ref{anti-chiral-condition_1}), 
the solution of the anti-chiral 
superfield should have the similar form as Abelian case: 
\begin{align}
\overline{\Psi}(x^\mu,\theta,\tilde{\theta},\theta^\mu) \ = \ 
U \overline{\Psi}^\prime (x^\mu,\theta,\tilde{\theta}) U^{-1} \ = \ 
\overline{\Psi}^\prime (\tilde{z}^\mu,\theta,\tilde{\theta}).
\label{nonab_solution_of_anti-chiral_condition_3}
\end{align}

Similar to the Abelian case, the super transformations for the component 
fields can be read off from 
\begin{eqnarray}
s_{A}\Phi'(x^\mu,\theta,\tilde{\theta},\theta^\mu)
&=&{Q'}_{A}\Phi'(x^\mu,\theta,\tilde{\theta},\theta^\mu) \\
s_{A}\overline{\Psi}'(x^\mu,\theta,\tilde{\theta}) 
&=&{Q''}_{A}\overline{\Psi}'(x^\mu,\theta,\tilde{\theta}),  
\end{eqnarray}
where ${Q'}_{A}$ and ${Q''}_{A}$ are, respectively, given by 
(\ref{susy_trasformation_4}) and (\ref{susy_trasformation_5}).
More explicitly we can equivalently write as 
\begin{eqnarray}
s\Psi'(x^\mu,\theta^\mu)&=&Q'\Psi'(x^\mu,\theta^\mu)+
i{\Psi'}^2(x^\mu,\theta^\mu),\ \ \ 
\tilde{s}\Psi'(x^\mu,\theta^\mu)=\tilde{Q}'\Psi'(x^\mu,\theta^\mu),
\nonumber \\
s_\mu\Psi'(x^\mu,\theta^\mu)&=&{Q'}_\mu\Psi'(x^\mu,\theta^\mu), \\
s\overline{\Psi}'(x^\mu,\theta,\tilde{\theta})
&=&Q''\overline{\Psi}'(x^\mu,\theta,\tilde{\theta}),\ \ \  
\tilde{s}\overline{\Psi}'(x^\mu,\theta,\tilde{\theta})
=\tilde{Q}''\overline{\Psi}'(x^\mu,\theta,\tilde{\theta}), \nonumber \\
s_\mu\overline{\Psi}'(x^\mu,\theta,\tilde{\theta})
&=&{Q''}_\mu\overline{\Psi}'(x^\mu,\theta,\tilde{\theta}), 
\end{eqnarray}
where $\Psi'(x^\mu,\theta^\mu)$ and 
$\overline{\Psi}'(x^\mu,\theta,\tilde{\theta})$ has the similar 
expansion form as the Abelian case (\ref{expansion_psi_psibar}).\\
The non-Abelian version of $N=2$ twisted SUSY transformation 
is given in Table \ref{NABFtrans}.
\begin{table}
\begin{center}
\begin{tabular}{c|c|c|c}
\hline
fields & $s$ & $s_{\rho}$ & $\tilde{s}$ \\
\hline
$c$ & $-c^2$ & $-i\omega_{\rho}$ & $0$ \\
$\omega_{\mu}$ & $\partial_{\mu}c+[\omega_{\mu},c]$ &
$-i\epsilon_{\rho\mu}\lambda$ & $-\epsilon_{\mu\nu}\partial_{\nu}c$ \\
$\lambda$ & $\epsilon_{\mu\nu}\partial_{\mu}\omega_{\nu}
+\epsilon_{\mu\nu}\omega_{\mu}\omega_{\nu}-\{c,\lambda\}$ &
$0$ & $-\partial_{\mu}\omega_{\mu}$\\
\hline
$\bar{c}$ & $-ib$ & 0 & $-i\phi$ \\
$b$ & $0$ & $\partial_{\rho}\bar{c}$ & $-i\rho$\\
$\phi$ & $i\rho$ & $-\epsilon_{\rho\nu}\partial_{\nu}\bar{c}$ & 0 \\
$\rho$ & 0 & $-\partial_{\rho}\phi-\epsilon_{\rho\nu}\partial_{\nu}b$ & 
0\\
\hline
\end{tabular}
\caption{$N=2$ twisted SUSY transformation of component fields for non-Abelian BF.}
\label{NABFtrans}
\end{center}
\end{table}
The non-Abelian version of $N=2$ twisted supersymmetric action can be 
constructed in the similar way as the Abelian case:
\begin{eqnarray}
 S_{BF}&=& \int d^2x \int d^4\theta \ \mathrm{Tr}\  
 ( i \overline{\Psi} \Phi ) =\int d^2x \ 
s\tilde{s}\frac{1}{2}\epsilon^{\mu\nu}s_\mu s_\nu \ \mathrm{Tr}\ (-i\bar{c}c)
 \nonumber \\
&=& \int d^2 x\ \mathrm{Tr}\ \big[
\phi(\epsilon_{\mu\nu}\partial_{\mu}\omega_{\nu}
+\epsilon_{\mu\nu}\omega_{\mu}\omega_{\nu}-\{c,\lambda\}) \nonumber \\
&& + b\partial_{\mu}\omega_{\mu} - i \bar{c}\partial_{\mu}D_{\mu}c + 
i \rho\lambda \big],
\label{nonab_BF}
\end{eqnarray}
with $D_{\mu}c\equiv \partial_{\mu}c+[\omega_{\mu},c]$. 
This action can be identified as the quantized non-Abelian BF model 
with Landau gauge fixing accompanied by auxiliary fields .

\section{Exact twisted SUSY on a lattice}
\label{exact_lat}

 The $N=2$ twisted SUSY algebra in two continuum dimensions has been
 reviewed in Section \ref{twist_cont}, and fully discussed in Ref ~\cite{KKU}.
In this section we formulate the twisted superspace on a lattice parallel 
to the continuum formulation. We introduce a mild noncommutativity 
to preserve the lattice Leibniz rule as we discussed in 
Section \ref{Leibniz_rule}.

\subsection{$N=D=2$ twisted superspace on a lattice}

 Twisted $N=D=2$ superspace is defined by the parameter space
 $(x^{\mu},\theta^A)$ where the label $A$ can take four values. 
 The differential operators  $Q_A$ that generate infinitesimal SUSY 
 transformations in superspace were given in Eq.(\ref{eq:Q-operator}) and the
 algebra they satisfy in Eq. (\ref{N=2 left operator algebra}).
 It will  sometimes be convenient in this section, rather than writing the
 individual equations, to use a  compact notation and write the differential 
 operators as
 \be
 Q_A = \frac{\partial}{\partial \theta^A} 
+ \frac{\ii}{2} f_{AB}^{\mu} \theta^B \partial_{\mu},
 \label{susydiffop}
 \eeq
 and the corresponding algebra as
 \be
 \{Q_A , Q_B \} = - \ii f_{AB}^{\mu} \partial_{\mu},
 \label{sumal}
 \eeq
 where the constants $f_{AB}^{\mu}$ are symmetric in $AB$ and can be read from 
 (\ref{N=2 left operator algebra}).

 As discussed in Section \ref{Leibniz_rule} the derivative operator $\partial_{\mu}$ is replaced
 on a square lattice by one of the finite difference operators $\Delta_{+\mu}$
 and $\Delta_{-\mu}$ defined in Eqs (\ref{deltapiu}) and (\ref{deltaminus}).
 So if we denote the generators of the SUSY algebra (\ref{sumal}) on 
 the lattice by the same $\Q_A$ then the algebra  becomes:
 \be
 \{\Q_A,\Q_B\} = - \ii f_{AB}^{\mu} \Delta_{\pm \mu}.
 \label{sumall}
 \eeq
 The  ambiguity at the r.h.s. of (\ref{sumall}) is inherent to the lattice 
 formulation and has to be removed by choosing, for each pair of values of $A$ 
 and $B$  corresponding  to a non zero value of $f_{AB}^{\mu}$,  
 either $\Delta_{+ \mu}$ or $\Delta_{- \mu}$.
 As we shall see shortly this can be done  on consistency grounds.
 In view of the algebra (\ref{sumall}) and of the commutators (\ref{newcomm})
 and (\ref{negcomm}) it is natural to assume that the supercharges
 $\Q_A$ act on a superfield $\Fc (x,\theta)$ according to Eq. (\ref{O}),
 namely:

 \be
 \left[ \Q_A, \Fc (x,\theta)\right\} = T(2 \a_A) ~\us_A \Fc (x,\theta) =
 T(\a_A) ~\us^{(s)}_A \Fc (x,\theta)~ T(\a_A),
 \label{supsupvar}
 \eeq
 where $\us_A$ and $\us^{(s)}_A$ are SUSY transformations (the latter
 symmetrized in the sense of Section \ref{Leibniz_rule}) and $T(2 \a_A)$ the shift operator 
 associated to $\Q_A$. The mixed bracket notation at the l.h.s. denotes a 
 commutator or an anticommutator according to the Grassmann grading of 
 $\Fc (x,\theta)$.

 Consistency of Eq. (\ref{supsupvar}) with the graded algebra (\ref{sumall})
 leads to a set of equations for the shifts $\a_A$ associated to the
 SUSY transformations. 
 In fact, consider the Jacobi identity\footnote{We take $\Fc (x,\theta)$ here to
 be a bosonic superfield. Obvious changes in the signs would apply in the
 fermionic case.}:
 \be
 \left[ \{\Q_A,\Q_B\}, \Fc (x,\theta)\right]-\left\{ [\Q_B ,\Fc (x,\theta)],
 \Q_A \right\}+\left\{ [\Fc (x,\theta),\Q_A ],\Q_B\ \right\} =0.
 \label{jacobi}
 \eeq
 From the super algebra (\ref{sumall}) and the commutators (\ref{supsupvar}) and
 (\ref{newcomm}) we obtain:
 \bea
 \lefteqn{
  -i  f^{\mu}_{AB} ~T(\pm 2 \m)~ \upar_{\pm \mu}\Fc (x,\theta)
 + T(2 \a_B) T(2 \a_A)~\us_B \us_A \Fc (x,\theta)}\hspace{10em}\nonumber\\
  &&{} + T(2 \a_A) T(2 \a_B)~ 
 \us_A \us_B \Fc (x,\theta)=0.
 \label{jacobi2}
 \eea
 As shifts are additive this implies the following relations:
 \be
 \a_A + \a_B =\pm \m~~~~~~~~~~\mbox{iff}~~~~~~~ f_{AB}^{\mu} \neq 0,
 \label{conshift}
 \eeq
 and
 \be
 \left( \us_B \us_A + \us_A \us_B \right)\Fc (x,\theta) =  i f^{\mu}_{AB}\upar_{\pm \mu}
 \Fc (x,\theta).
 \label{susycomm}
 \eeq
 In both (\ref{conshift}) and (\ref{susycomm}) the plus (resp. minus) sign at 
 the r.h.s. is chosen if $\Delta_{+ \mu}$ (resp. $\Delta_{- \mu}$) appears at 
 the r.h.s. of (\ref{sumall}).
 Consider now the shift equations (\ref{conshift}) in the specific case of the 
 super algebra (\ref{N=2 left operator algebra}). 
 By setting in (\ref{conshift}) the values of $A$ and $B$ for which 
 $f_{AB}^{\mu} \neq 0$ we obtain the following  equations:
 \ba
 i)~~~~ \a + \a_1 =\pm \hat{n}_1,&~~~~~~~~~~~~~&ii)~~~~\a + \a_2 =
 \pm \hat{n}_2 ,  \nonumber \\
 iii)~~~~\at + \a_1 = \pm \hat{n}_2,&~~~~~~~~~~~~~&iv)~~~~ \at + \a_2 =
 \pm \hat{n}_1 .
 \label{cond}
 \ea
 In principle the signs at the r.h.s. of Eqs (\ref{cond}) can be chosen in an
 arbitrary way. However  by comparing the linear combinations $i) + iv)$ and 
  $ii) + iii)$ one finds that the resulting equations are 
 compatible only if the signs in front of $\hat{n}_1$ in $i)$ and $iv)$ are
 opposite and at the same time the signs in front of $\hat{n}_2$ in $ii)$ and 
 $iii)$ are also opposite. 
 If these conditions are satisfied, the linear combinations $i) + iv)$ and 
  $ii) + iii)$ are compatible and are equivalent to a unique condition, namely
  the vanishing of the sum of all shifts:
 \be
 \a + \a_1 + \a_2 +\at =0 .
 \label{zerosum}
 \eeq
  The system (\ref{cond}) is then replaced by the condition (\ref{zerosum}) plus
 two of the equations (\ref{cond}), for instance $i)$ and $ii)$, that form with
 (\ref{zerosum}) a system of linearly independent equations.
 As the linearly independent equations are now only three, the solution  will 
 depend on an arbitrary vector. In other words, one of the shifts,
 for instance $\a$, is not determined and can be chosen arbitrarily.
 Besides this arbitrariness, there are four possible sign choices in Eqs~$i)$ 
 and $ii)$ that will give four distinct solutions. 
 Let us postpone the discussion about the meaning of such multiplicity of
 solutions and concentrate for the moment on a specific sign choice, choosing 
 for instance the plus sign at the r.h.s. of both  $i)$ and $ii)$. 
 The linearly independent shift equations are then
 given by Eq. (\ref{zerosum}) and by
 \be
 i)~~~~ \a + \a_1 =+ \hat{n}_1,~~~~~~~~~~~~~ii)~~~~\a + \a_2 =
 + \hat{n}_2 .
 \label{indcond}
 \eeq

 As already remarked this system does not determine $\a_A$ completely and one
 shift, say $\a$ can be treated as a free parameter.
 The whole formalism can be developed without ever specifying this free
 parameter and in what follows, unless otherwise specified,  the formulas 
 (such as for instance the different superfield expansions) will be valid for an
 arbitrary  $\a$.
 On the other hand it will be convenient, in order to represent the different 
 component fields on a two dimensional square lattice, to make a particular
 choice for  $\a$.  In this respect two choices appear to be particularly
 convenient: the first one is the most symmetric, in the sense that all vectors 
 $\a,\a_1,\at,\a_2$ have the same length and can be obtained from any of them by successive
 rotations of $\frac{\pi}{2}$:  
 \be
 A)~~\a \equiv (1/2,1/2),~~~~\at \equiv (-1/2,-1/2),~~~~\a_1 \equiv (1/2,-1/2),
 ~~~~\a_2 \equiv (-1/2,1/2).
 \label{a}
 \eeq
 The second choice, which we shall call ``asymmetric'', corresponds to putting 
 simply $\a=0$ and gives:
 \be
 B)~~\a \equiv (0,0),~~~~\at \equiv (-1,-1),~~~~\a_1=\hat{n}_1 \equiv (1,0),~~~~
 \a_2 =\hat{n}_2 \equiv (0,1).
 \label{a-asym}
 \eeq
 
 Let us consider now the three extra solutions that  correspond to the three
 remaining choices of the signs in $i)$ and $ii)$. It is easy to check that 
 these can be obtained from (\ref{a}) by 
 the mirror image reflection with respect to 1- and/or 2- axis.
  Now the problem is: why four distinct solutions?
 As long as one restricts oneself  to the SUSY algebra 
 (\ref{N=2 left operator algebra}), then one can consistently choose one 
 solution, say (\ref{a}), and forget about the others.
  However if one wants to implement  on  the lattice the $R$-symmetry and the 
  discrete Lorentz rotations  of multiples of $\pi/2$ (which is beyond the 
  scope of the present paper) then all four solutions will have to be
  considered. In fact $R$-symmetry and discrete Lorentz rotations would mix the 
  SUSY algebras on the lattice that correspond to the four different 
  solutions.
  With the sign choice (\ref{indcond}) all ambiguity in (\ref{sumall}) is
  removed and we can explicitly write the non trivial anticommutators  
  of the super algebra (\ref{N=2 left operator algebra}) on the lattice:
  \be
  \{\Q,\Q_{\mu}\} =  \ii~ \Delta_{+\mu},~~~~~~~~~~\{\tilde{\Q},\Q_{\mu}\} 
  = -\ii~ \epsilon_{\mu\nu}~ \Delta_{-\nu}.
  \label{la}
  \eeq
  
  An explicit representation of the algebra (\ref{la}) in terms of the 
  $\theta^A$ variables can easily be found:
   \ba
  &\Q= \frac{\partial}{\partial \theta} +\frac{\ii}{2} \theta^{\mu} 
  \Delta_{+\mu},~~~~~~\tilde{\Q} = \frac{\partial}{\partial \tilde{\theta}} 
  -\frac{\ii}{2} \epsilon_{\mu \nu} \theta^{\mu}\Delta_{-\nu}, & 
  \nonumber \\
  &~~~~~~~~\Q_{\mu}= \frac{\partial}{\partial \theta^{\mu}}+\frac{\ii}{2} \left(  
  \theta \Delta_{+\mu} - \tilde{\theta} \epsilon_{\mu \nu} 
\Delta_{-\nu}\right).~~~~~~&
  \label{susygen}
  \ea

  The supercharges (\ref{susygen}) are consistent with 
  the SUSY variations (\ref{supsupvar}) only if the superfield 
   $\Fc (x,\theta)$ itself depends on the shift operators $T(2 \a_A)$.
   It is not difficult to check that the correct dependence is obtained by
   simply replacing in $\Fc (x,\theta)$ the Grassmann variables $\theta^A$ with 
   ``dressed'' Grassmann variables $\th^A$ defined as:
  \be
  \th^A = T(2 \a_A) \theta^A.
  \label{thetashift}
  \eeq
   Notice that the new variables $\th_A$ do not (anti)commute anymore with a 
   field $\Psi(x)$:
   \be
   \th^A \Psi(x+\a_A) = (-1)^{|\Psi|} \Psi(x-\a_A) \th^A.
   \label{thetacomm1}
   \eeq
   The expansion of a superfield  $\Fc (x,\th_A)$ into
   its component fields  has to now take into account the noncommutativity of
   the Grassmann variables $\th_A$ shown in (\ref{thetacomm1}). As a result the
   arguments of the component fields are sensitive to their relative order with
   respect to the Grassmann $\th$ variables:
   \ba
   \uFc (x,\th_A)&=& \uvarphi(x) + \th_A \uvarphi_A(x+\a_A)+\frac{1}{2} 
   \th_A \th_B
   \uvarphi_{AB}(x+\a_A+\a_B)+ \cdots  \nonumber \\
   &=&\uvarphi(x) +(-1)^{|\varphi|} \uvarphi_A(x-\a_A) \th_A +\frac{1}{2}
   \uvarphi_{AB}(x-\a_A-\a_B) \th_A \th_B+ \cdots. \nonumber \\
   \label{superfield}
   \ea
   The superfield $ \uFc (x,\th_A)$ and its leading component $\uvarphi(x)$ are
   defined on a lattice $L$ ( $x \in L$) which will be specified later.
   It is clear from (\ref{superfield}) that the support of the other components
   are lattices shifted with respect to $L$. For instance the field $\uvarphi_A$
   is defined on a lattice $L_A$ which is shifted of $\a_A$ with respect to $L$.
   
   So it is apparent from (\ref{superfield}) that a superfield on the 
   lattice is a non local object since the component fields of  
   $\uFc (x,\th_A)$ are spread on a cluster of points around $x$. It is also
   true the vice versa in the sense that a component field on a definite site, 
   for instance  $\uvarphi_A(x-\a_A)$, is not linked to a definite superfield
   but can appear as a component field either in the expansion of 
   $  \uFc (x,\th_A)$ or of $ \uFc (x-2\a_A,\th_A)$
   according to which of the two expansions in (\ref{superfield}) is used.
   Let us now determine the lattice $L$. Consider first 
   that the multiplication of a superfield by $\th_A$ has to be well defined in
    order to have a well defined multiplication between two superfields.
    Hence in the relation
   \be
   \th_A ~\uFc (x,\th) = (-1)^{|\uFc|}\uFc (x- 2 \a_A,\th )~~ \th_A ,
   \label{norah}
   \eeq
   both sides have to be well defined, and the r.h.s. makes sense only if from 
   $x \in L$ it follows $x-2 \a_A \in L$ for any $A$. 
   Suppose now that the origin of the axis is a point of on $L$, 
   then all points of $L$ are obtained as linear combinations, with integer even
   coefficients, of the fundamental shifts $\a_A$. As a matter of fact, because
   of (\ref{zerosum}), these are not linearly independent and any three of them,
   for instance $\a$ and $\a_{\mu}$, can be chosen as basis:
   \be
   x \in L~~~~~~~~~{\rm iff}~~~~~~~~~~~x= 2 k \a +2 k_1 \a_1 +2 k_2 \a_2 ,
   \label{elle}
   \eeq
   which can also be written as
   \be
    x= 2 (k - k_1 -k_2) \a + 2 k_1 \hat{n}_1 + 2 k_2 \hat{n}_2.
    \label{elle2}
    \eeq
    Eq. (\ref{elle}) shows an interesting and unexpected feature. Although the
    original theory is two-dimensional the resulting lattice, corresponding to
    the most general choice of the shifts $\a_A$, naturally lives in three
    dimensions as it is parameterized by the three integers  $\{k,k_1,k_2\}$.
    See Fig.~\ref{fig:three-dim}.
    Of course only two of the dimensions are dynamical, in the sense that only
    finite differences in the $\hat{n}_1 $ and $ \hat{n}_2$ directions appear in
    the algebra and eventually in the Lagrangian. The third dimension, defined
     by the $\a$ direction in (\ref{elle2}), appears to be essentially related
     to the realization of SUSY although a deeper understanding of this
     point is probably needed.
   The lattices $L_A$, $L_{AB}$ etc. on which the different components of the
   superfield are defined can be easily obtained from 
   (\ref{superfield}) by inserting $x$ given in (\ref{elle}). 
   It is convenient to  relabel the component fields in the expansion of 
   $\uFc (x,\th)$ according to the following labeling of the expansion 
   parameters: $\th_A=(\th_S,\th_\mu,\th_P)=(\th,\th_\mu,\tilde{\th})$. 
   For instance the coefficient of $\tilde{\th}~\th_1$ in the expansion of  
   $\uFc (x,\th)$ will be denoted by $\uvarphi_{P1}$ and so on.
   We find then that: 
   i) Due to the vanishing sum of $\a_A$, Eq. (\ref{zerosum}),
   component fields 
   with complementary sets of labels are defined on the same lattice. For 
   instance $\uvarphi_{S12P}$ has the same support $L$ as $\uvarphi$, 
   $\uvarphi_{S2P}$ the same as $\uvarphi_{1}$ and so on. 
   Without loss of generality we can then restrict ourselves to
   component fields whose labels do not contain the index $P$.   
   ii) Consider then the field $\uvarphi_i$ with $i \in \{ \emptyset, \{S\},
   \{1\},\{2\},\{S1\}, \{S 2\},\{1 2\},\{S 1 2\} \}$. It is defined on the 
   lattice $L_i$ whose points $y$ are of the form  
   \be
   y = m_S ~\a + m_1 ~\a_1 + m_2 ~\a_2,
   \label{generallatt}
   \eeq
   where $m_S,m_1,m_2$ are odd integers if the label $i$ of the field contains 
   the corresponding index and even integers otherwise. For instance if $i 
   \equiv \{S1\}$ then $m_S$ and $m_1$ are odd and $m_2$ even.
   The union $L_0$ of all $L_i$'s , namely 
   $L_0  = \bigcup_i L_i$ 
   is made of all
   the points given in (\ref{generallatt}) with arbitrary integers
   $m_S,m_1,m_2$. A particular superfield can be associated with a fundamental
   cell of $L_0$, its component fields being associated in pairs its vertices.
   Moving of one unit in any of the three directions of the lattice (namely
   increasing of one unit any of the integers  $m_S,m_1,m_2$)
   leads from a site 
   occupied by a bosonic component field to one occupied by a fermionic one 
   or vice versa. 
   See Fig.~\ref{fig:three-dim-cell}.
   \begin{figure}
    \begin{minipage}[b]{.5\linewidth}
      \scalebox{.7}{
       \input{three-dim.pstex_t}
      }
   \caption{Three-dimensional lattice sites of $L$.}
   \label{fig:three-dim}
    \end{minipage}     
    \hspace{.02\linewidth}
    \begin{minipage}[b]{.45\linewidth}
      \scalebox{.9}{
       \input{three-dim-cell.pstex_t}
      }
   \caption{Configuration of component fields in a fundamental cell of $L_0$.}
   \label{fig:three-dim-cell}
    \end{minipage}
   \end{figure}

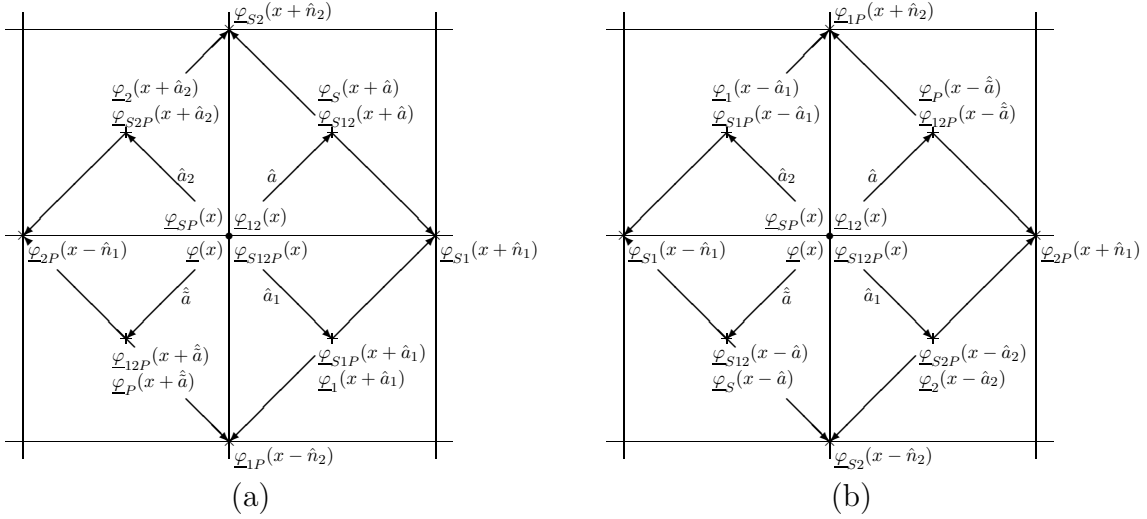
\begin{figure}
\begin{minipage}[b]{.47\linewidth}
 \begin{center}
 \scalebox{.65}{
 \input{gen_sf_sym_a.tex}
 }

 (a)
 \end{center}
\end{minipage}
\hspace{.05\linewidth}
\begin{minipage}[b]{.47\linewidth}
 \begin{center}
 \scalebox{.65}{
  \input{gen_sf_sym_b.tex}
 }

 (b)
 \end{center} 
\end{minipage}
 \caption{Component fields on the lattice with the symmetric choice
 of shift parameter A) 
 for $\uFc (x,\th_A)$ where $\th_A$ is located on (a) the left and (b)
 the right of the component fields.}
 \label{fig:sufield-shiftsym-a}
\end{figure}

   The symmetric and asymmetric choices A) and B) of the shift parameters amount
   to impose an extra linear relation (respectively $\a=\frac{1}{2} (\hat{n}_1 
   + \hat{n}_2)$ and $\a=0$) among the shifts, and hence to effectively
   reduce the dimension of the lattice from three to two.
   Consider for instance the symmetric parameter choice A). If we insert in 
   (\ref{elle2}) the relation $\a=\frac{1}{2} (\hat{n}_1 + \hat{n}_2)$ we find
   for the points $x$ of the lattice $L$ on which the superfield is defined:
   \be
   x=  (2 k_1+h) \hat{n}_1 + (2 k_2+h) \hat{n}_2,
   \label{elle3}
   \eeq
   where $h=k-k_1-k_2$ is an arbitrary integer. Hence the coordinates of $x$ on
   the two dimensional plane are given by either two even integers (if $h$ is
   even) or two odd integers (if $h$ is odd). In this choice all shifts $\a_A$ 
   have half integer coordinates (see Eq. (\ref{a})), it is then clear from 
   (\ref{superfield}) that the field $\uvarphi_A$ is defined on a lattice $L_A$
    which is shifted with respect to $L$ and whose sites have half-integer 
    coordinates.
    In general, as clearly shown in the expansion (\ref{superfield}), with the
    symmetric choice A) if bosonic fields are defined on sites with integer 
    coordinates fermionic fields have half-integer coordinates and vice versa. 
   The position on the lattice of the different component fields of 
   $\uFc (x,\th_A)$ for the symmetric parameter choice A) is shown in 
   Fig.\ref{fig:sufield-shiftsym-a}, where 
   (a) and (b) refer to  the two expansions of Eq. (\ref{superfield}).

  In the asymmetric parameter choice B) ($\a=0$) we have simply from
  (\ref{elle2}): 
  \be
  x = 2 k_1 \hat{n}_1 + 2 k_2 \hat{n}_2,
  \label{elle4}
  \eeq
  hence the sites of $L$ are all of the (even,even) type. As all shifts $\a_A$
  have integer coordinates bosonic and fermionic component fields are all
  defined on integer sites and the same site has in general both bosonic and
  fermionic fields.
  This is shown in Fig.\ref{fig:sufield-shiftnat-a}
 where the position on the lattice of the 
different component fields of $\uFc (x,\th_A)$ is shown.

\begin{figure}
\begin{minipage}[b]{.47\linewidth}
 \begin{center}
  \scalebox{.65}{
  \input{gen_sf_nat_a.tex}
  }

 (a)
 \end{center} 
\end{minipage}
\hspace{.05\linewidth}
\begin{minipage}[b]{.47\linewidth}
 \begin{center}
  \scalebox{.65}{
  \input{gen_sf_nat_b.tex}
  }

 (b)
 \end{center} 
\end{minipage}
\caption{Component fields on the lattice with the asymmetric choice of shift 
parameter B) for $\uFc (x,\th_A)$ where $\th_A$ is located on (a) the
 left and (b) the right of the component fields.}
 \label{fig:sufield-shiftnat-a}
\end{figure}

  Following the pattern already introduced in Section \ref{Leibniz_rule}, Eq. (\ref{Odef}), it is
  convenient to define the ``arrowed'' supercharges $\auQ_A$:
  \be
  \Q_A = T(2 \a_A)~ \auQ_A,
  \label{arrsupch}
  \eeq
  which have an 
 expression in terms of $\th^A$:
   \ba
  &\auQ=  \frac{\partial}{\partial \th} +\frac{\ii}{2} 
  \th^{\mu}\arrowdel_{+\mu}
   ,~~~~~~\overrightarrow{\tilde{\uQ}} = \frac{\partial}{\partial \tilde{
  \th}} -\frac{\ii}{2} \epsilon_{\mu \nu} \th^{\mu}
  \arrowdel_{-\nu}, & 
  \nonumber \\
  &~~~~~~~~\auQ_{\mu}=\frac{\partial}{\partial \th^{\mu}}+
  \frac{\ii}{2} \left( \th\arrowdel_{+\mu} - \epsilon_{\mu \nu}
  \tilde{\th} \arrowdel_{-\nu}  \right).~~~~~~&
  \label{susygen2}
  \ea
  Using the $\auQ_A$ charges SUSY transformations $\us_A$ on a lattice
  can be written as shifted commutators as in (\ref{arrowO}):
  \be
  \us_A \uFc (x,\th)= \auQ_A ~\uFc (x,\th)- (-1)^{|\uFc|} \uFc (x+ 2 \a_A,\th)~
  \auQ_A,
  \label{sa}
  \eeq
  while for the ``symmetrized'' variations $\us^{(s)}_A$ we have:
  \be
  \us^{(s)}_A \uFc (x,\th)= \auQ_A ~\uFc (x-\a_A,\th)- (-1)^{|\uFc|} 
  \uFc (x+  \a_A,\th)~\auQ_A.
  \label{symmsa}
  \eeq
It is worth to note here that the location of the lattice position of 
the component fields in $\us^{(s)}_A \uFc (x,\th)$ is not necessary on $x$.
  Eqs (\ref{sa}) or (\ref{symmsa}) can be used to calculate the SUSY
  transformations of the component fields of the superfield $\uFc (x,\th)$ by
  calculating both side of the equation and equating the coefficients of the
  expansion in $\th_A$.
  Let us write the expansion of $\uFc (x,\th)$ using the same notations
  as in the
  continuum case ( Eq. (\ref{eq:F})):
  \begin{equation}
\begin{split}
  \uFc (x,\th)
 =&\uphi(x)+\th^\mu\uphi_\mu(x+\a^{\mu})+\th^2 \tilde{\uphi}(x+\a_1+\a_2)\\
 &+\th\Bigl( \upsi(x+\a)+\th^\mu\upsi_\mu(x+\m)
   +\th^2 \tilde{\upsi}(x-\at)\Bigr)\\
 &+\tilde{\th}\Big(\uchi(x+\at)+\th^\mu\uchi_\mu(x-|\epsilon_{\mu \nu}|\n)
   +\th^2 \tilde{\uchi}(x-\a)\Bigr)\\
 &+\th\tilde{\th}\Bigl(\ulambda(x+\a+\at)+\th^\mu\ulambda_\mu(x+\m +\at)
   +\th^2 \tilde{\ulambda}(x)\Bigr).\\
\end{split}
\label{eq:Fl}
\end{equation}
The position on the lattice of the different component fields of $\uFc (x,\th)$ 
can be recognized by the same figures for (\ref{superfield}): 
Figs; \ref{fig:sufield-shiftsym-a}, 
\ref{fig:sufield-shiftnat-a} 
with the identifications:  $(\uvarphi)=(\uphi)$, 
$(\uvarphi_{S},\uvarphi_{1},\uvarphi_{2},\uvarphi_{P}) = 
(\upsi,\uphi_1,\uphi_2,\tilde{\uphi})$, 
$(\uvarphi_{S1},\uvarphi_{S2},\uvarphi_{SP},\uvarphi_{12},\uvarphi_{1P},\uvarphi_{2P}) = 
(\upsi_1,\upsi_2,\ulambda,\tilde{\uphi},\uchi_1,\uchi_2)$, 
$(\uvarphi_{S12},\uvarphi_{S1P},\uvarphi_{S2P},\uvarphi_{12P})= 
(\tilde{\upsi},\ulambda_1,\ulambda_2,\tilde{\uchi})$, 
$(\varphi_{S12P})=(\tilde{\ulambda})$.

By inserting the expansion above into Eq. (\ref{symmsa}) and using the explicit
form of the supercharges $\auQ_A$, one obtains the SUSY 
transformations of the different components of $\uFc (x,\th)$. These are
given in Table \ref{T-laws_F_lat}.
\begin{table}[htbp]
\begin{center}
\renewcommand{\arraystretch}{1.3}
\begin{tabular}{c|c|c|c}
\hline
$\uphi^A$ & $\us\uphi^A$ & $\us_\mu\uphi^A$ & $\tilde{\us}\uphi^A$ \\
\hline
$\uphi$
 & $\upsi$
 & $\uphi_\mu$
 & $\uchi$ \\
$\uphi_\rho$
 & $-\upsi_\rho-\frac{i}{2}\upar^{(s)}_{\rho}\uphi$
 & $-\epsilon_{\mu\rho}\tilde{\uphi}$
 & $-\uchi_\rho+\frac{i}{2}\epsilon_{\rho\sigma}\upar^{(s)}_{\sigma}\uphi$ \\
$\tilde{\uphi}$
 & $\tilde{\upsi}+\frac{i}{2}\epsilon^{\rho\sigma}
   \upar^{(s)}_{\rho}\uphi_\sigma$
 & $0$
 & $\tilde{\uchi}-\frac{i}{2}\upar^{(s)}_{\rho}\uphi_\rho$ \\
\hline
$\upsi$
 & $0$
 & $\upsi_\mu-\frac{i}{2}\upar^{(s)}_{\mu}\uphi$
 & $\ulambda$ \\
$\upsi_\rho$
 & $-\frac{i}{2}\upar^{(s)}_{\rho}\upsi$
 & $-\epsilon_{\mu\rho}\tilde{\upsi}+\frac{i}{2}\upar^{(s)}_{\mu}\uphi_\rho$
 & $-\ulambda_\rho+\frac{i}{2}
\epsilon_{\rho\sigma}\upar^{(s)}_{\sigma}\upsi$ \\
$\tilde{\upsi}$
 & $\frac{i}{2}\epsilon^{\rho\sigma}\upar^{(s)}_{\rho}\upsi_\sigma$
 & $-\frac{i}{2}\upar^{(s)}_{\mu}\tilde{\uphi}$
 & $\tilde{\ulambda}-\frac{i}{2}\upar^{(s)}_{\rho}\upsi_\rho$ \\
\hline
$\uchi$
 & $-\ulambda$
 & $\uchi_\mu+\frac{i}{2}\epsilon_{\mu\nu}\upar^{(s)}_{\nu}\uphi$
 & $0$ \\
$\uchi_\rho$
 & $\ulambda_\rho-\frac{i}{2}\upar^{(s)}_{\rho}\uchi$
 & $-\epsilon_{\mu\rho}\tilde{\uchi}-\frac{i}{2}
\epsilon_{\mu\nu}\upar^{(s)}_{\nu} \uphi_\rho$
 & $+\frac{i}{2}\epsilon_{\rho\sigma}\upar^{(s)}_{\sigma}\uchi$ \\
$\tilde{\uchi}$
 & $-\tilde{\ulambda}+\frac{i}{2}\epsilon^{\rho\sigma}
\upar^{(s)}_{\rho}\uchi_\sigma$
 & $\frac{i}{2}\epsilon_{\mu\nu}\upar^{(s)}_{\nu}\tilde{\uphi}$
 & $-\frac{i}{2}\upar^{(s)}_{\rho}\uchi_\rho$ \\
\hline
$\ulambda$
 & $0$
 & $\ulambda_\mu+\frac{i}{2}\upar^{(s)}_{\mu}\uchi+
 \frac{i}{2}\epsilon_{\mu\nu}\upar^{(s)}_{\nu}\upsi$
 & $0$ \\
$\ulambda_\rho$
 & $-\frac{i}{2}\upar^{(s)}_{\rho}\ulambda$
 & $-\epsilon_{\mu\rho}\tilde{\ulambda}
  -\frac{i}{2}\upar^{(s)}_{\mu}\uchi_\rho
  -\frac{i}{2}\epsilon_{\mu\nu}\upar^{(s)}_{\nu}\upsi_\rho$
 & $\frac{i}{2}\epsilon_{\rho\sigma}\upar^{(s)}_{\sigma}\ulambda$ \\
$\tilde{\ulambda}$
 & $\frac{i}{2}\epsilon^{\rho\sigma}\upar^{(s)}_{\rho}\ulambda_\sigma$
 & $\frac{i}{2}\upar^{(s)}_{\mu}\tilde{\uchi}
  +\frac{i}{2}\epsilon_{\mu\nu}\upar^{(s)}_{\nu}\tilde{\upsi}$
 & $-\frac{i}{2}\upar^{(s)}_{\rho}\ulambda_\rho$ \\
\hline
\end{tabular}
\caption{$N=2$ twisted SUSY transformation of the component fields of the superfield
 $\uFc (x,\th)$ on a lattice.}
\label{T-laws_F_lat}
\end{center}
\end{table}
The arguments of the fields in Table \ref{T-laws_F_lat}
 should be the appropriate ones corresponding
to the expansion (\ref{eq:Fl}). Consider for example in 
Table \ref{T-laws_F_lat}
 the field $\upsi$ whose argument in the expansion (\ref{eq:Fl}) is
$x+ \a$. The argument of $\us_\mu \upsi = \upsi_\mu -\frac{i}{2}\upar_\mu \uphi$
is then $x+\a+\a_\mu=x+\m$ which is consistent again with the expansion 
(\ref{eq:Fl}). More explicitly, 
\bea
\us_\mu \upsi(x+\a) &=& \upsi_\mu(x+\a+\a_\mu) -\frac{i}{2}\upar_{+\mu} \uphi(x) 
\nonumber \\
 &=&\upsi_\mu(x+\a+\a_\mu) -\frac{i}{2}\upar^{(s)}_{\mu} \uphi(x+\a+\a_\mu), 
\eea
where the symmetric difference operator $\upar^{(s)}_{\mu}$ 
is originated from 
the positive difference operator $\upar_{+\mu}$ and thus carries $+2\m$ shift. 
It should be remembered that if a field
$\uPhi(x)$ is bosonic (resp. fermionic)
then $\us_A \uPhi$ is fermionic
(resp. bosonic) and is defined on the sites $x+ \a_A$. 

The SUSY transformations on the lattice of Table \ref{T-laws_F_lat}
 coincide with the
ones obtained  in the continuum (see Table \ref{T-laws_F}) by defining $s_A F(x^\mu,\theta,
\theta^\mu,\tilde{\theta}) = \{ Q_A,F(x^\mu,\theta,\theta^\mu,\tilde{\theta}) \}
$ provided we replace the fields in the continuum with the corresponding 
underlined fields on the lattice, we make the appropriate shifts in the fields'
 arguments and we replace the partial derivative $\partial_\mu$ with the 
 symmetrized finite difference $\upar^{(s)}_{\mu}$.
  
  Finally let us consider the Leibniz rule, which follows from the Leibniz
  rule for the (anti)commutator (\ref{supsupvar}) or more directly from Eqs
  (\ref{sa}) or (\ref{symmsa}):
  \ba
  \us_A (\uFc_1 (x,\th)~ \uFc_2 (x,\th))& = &  (\us_A \uFc_1 (x,\th))~ 
  \uFc_2 (x,\th)  \nonumber \\
  &+& (-1)^{|\uFc_1|} \uFc_1(x+2 \a_A,\th) ~(\us_A \uFc_2(x,\th)), \label{llbrule} \\
  \us^{(s)}_A (\uFc_1 (x,\th) ~\uFc_2 (x,\th))&=& (\us^{(s)}_A \uFc_1(x,\th) )~
  \uFc_2(x- \a_A,\th) \nonumber \\&+&
  (-1)^{|\uFc_1|} \uFc_1(x+ \a_A,\th) ~(\us^{(s)}_A \uFc_2(x,\th)). 
  \label{llbrules}
  \ea

  \subsection{Chiral superfields}

  Chiral and anti-chiral conditions were defined in the continuum using the 
  differential operators which generate the shift transformation induced by
  right multiplication in superspace. They are given explicitly in Eq. 
  (\ref{eq:D-operator}) 
  which can be summarized in the compact
  notation of Eq. (\ref{susydiffop}) as:
  \be
  D_A = \frac{\partial}{\partial \theta^A} 
- \frac{\ii}{2} f_{AB}^{\mu} \theta^B \partial_{\mu}.
 \label{susydiffrightop}
 \eeq
 They anticommute with the SUSY generators $Q_A$ and satisfy the same
 algebra, up to a sign: $\{D_A,D_B\}=-\{Q_A,Q_B \}= \ii f_{AB}^{\mu}
 \partial_{\mu}$.
 The definition of $D_A$ on a square lattice proceeds in the same way as for
 $Q_A$. If we denote by $\D_A$ the corresponding lattice operators we have, in
 analogy with (\ref{sumall})
 \be
 \{ \D_A, \D_B \} = \ii~ f_{AB}^{\mu} \Delta_{\pm \mu}.
 \label{dimall}
 \eeq
 The sign ambiguity is resolved in exactly the same way as for the SUSY
 generators $\Q_A$, and in analogy with Eq. (\ref{la}) we find:
 \be
 \{\D,\D_{\mu}\} = -\ii~ \Delta_{+\mu},~~~~~~~~~~\{\tilde{\D},\D_{\mu}\} = 
 \ii~ \epsilon_{\mu \nu}~ \Delta_{-\mu},
  \label{laD}
  \eeq
  with all the other anticommutators vanishing.
  Again an explicit expression for $\D_A$ satisfying (\ref{laD}) can easily be
  found:
   \ba
  &\D= \frac{\partial}{\partial \theta} -\frac{\ii}{2} \theta^{\mu} 
  \Delta_{+\mu},~~~~~~\tilde{\D} = \frac{\partial}{\partial \tilde{\theta}} 
  +\frac{\ii}{2} \epsilon_{\mu \nu} 
\theta^{\mu} \Delta_{-\nu},&   \nonumber \\
  &~~~~~~~~\D_{\mu}= \frac{\partial}{\partial \theta^{\mu}}-\frac{\ii}{2} \left(  
  \theta  \Delta_{+\mu}- \tilde{\theta} 
\epsilon_{\mu \nu} \Delta_{-\nu}\right).~~~~~~&
  \label{Dgen}
  \ea
  Alternatively, as in the case of the supercharges $\Q_A$ it is
  possible to introduce the arrowed differential operators $\auD_A$ defined by
  \be
  \D_A = T(2 \a_A) \auD_A.
  \label{uaD}
  \eeq
  Their explicit form is quite similar to the one given for  $\Q_A$ in
  (\ref{susygen2}), namely:
   \ba
  &\auD=  \frac{\partial}{\partial \th} -\frac{\ii}{2} 
  \th^{\mu}\arrowdel_{+\mu}
  ,~~~~~~\overrightarrow{\tilde{\uD}} = \frac{\partial}{\partial \tilde{
  \th}} +\frac{\ii}{2} \epsilon_{\mu \nu} \th^{\mu} 
  \arrowdel_{-\nu}, & 
  \nonumber \\
  &~~~~~~~~\auD_{\mu}= \frac{\partial}{\partial \th^{\mu}}-
  \frac{\ii}{2} \left( \th\arrowdel_{+\mu} - \epsilon_{\mu \nu}
  \tilde{\th} \arrowdel_{-\nu}  \right).~~~~~~&
  \label{uaD2}
  \ea
  
  It is convenient, as in the continuum case, to write the differential
  operators $D_A$ as
  \ba
  &\U\D\U^{-1}=  \frac{\partial}{\partial \theta}~ ,~~~~~~ 
  \U\tilde{\D}\U^{-1} 
  = \frac{\partial}{\partial \tilde{\theta}},~ &\nonumber \\
  &~~~~~~~~\U^{-1}\D_{\mu}\U = \frac{\partial}{\partial \theta^{\mu}},~~~~~~&
  \label{Dgen2}
  \ea
  where the operator $\U$ is given by
  \be
  \U = \ee^{-\frac{\ii}{2} \left( \theta \theta^{\mu} 
  \Delta_{+\mu} -
   \tilde{\theta} \epsilon_{\mu \nu} \theta^{\mu} \Delta_{-\nu}
   \right) }= \ee^{-\frac{\ii}{2} 
  \left( \th \th^{\mu}\arrowdel_{+\mu} -  \tilde{\th} 
  \epsilon_{\mu \nu} \th^{\mu}\arrowdel_{-\nu} \right) }.
  \label{Uop}
  \eeq
  We shall now consider chiral and anti-chiral superfields on the lattice, which
  we shall denote respectively $\uPsi(x,\th)$ and $\overline{\uPsi}(x,\th)$.
  They are defined by imposing on a generic superfield $\uFc (x,\th)$ the chiral
  and anti-chiral conditions, namely:
  
  \ba
  &&\D~\uPsi(x,\th) \equiv \left[ \D , \uPsi(x,\th)\right\}=0,~~~~~~\tilde{\D}~
  \uPsi(x,\th) \equiv \left[ \tilde{\D} , \uPsi(x,\th)\right\}=0,
 \label{chcond}
  \\ &&~~~~~~~~~~~~~~~~~\D_\mu~\overline{\uPsi} \equiv \left[ \D_\mu ,
  \overline{\uPsi}(x,\th)\right\}=0.
  \label{antichcond}
  \ea
  The general solution of the chiral and anti-chiral conditions (\ref{chcond})
  and (\ref{antichcond}) can be found by using the form (\ref{Dgen2}) of the
  differential operators $\D_A$. In fact if we define 
  \ba
  \U \uPsi(x,\th) \U^{-1}&=& \uPsi^{'}(x,\th),  \label{chf} \\
  \U^{-1}\overline{\uPsi}(x,\th)\U  &=& \overline{\uPsi}^{'}(x,\th),
  \label{antichf}
  \ea
  the conditions (\ref{chcond}) and (\ref{antichcond}) respectively become:
  \be
  \frac{\partial}{\partial \theta}~ \uPsi^{'}(x,\th)=0,~~~~~~~~~\frac{\partial}
  {\partial \tilde{\theta}}~\uPsi^{'} (x,\th)=0,
  \label{chcond2}
  \eeq
  and
  \be
  \frac{\partial}{\partial \theta^\mu}~\overline{\uPsi}^{'}(x,\th)=0.
  \label{antichcond2}
  \eeq
 Hence the reduced superfields $\uPsi^{'} (x,\th)$ and $\overline{\uPsi}^{'}
 (x,\th)$ are functions only of $(x,\th^\mu)$ and of $(x,\th,\tilde{\th})$
 respectively:
 \ba
 \uPsi^{'} (x,\th)&=& \uphi(x)+ \th^\mu \upsi_\mu(x+\a_\mu) +  
 \th^1 \th^2 \uphit(x+\a_1 +\a_2), \label{csc} \\
 \overline{\uPsi}^{'}(x,\th)&=& \uvarphi(x) + \th \uchi(x+\a) + \tilde{\th}
 \uchit(x + \at) + \th \tilde{\th} \uvarphit(x+\a+\at). \label{acsc}
 \ea
 Finally, by inserting (\ref{csc}) and (\ref{acsc}) respectively into (\ref{chf})
 and (\ref{antichf}) a straightforward calculation gives the explicit expansion 
 of a chiral and anti-chiral superfield on the lattice:
  \ba
  \uPsi(x,\th)& =&  \uphi(x) + \th^{\mu} \upsi_{\mu} ( x+ \a_{\mu}) 
  + \frac{1}{2} 
  \epsilon_{\mu \nu} \th^{\mu} \th^{\nu} \uphit(x+\a_1+\a_2) 
  + \frac{\ii}{2} \th\th^{\mu} \upar^{(s)}_{\mu} \uphi(x+\m)  \nonumber \\
  & & -\frac{\ii}{2} \tilde{\th} \epsilon_{\mu \nu} \th^{\mu} 
  \upar^{(s)}_{\nu} \uphi(x-\n) 
+ \frac{\ii}{4} \th  \epsilon_{\mu \nu} \th^{\mu} 
  \th^{\nu} \epsilon_{\rho \sigma} \upar^{(s)}_{\rho} 
  \upsi_{\sigma}(x  +\hat{n}_1 +\a_2)
  \nonumber \\ & & -\frac{\ii}{4} \tilde{\th} \epsilon_{\mu \nu} \th^{\mu} 
  \th^{\nu} \upar^{(s)}_{\rho} \upsi_{\rho}(x - \a) +  \frac{1}{4} 
  \th \tilde{\th} \epsilon_{\mu \nu} \th^\mu \th^\nu 
  \upar^{(s)}_{\rho} \upar^{(s)}_{\rho} \uphi(x), 
  \label{chsupex}
  \ea
  \ba
  \overline{\uPsi}(x,\th)&=& \uvarphi(x) + \th \uchi(x+\a) + \tilde{\th}
  \uchit(x+\tilde{\a}) + \th \tilde{\th} \uvarphit(x+\a+\at) \nonumber \\
  & &-\frac{\ii}{2} \th \th^{\mu} \upar^{(s)}_{\mu} \uvarphi(x+\m) 
  +  \frac{\ii}{2} 
  \tilde{\th} \th^{\mu} \epsilon_{\mu \nu} \upar^{(s)}_{\nu}\uvarphi(x-\n)+
   \frac{\ii}{2} \th \tilde{\th} \epsilon_{\mu \nu} \th^{\mu}
  \upar^{(s)}_{\nu} \uchi(x+\m+\at)  \nonumber \\& & + \frac{\ii}{2} \th 
  \tilde{\th} \th^\mu 
  \upar^{(s)}_{\mu} \uchit(x+\m+\at) + \frac{1}{4} \th \tilde{\th} 
  \epsilon_{\mu \nu} \th^\mu \th^\nu \upar^{(s)}_{\rho} \upar^{(s)}_{\rho} 
  \uvarphi(x).
  \label{antichsupex}
  \ea
   Again the above expansion of chiral and anti-chiral superfields are exactly 
   the same as one would obtain in the continuum, with the partial derivative
   replaced by the symmetric finite difference on the lattice and suitable
   shifts in the arguments of the fields. 



\section{$N=2$ SUSY invariant BF and Wess-Zumino actions on the lattice}
\label{BF_WZ_lat}


Based on the formulation of $N=2$ twisted superspace formalism on the lattice, 
we can explicitly construct SUSY invariant BF and Wess-Zumino actions on the lattice. 
The construction proceeds quite parallel to the continuum formulation with 
the introduction of suitable noncommutative shifts for the Abelian case. 
In the case of non-Abelian extension we need to take care of the nontrivial 
shift dependence of the superfield.

\subsection{Abelian super BF in two dimensions}

We first consider the case of fermionic chiral and anti-chiral superfield which 
leads to the twisted $N=2$ SUSY invariant quantized BF action on the lattice. 
We expand the reduced superfield $\uPsi'(x,\th)$ and $\overline{\uPsi}'(x,\th)$ 
given in (\ref{csc}) and (\ref{acsc}) with the same renaming of the fields as 
the continuum expression (\ref{expansion_psi_psibar}):
  \bea
  \uPsi' (x^\mu,\th^\mu)  
&=& U\uPsi(x^\mu,\th,\th^\mu,\tilde{\th})U^{-1} 
= \ee^{\th^\mu \us_\mu}i\uc(x) \nonumber \\
&=&i\uc(x)+i\th^\mu \us_\mu \uc(x) +i\th^{1} \th^{2} \us_2 \us_1 \uc(x)
 \nonumber \\
&=& i\uc(x) + \th^{\mu} \uomega_{\mu}(x+\a_{\mu}) + 
  i\th^{1} \th^{2} \ulambda(x+\a_1+\a_2), \nonumber \\
  \overline{\uPsi}'(x^\mu,\th,\tilde{\th}) 
&=&U^{-1}\overline{\uPsi}(x^\mu,\th,\th^\mu,\tilde{\th})U 
= \ee^{\th \us + \tilde{\th} \tilde{\us}}i\bar{\uc}(x) \nonumber \\
 &=& i\bar{\uc}(x)+i\th \us \bar{\uc}(x)
  +i\tilde{\th}\tilde{\us}\bar{\uc}(x)
  +i\th \tilde{\th} \tilde{\us} \us \bar{\uc}(x) \nonumber \\
  &=& i\bar{\uc}(x) + \th \ub(x+\a) + \tilde{\th}\uphi(x+\at)
       -i \th ~\tilde{\th} \urho(x+\a+\at).
  \label{chilat}
  \eea
These are the lattice counterpart of the relations in (\ref{chi_anti-chi_superfield}) 
and provide additional reasonings why the component fields in the chiral and 
anti-chiral superfields need the corresponding shifts.   
Since the supercharge $s_A$ carries the noncommutative shift $2\a_A$, 
the lattice coordinate of $s_A\uvarphi(x)$ is naturally correlated with 
the $\a_A^\mu$ direction. In order to keep the symmetric nature of the formulation 
as we have shown in the last section of (\ref{superfield}), we define the new field 
$\uvarphi(x+\a_A)=s_A\uvarphi(x)$ on the center of the coordinate between $x^\mu$ and 
$x^\mu+2a^\mu_A$. See Fig. \ref{fig:field-shift}. 
\begin{figure}
 \begin{center}
 \input{transformation.tex}
 \end{center}
  \caption{Relative location of $\uvarphi(x)$ and $\us_A\uvarphi(x)$ on the lattice.}
  \label{fig:field-shift}
\end{figure}

Compared with the general superfield of (\ref{eq:Fl}), the chiral and 
anti-chiral superfields have smaller number of elementary components which are 
again nonlocally scattered within the double size lattice. 
For the cases of the symmetric choice of the shift parameter A) in (\ref{a}) 
and the asymmetric choice B) in (\ref{a-asym}), we show how the component fields of 
the chiral and anti-chiral superfields are scattered on the lattice in 
Fig. \ref{fig:superfield-sym} and Fig. \ref{fig:superfield-asym}, respectively.
For the case of the symmetric choice of the parameter A), 
the fermionic fields are 
on the integer lattice sites or equivalently the original lattice sites while 
the bosonic fields are on the half integer lattice namely on the dual 
lattice sites. For the case of the asymmetric parameter choice B), fermions and 
bosons are mixed up on the integer lattice sites. It is interesting to note that 
the nonlocal extension of the fields is limited in the double size lattice for both 
cases. This structure of the locations of component fields on the lattice is expected 
from the general arguments of Section \ref{exact_lat}.
 
\begin{figure}
\begin{minipage}[b]{.47\linewidth}
  \begin{center}
   \hspace*{-2em}
  \scalebox{.8}{
 \input{superspace_sym.tex}
 }
 \end{center}
 \caption{Component fields of fermionic chiral and anti-chiral superfields for 
the symmetric choice of shift parameters A).}
 \label{fig:superfield-sym}
\end{minipage}
\hspace{.05\linewidth}
\begin{minipage}[b]{.47\linewidth}
  \begin{center}
  \hspace*{-2em}
  \scalebox{.8}{
  \input{superspace_nat.tex}
  }
 \end{center}
 \caption{Component fields of fermionic chiral and anti-chiral superfields for 
the asymmetric choice of shift parameters B).}
 \label{fig:superfield-asym}
\end{minipage}
\end{figure}


The SUSY transformation of the component fields of the chiral 
and anti-chiral superfields on the lattice can be obtained from 
\begin{align}
\us_A \uPsi^{\prime} (x^\mu,\th^\mu) \ &= \ 
\auQ^{\prime}_A \uPsi^{\prime} (x^\mu,\th^\mu)
+\uPsi^{\prime} (x^\mu+2\a_A,\th^\mu)\auQ^{\prime}_A ,\nonumber \\  
\us_A \overline{\uPsi}^\prime (x^\mu,\th,\tilde{\th}) \ &= \ 
\auQ^{\prime\prime}_A \overline{\uPsi}^\prime (x^\mu,\th,\tilde{\th})
+\overline{\uPsi}^\prime (x^\mu+2\a_A,\th,\tilde{\th})\auQ^{\prime\prime}_A , 
\label{lattsusy_trasformation_1}
\end{align}
where the lattice version of $\auQ^{\prime}_A$ and $\auQ^{\prime\prime}_A$ 
in (\ref{susy_trasformation_5}) are given by
\begin{align}
\auQ^{\prime}  &=  \frac{\partial}{\partial \th}
  + i \th^\mu\arrowdel_{+\mu}, \ 
&\overrightarrow{\tilde{\uQ}}^{\prime}&=  \frac{\partial}{\partial \tilde{\th}}
  - i \th^\mu\epsilon_{\mu\nu}\arrowdel_{-\nu}, \ 
&\auQ^{\prime}_\mu &=  \frac{\partial}{\partial \th^\mu}, 
\label{susy_trasformation_4a}\\
\auQ^{\prime\prime}  &=  \frac{\partial}{\partial \th}, \ 
&\overrightarrow{\tilde{\uQ}}^{\prime\prime}
&=  \frac{\partial}{\partial \tilde{\th}}, \ 
&
\lefteqn{\hspace*{-4em}
 \auQ^{\prime\prime}_\mu =  \frac{\partial}{\partial \th^\mu} 
 + i \th\arrowdel_{+\mu} - i \tilde{\th}\epsilon_{\mu\nu}\arrowdel_{-\nu}}
&\hspace*{10em}.\nonumber\\
\label{lattsusy_trasformation_2}
\end{align}
$N=2$ twisted SUSY transformation for the component fields of 
chiral and anti-chiral fields on the lattice is given in Table \ref{ABFtrans}, where the 
symmetric difference operator defined in (\ref{sdiff}) is used. 
Each of the term in the table has natural geometrical meaning. For example 
$\us_\mu \uc(x)$ defines a new field $-i\uomega_\mu(x+\a_{\mu})$ on the lattice 
site $x+\a_{\mu}$ as we can see in Fig.\ref{fig:superfield-sym}. Then 
$\us \uomega_\nu(x+\a_{\nu})$ defines the symmetric difference of the field $\uc$ 
on the site $x+\a_{\nu}+\a=x+\n$ in $\nu$ direction. 

\begin{table}
\renewcommand{\arraystretch}{1.3}
\renewcommand{\tabcolsep}{5pt}
 \begin{tabular}{c|c|c|c}
\hline
 $\uphi^{A}$ & $\us\uphi^{A}$ & $\us_\mu\uphi^{A}$ & $\tilde{\us}\uphi^{A}$  \\
  \hline
  $\uc(x)$ & $0$ &  $-i\uomega_\mu(x+\a_{\mu})$ & $0$\\
  $\uomega_\nu(x+\a_{\nu})$& $\upar^{(s)}_{\nu}\uc(x+\n)$ &
  $-i\epsilon_{\mu\nu}\ulambda(x+\a_1+\a_2)$ & 
 $-\epsilon_{\nu\rho}\upar^{(s)}_{\rho}c(x-\hat{n}_{\rho})$  \\
 $\ulambda(x+\a_1+\a_2)$ &
  $\epsilon_{\rho\sigma}\upar^{(s)}_{\rho}
  \uomega_{\sigma}(x+\hat{n}_\rho+\a_\sigma)$
 & $0$ & $-\upar^{(s)}_{\rho}\uomega_{\rho}(x-\a)$ \\
  \hline
  $\bar{\uc}(x)$ & $-i\ub(x+\a)$ & $0$ & $-i\uphi(x+\at)$ \\
  $\ub(x+\a)$ & $0$ & $\upar^{(s)}_{\mu}\bar{\uc}(x+\m)$ 
  & $-i\urho(x+\a+\at)$ \\
  $\uphi(x+\at)$ & $i\urho(x+\a+\at)$ 
& $-\epsilon_{\mu\rho}\upar^{(s)}_{\rho}\bar{\uc}(x-\hat{n}_{\rho})$ 
& $0$\\
  $\urho(x+\a+\at)$ & $0$ &
  $-\upar^{(s)}_{\mu}\uphi(x+\m+\at)$
  & $0$ \\[-3pt]
 & & $-\epsilon_{\mu\rho} \upar^{(s)}_{\rho}\ub(x-\hat{n}_{\rho}+\a)$ &
\\[2pt]
\hline
 \end{tabular}
\caption{N=2 twisted SUSY transformation for Abelian BF model.}
\label{ABFtrans}
\end{table}

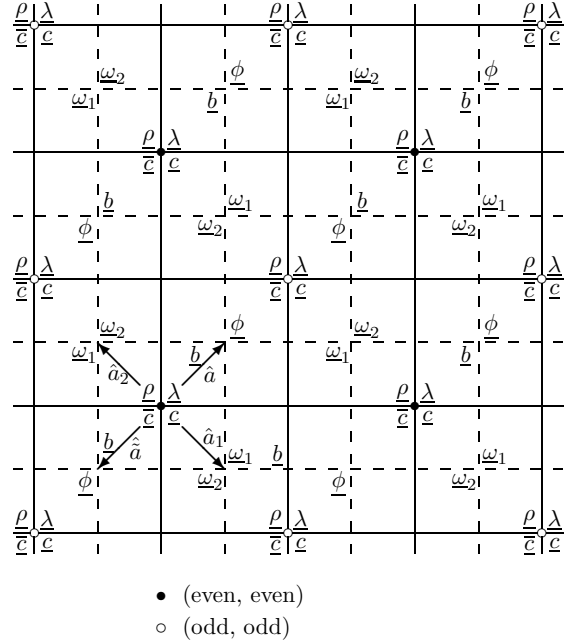
\begin{figure}
 \begin{center}
  \scalebox{.8}{
  \input{superspace_sym2.tex}
  }
 \end{center}
\caption{Component fields of Abelian super BF action on the lattice for 
the symmetric choice of the shift parameters A).}
\label{fig:lattaction-sym}
\end{figure}

The twisted $N=2$ SUSY invariant BF action is obtained from a bilinear form of the 
lattice version of the chiral and anti-chiral fields as in the continuum: 
 \begin{eqnarray}
  S_{BF}^{lat} &=& \int d^4\th ~ \sum_x ~ 
i\overline{\uPsi}(x,\th,\th_\mu,\tilde{\th})~\uPsi(x,\th,\th_\mu,\tilde{\th}) \\
   &=& 
    \sum_x \us \tilde{\us} \us_1 \us_2  \bigl(-i\bar{\uc}(x)\uc(x)\bigr)\\
   &=& \sum_x\Bigl[
        \uphi(x-\at)\epsilon_{\mu\nu}\upar^{(s)}_{\mu}\uomega_\nu(x-\at)
   + \ub(x-\a) \upar^{(s)}_{\mu}\uomega_\mu(x-\a)\nonumber\\
   & & 
   -i\bar{\uc}(x) \upar^{(s)}_{\mu}\upar^{(s)}_{\mu} \uc(x)
    +i\urho(x-\a-\at)\ulambda(x-\a-\at)
	  \Bigr]\\ [5pt]
\label{BFlattaction-sym}
   &=& \sum_x\Bigl[
        \uphi(x-\at)\epsilon_{\mu\nu}\upar_{+\mu}\uomega_\nu(x+\a_{\nu})
   + \ub(x-\a) \upar_{-\mu}\uomega_\mu(x+\a_{\mu})\nonumber\\
   & & 
   -i\bar{\uc}(x) \upar_{+\mu}\upar_{-\mu} \uc(x)
    +i\urho(x-\a-\at)\ulambda(x-\a-\at)
	  \Bigr],
\label{BFlattaction}
 \end{eqnarray}
where two equivalent expressions with symmetric difference operators and with forward 
and backward difference operators are given. In the case of the symmetric difference 
operator the assigned field coordinate is located in the middle of the coordinate of 
the difference fields while 
in the case of the forward and backward difference operator the assigned field 
coordinate is the ending and starting site of the corresponding difference. 
The twisted $N=2$ SUSY invariance up to the surface terms is obvious since the action 
has an exact form with respect to the twisted lattice supercharges. 
We show the component fields of super BF action on the lattice for the 
symmetric choice of the shift parameters A) 
in Fig.\ref{fig:lattaction-sym}. 

Component fields except for the leading one are generated by operating 
supercharges on the leading and carry the same noncommutativity as 
supercharges.
For example, $\uomega_\mu(\a_\mu) = i\us_\mu \uc(x)$ carries the
noncommutative shift $2\a_\mu$ in addition to a possible shift carried by 
$\uc(x)$. Here we assume $\uc(x)$ and $\bar{\uc}(x)$ do not carry a shift. 
Then we should be careful when
interchanging the order of product.  In the action, however, we can freely 
interchange the field cyclicly without shift of the arguments:
\begin{eqnarray}
 S_{BF}^{lat}
 &\sim&
 \sum_x \left[\us \bar{\uc}(x+2\at+2\a_1+2\a_2) \tilde{\us}\us_1\us_2 \uc(x)
  +\cdots \right]\\
 &=&
  \sum_x \left[\tilde{\us}\us_1\us_2 \uc(x+2\a) \us \bar{\uc}(x)
  +\cdots \right]\\
 &=&
  \sum_x \left[\tilde{\us}\us_1\us_2 \uc(x) \us \bar{\uc}(x-2\a)
  +\cdots \right].
 \label{cyclic}
\end{eqnarray}
In the first equality, we change the order of the product by taking 
into account the noncommutativity.  Next we shift the  argument in the 
summation. Since the shift parameters satisfy eq.(\ref{zerosum}),
\begin{equation}
 \a+\a_1+\a_2+\at=0,
\end{equation}
$-2\a$ in eq.(\ref{cyclic}) can be replaced by $2\at+2\a_1+2\a_2$ and thus we
recover the original argument of $\bar{\uc}$.
It is straightforward to generalize this argument into a cyclic interchange of 
three or more fields.

It is interesting to note that each term in the action has geometrical meaning. 
For example the field $\uphi(x-\at)$ is located in the middle of the surrounding fields 
$\uomega_2(x-\at\pm \hat{n}_1)$ and $\uomega_1(x-\at\pm \hat{n}_2)$ which define the 
rotation $\epsilon_{\mu\nu}\upar^{(s)}_{\mu}\uomega_\nu(x-\at)$. 
The field $\ub(x-\a)$ is located at the center of a cross composed by 
$\uomega_1(x-\a\pm \hat{n}_1)$ and $\uomega_2(x-\a\pm \hat{n}_2)$ which define 
the divergence $\upar_{-\mu}\uomega_\mu(x+\a_{\mu})$.
In the case of the symmetric parameter choice A), $\bar{\uc}(x),\uc(x)$ and 
$\urho(x-\a-\at),\ulambda(x-\a-\at)$ are located in the same lattice site $x$ 
since $\a+\at=0$. 

It is important to recognize that $\bar{\uc}(x)$ and 
($\uphi(x-\at)$, $\ub(x-\a)$) are in the same coordinate sum of the 
action (\ref{BFlattaction-sym}) 
while ($\uphi(x-\at)$, $\ub(x-\a)$) are not included in the component expansion 
of the anti-chiral superfields $\overline{\uPsi}'(x,\th,\tilde{\th})$ in (\ref{chilat}) where 
$\bar{\uc}(x)$ is the leading component. 
Instead $\uphi(x-\at)$ and $\ub(x-\a)$ belong to the component expansion of 
$\overline{\uPsi}'(x^\mu+\hat{n}_1+\hat{n}_2,\th,\tilde{\th})$ and 
$\overline{\uPsi}'(x^\mu-\hat{n}_1-\hat{n}_2,\th,\tilde{\th})$, respectively. 
Therefore for the symmetric choice A) the summation of the action should cover 
the coordinates of (even,even)=$(2n_1,2n_2)$ and (odd,odd)=$(2n_1+1,2n_2+1)$ 
with $n_1,n_2 \in {\bf Z}$:
\be
\sum_x=\sum_{\hbox{(even,even)}} + \sum_{\hbox{(odd,odd)}}.
\label{even-odd-sum-1}
\eeq
As we have explained in Section \ref{exact_lat}, 
this structure of summation region of lattice 
sites naturally appears from the general arguments as well. 

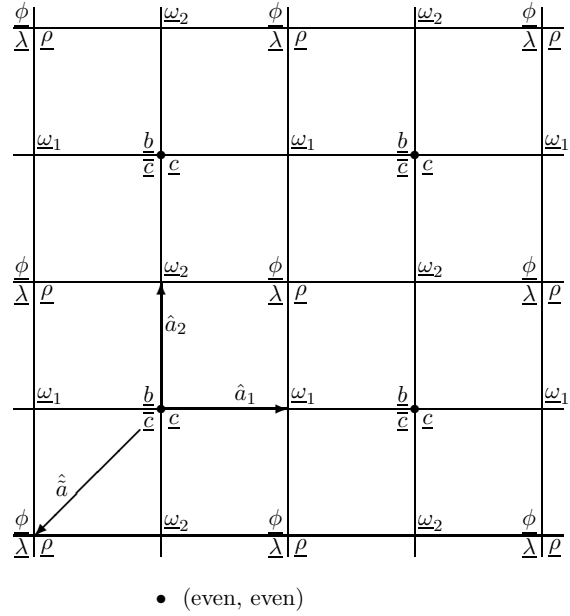
\begin{figure}
 \begin{center}
  \scalebox{.8}{
  \input{superspace_nat2.tex}
  }
 \end{center}
\caption{Component fields of Abelian super BF action on the lattice for 
the asymmetric choice of the shift parameters B).}
\label{fig:lattaction-asym}
\end{figure}

We also show the component fields of super BF action on the lattice for 
the asymmetric choice of the shift parameters B) 
in Fig.\ref{fig:lattaction-asym} 
where only the summation of (even,even) sites appears for coordinate sum of 
the action, which can be understood from the general arguments in the last section.


\subsection{$N=2$ supersymmetric Wess-Zumino action in two dimensions}

We next consider the bosonic version of the chiral and anti-chiral superfields 
(\ref{csc}) and (\ref{acsc}):
\begin{eqnarray}
  \underline{\Phi}'(x^\mu,\th^\mu)
&=& U\uPhi(x^\mu,\th,\th^\mu,\tilde{\th})U^{-1} \nonumber \\
   &=& \uphi(x) + \th^\mu\upsi_\mu(x+\a_\mu) + \th^2 \uphit(x+\a_1+\a_2) \nonumber \\
 \overline{\underline{\Phi}}'(x^\mu,\th,\tilde{\th})
&=&U^{-1}\overline{\uPhi}(x^\mu,\th,\th^\mu,\tilde{\th})U  \nonumber \\
   &=& \uvarphi(x) + \th\uchi(x+\a) + \tilde{\th}\uchit(x+\at)
  + \th\tilde{\th}\uvarphit(x+\a+\at),
\end{eqnarray}
where we rename $\uPsi'=\underline{\Phi}'$, 
$\overline{\uPsi}'=\overline{\underline{\Phi}}'$ to show the bosonic nature of the 
superfields. 
Compared with the fermionic superfields, the Grassmann nature of the component fields 
in superfields is interchanged.
Similar to the fermionic superfields, bosonic chiral and anti-chiral superfields include 
fermionic and bosonic component fields scattered within a double size lattice 
as shown in Fig.\ref{fig:bosonic-sufield-sym} for the symmetric shift parameter 
choice A), and in Fig.\ref{fig:bosonic-sufield-asym} for the asymmetric shift 
parameter choice B). 
In contrast with the component fields of the fermionic superfields the bosonic 
fields, $\uphi(x), \uphit(x), \uvarphi(x)$ and $\uvarphit(x)$ are located 
on the integer sites or equivalently on the original lattice sites while the 
fermionic fields are on the dual sites for the symmetric shift parameter choice A). 
The interchange of relative position of fermionic and bosonic fields on the lattice 
for the asymmetric shift parameter choice B) works similar to the symmetric 
case.

\begin{figure}
 \begin{minipage}[b]{.47\linewidth}
  \begin{center}
  \hspace*{-2em}
  \scalebox{.8}{
  \input{b_superspace_sym.tex}
  }
 \end{center}
 \caption{Component fields of bosonic chiral and anti-chiral superfields for the
 symmetric parameter choice A).}
 \label{fig:bosonic-sufield-sym}
 \end{minipage}
\hspace{.04\linewidth}
\begin{minipage}[b]{.48\linewidth}
  \begin{center}
   \hspace*{-2em}
   \scalebox{.8}{
  \input{b_superspace_nat.tex}
  }
 \end{center}
 \caption{Component fields of bosonic chiral and anti-chiral superfields for the
 asymmetric choice of shift parameter B).}
 \label{fig:bosonic-sufield-asym}
\end{minipage}
\end{figure}

The $N=2$ twisted SUSY transformation of component fields of superfields can be 
obtained similar to the fermionic version by 
(\ref{lattsusy_trasformation_1}) with the same lattice supercharge operators 
as (\ref{susy_trasformation_4a}) and (\ref{lattsusy_trasformation_2}). 
We list the SUSY transformation of the component fields on the lattice 
in Table \ref{tab:trans_bosonic_gen}. 
The relative location between the original field and the transformed field has a 
natural geometrical interpretation due to the shifting nature of the twisted 
supercharge operators like the fermionic superfields. 

\begin{table}
\renewcommand{\arraystretch}{1.3}
 $
 \begin{array}{c|c|c|c}
\hline
\uphi^A  & \us\uphi^A & \us_\mu\uphi^A & \tilde{\us}\uphi^A \\
\hline
  \uphi(x)  & 0 & \upsi_\mu(x+\a_\mu)  & 0\\
  \upsi_\nu(x+\a_\nu) & -i\upar_{+\nu}\uphi(x) & -\epsilon_{\mu\nu}\uphit(x+\a_1+\a_2) & i\epsilon_{\nu\mu}\upar^{-\mu}\phi(x)  \\
  \uphit(x+\a_1+\a_2) & i\epsilon^{\mu\nu}\upar_{+\mu}\psi_\nu(x+\a_\nu)  & 0 &  -i\upar_{-\mu}\psi^\mu(x+a_\mu) \\
\hline
  \uvarphi(x) & \uchi(x+\a)  & 0 &  \uchit(x+\at) \\
  \uchi(x+\a)  & 0 &  -i\upar_{+\mu}\uvarphi(x) &  \uvarphit(x+\a+\at) \\
  \uchit(x+\at) & -\uvarphit(x+\a+\at) & i\epsilon_{\mu\nu}\upar^{-\nu}\uvarphi(x) & 0\\
  \uvarphit(x+\a+\at)  & 0 &  i\epsilon_{\mu\nu}\upar^{-\nu}\uchi(x+\a)+i\upar_{+\mu}\uchit(x+\at) & 0\\
\hline
 \end{array}
 $
\caption{N=2 twisted SUSY transformation of component fields for the bosonic superfield.}
\label{tab:trans_bosonic_gen}
\end{table}

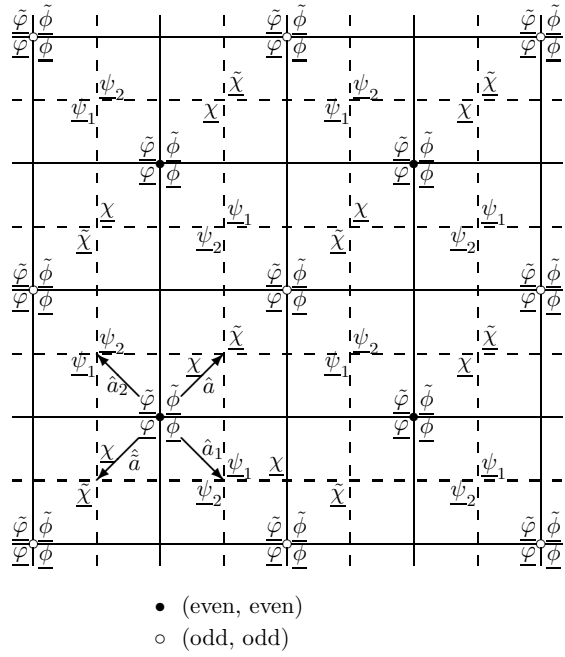
\begin{figure}
 \begin{center}
  \scalebox{.8}{
  \input{b_superspace_sym2.tex}
  }
 \end{center}
\caption{Component fields of Wess-Zumino action for the
 symmetric choice of the shift parameters A).}
\label{fig:Wess-Zumino-sym}
\end{figure}

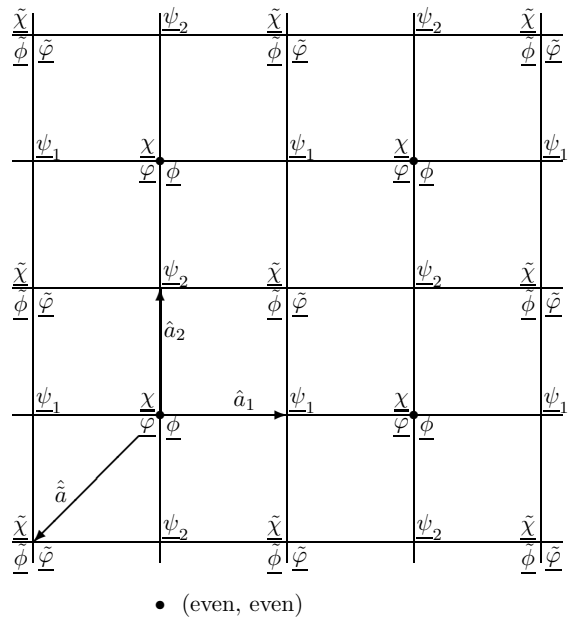
\begin{figure}
 \begin{center}
   \scalebox{.8}{
   \input{b_superspace_nat2.tex}
   }
 \end{center}
\caption{Component fields of Wess-Zumino action for the
 asymmetric choice of the shift parameters B).}
\label{fig:Wess-Zumino-asym}
\end{figure}

We can now obtain $N=2$ twisted super symmetric action by the bilinear product 
of the chiral and anti-chiral bosonic superfields which has an exact form with 
respect to all $N=2$ twisted supercharges:
\begin{eqnarray}
 S_{WZ}^{lat} &=& \int d^4\th ~ \sum_x ~ 
\overline{\uPhi}(x,\th,\th_\mu,\tilde{\th})~\uPhi(x,\th,\th_\mu,\tilde{\th}) 
\nonumber \\
&=& \us \tilde{\us} \us_1 \us_2   \sum_x\uvarphi(x)\uphi(x) \nonumber\\
 &=& \sum_x \Bigl[
    \uvarphi(x)\upar^{+\mu}\upar_{-\mu} \uphi(x)
   + \uvarphit(x+\a_1+\a_2)\uphit(x+\a_1+\a_2) \nonumber\\
 &&
    +i  \uchi(x-\a)\upar_{-\mu} \upsi_\mu(x+\a_\mu)
      +i \epsilon_{\mu\nu}\uchit(x-\at)\upar_{+\mu}
       \upsi_\nu(x+\a_\mu) \Bigr]  \nonumber\\
 &=& \sum_x \Bigl[
   -\uphi_i(x) \upar^{+\mu}\upar_{-\mu}\uphi_i(x)
   -\uF_i(x+\a_1+\a_2) \uF_i(x+\a_1+\a_2)\nonumber\\
 && +i \uxibar_{i\alpha}(x)(\gamma_\mu)_{\alpha\beta}
     \frac{\upar_{+\mu}+\upar_{-\mu}}{2}\uxi_{\beta i}(x)
    -i \uxibar_{i\alpha}(x)(\gamma_5)_{\alpha\beta}
     \frac{\upar_{+\mu}-\upar_{-\mu}}{2}\uxi_{\beta j}(x)
    (\gamma_5\gamma_\mu)_{ji}
    \Bigr]. \nonumber\\
\end{eqnarray}
Here we redefine new bosonic fields by the linear combination of 
the bosonic component fields of superfields while 
we recombine the scalar, vector and tensor fermionic fields into the 
Dirac fields (to be precise Majorana field) by the Dirac-K\"ahler fermion 
mechanism as follows:
\begin{eqnarray}
 \uF_1(x+\a_1+\a_2)
  &=& \frac{1}{2}\left(\uphit(x+\a_1+\a_2)
      -\uvarphit(x+\a_1+\a_2)\right),   \nonumber \\
 \uF_2(x+\a_1+\a_2)
  &=& \frac{i}{2}\left(\uphit(x+\a_1+\a_2)
      +\uvarphit(x+\a_1+\a_2)\right),   \nonumber \\
 \uphi_1(x) &=& \frac{1}{2}\left(\uphi(x)-\uvarphi(x)\right),  \nonumber \\
 \uphi_2(x) &=& \frac{i}{2}\left(\uphi(x)+\uvarphi(x)\right),  \nonumber \\
 \uxi_{\alpha i}(x)
  &=& \frac{1}{2} \left(\uchi(x-\a)+\gamma_\mu\upsi_\mu(x-\a+\m)
       + \gamma_5\uchit(x-\a+\hat{n}_1+\hat{n}_2)\right)_{\alpha i}.
  \nonumber\\
\label{WZ-component-fields}
\end{eqnarray}
We can recognize that the first three terms in the last action are the 
two copies of the Wess-Zumino action in two dimensions. 
On the other hand the last term in the action mixes the components of 
Dirac fermions by $N=2$ SUSY transformation, which is necessary to keep 
the exact $N=2$ SUSY on the lattice for the whole action.
We thus recognize that this action is the Wess-Zumino action on the lattice.  
We show the configuration of the component fields for the Wess-Zumino action 
on the lattice in Fig.\ref{fig:Wess-Zumino-sym} for the symmetric shift 
parameter choice A), where the summation of the action should again cover 
the coordinates of (even,even) and (odd,odd) sites:  
\be
\sum_x=\sum_{\hbox{(even,even)}} + \sum_{\hbox{(odd,odd)}}.
\label{even-odd-sum-2}
\eeq
We show the case of the asymmetric shift parameter choice B) in 
Fig. \ref{fig:Wess-Zumino-asym}, where the coordinate sum of the action 
covers (even,even) sites.  

The third term in the action is the standard kinetic term of Dirac fermion with first 
derivative while the last term includes second derivative for the 
fermionic fields and thus higher order with respect to the lattice constant. 
Those terms stem from the well-known relations between the Dirac-K\"ahler fermion, 
staggered fermion and Kogut-Susskind fermion formulation on the lattice
\cite{G,KMN,KK}. 
It is interesting to recognize that to keep the exact SUSY invariance 
for the Wess-Zumino model on the lattice naive lattice version of the 
Wess-Zumino action is not enough to ensure the SUSY invariance on the lattice. 
The Dirac-K\"ahler fermion mechanism should be fundamentally introduced with 
extended SUSY of $N=2$, which is twisted $N=2$ SUSY in two dimensions. 
In the current formulation the component fields: $\uchi(x-\a), \upsi_\mu(x-\a+\m)$ 
and $\uchit(x-\a+\hat{n}_1+\hat{n}_2)$ can be recognized from 
(\ref{WZ-component-fields}) as the differential form 
of 0-, 1- and 2-form, respectively, and are defined on the site, link and plaquette 
of the dual lattice for the symmetric shift parameter choice A) which can 
be seen in Fig.\ref{fig:Wess-Zumino-sym}, while it has similar structure on the 
double size original lattice for the asymmetric shift parameter choice B), which 
we can recognize in Fig.\ref{fig:Wess-Zumino-asym}.   

In the above expression of the Wess-Zumino action we have used forward and 
backward difference operators to avoid unnecessary confusion instead of using 
the symmetric difference operator. 
In using the symmetric difference operator the last term of the action includes 
the difference of the symmetric difference operator which superficially vanishes. 
We need some care to translate $\upar_{+\mu}-\upar_{-\mu}$ into the difference 
of symmetric difference operator. 
In defining the new bosonic fields we have introduced imaginary component 
in the definitions of $\uF_2(x+\a_1+\a_2)$ and $\uphi_2(x)$ to make the final 
form the same as the Wess-Zumino action. This identification is related to do with 
the indefinite metric nature of the quantized ghost, which is more clearly 
stated in the formulation of $N=2$ twisted superspace formalism in the
continuum\cite{KKU}.

The standard $N=2$ SUSY transformation for the bosonic fields, 
auxiliary fields and fermion fields for the Wess-Zumino model on the lattice 
can be obtained from the corresponding twisted SUSY transformation 
of the component fields. The result is given in Appendix
\ref{app:transformation}.

\subsection{Non-Abelian extension of super BF}

The Abelian BF model constructed in the previous subsection
has no requirement for the shifting properties 
of $\uPsi(x,\th)$ and $\overline{\uPsi}(x,\th)$, 
which means that one can simply take
$\uPsi(x,\th)$ and $\overline{\uPsi}(x,\th)$ as shift-less
superfields $\Psi(x,\th)$ and $\overline{\Psi}(x,\th)$, 
respectively.
On the other hand, if one wishes
to construct a non-Abelian BF model on the lattice,
a more careful treatment would be needed and
we will actually see that the distinction between 
$\uPsi(x,\th)$ and $\Psi(x,\theta)$ 
is crucial to construct a non-Abelian model in a 
consistent way on the lattice. 

Keeping the above points in mind, one may begin with  
the following lattice counterparts of non-Abelian chiral and 
anti-chiral conditions (\ref{nonab_chiral_conditions1}) and 
(\ref{nonab_anti-chiral_conditions}),
\begin{eqnarray}
\D\Phi(x,\theta) 
-i\Phi(x,\theta)^2 &\equiv&
\frac{i}{2}\{\D-i\Phi(x,\theta),\D-i\Phi(x,\theta)\} = 0, \label{NAchiral_1} \\
\tilde{\D}\Phi(x,\theta) &\equiv&
\{\tilde{\D},\Phi(x,\theta)\} = 0, \label{NAchiral_2}\\
\D_{\mu}\overline{\Psi}(x,\theta) &\equiv&
\{\D_{\mu},\overline{\Psi}(x,\theta)\} = 0\label{NAchiral_3},
\end{eqnarray}
which are expressed with $\D_A$'s defined in (\ref{Dgen})
and shift-less superfields $\Phi(x,\theta)$ and
$\overline{\Psi}(x,\theta)$.
The crucial point here
is to notify from 
(\ref{NAchiral_1}) that 
if one wishes to rewrite the chiral and anti-chiral
conditions  
with the arrowed operators as in (\ref{uaD}), 
the shifting property of $\uD$ and $\uPhi(x,\theta)$ should coincide. 
In other words the shift operator to relate $\D$ and $\uD$ or $\Phi(x,\theta)$ 
and $\uPhi(x,\theta)$ should be matched and thus 
one needs to introduce $\uPhi(x,\th)$ as
\begin{eqnarray}
\Phi(x,\theta)=T(\a)\uPhi(x,\th)T(\a),
\label{Phi-shift}
\end{eqnarray}
and in a similar way for $\uPsi(x,\th)$,
\begin{eqnarray}
\Psi(x,\theta)=T(\a)\uPsi(x,\th)T(\a),
\end{eqnarray}
where $\th_A$'s are defined
in(\ref{thetashift}).
Furthermore, in order to ensure 
the shift-less property of resulting action, one also needs
to introduce $\overline{\uPsi}(x,\th)$ with a opposite
sign of shifting parameter $\a$,
\begin{eqnarray}
\overline{\Psi}(x,\theta)=T(-\a)\overline{\uPsi}(x,\th)T(-\a),\label{T(-a)}
\end{eqnarray}
so that the shift operator for the combination
$\overline{\Psi}(x,\theta)\Phi(x,\theta)$ vanishes.
Note that in the above definitions the underlined superfields 
are introduced symmetrically w.r.t. $T(\a_A)$ which reproduces the 
similar coordinate dependence as the component fields of the superfields 
in the previous subsection. 

With using the above definitions, the non-Abelian chiral conditions
(\ref{NAchiral_1}),(\ref{NAchiral_2}) and (\ref{NAchiral_3})
now can be rewritten as 
\begin{eqnarray}
\auD\uPhi(x-\a,\th)+\uPhi(x+\a,\th)\auD
-i\uPhi(x+\a,\th)\uPhi(x-\a,\th) &=& 0, \label{NAchiral_u1} \\
\overrightarrow{\tilde{\uD}}\uPhi(x-\at,\th)
+\uPhi(x+\at,\th)\overrightarrow{\tilde{\uD}}&=&0, \label{NAchiral_u2}\\
\auD_{\mu}\overline{\uPsi}(x-\a_{\mu},\th)
+\overline{\uPsi}(x+\a_{\mu},\th)\auD_{\mu}&=&0,\label{NAchiral_u3}
\end{eqnarray}
where the lattice coordinate of the underlined superfields are shifted to 
$x\pm \a_A$ by taking off
the shift operators $T(a_A)$ symmetrically from (\ref{NAchiral_1}),
(\ref{NAchiral_2}) and (\ref{NAchiral_3}). 
This symmetric coordinate dependence of the above chiral conditions 
has the same structure as that of the symmetric operator acting on 
the superfield as in (\ref{symmsa}). 

\begin{figure}
  \begin{center}
   \scalebox{.8}{
   \input{NABFfig_sym01.tex}
   }
 \end{center}
 \caption{Non-Abelian field configurations in the superfield for the
 symmetric choice of the shift parameters A).}
 \label{fig:NABFsym01}
\end{figure}
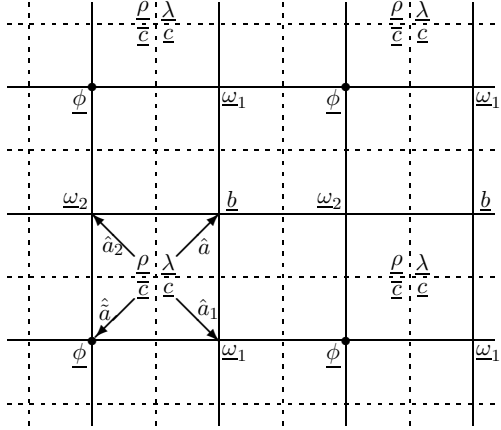

The solution for $\uPhi(x)$ can be found 
in a parallel way as in the continuum case,
\begin{eqnarray}
\uPhi(x-\a,\th)=\uPsi(x-\a,\th)+i\th\uPsi(x+\a,\th)\uPsi(x-\a,\th),
\end{eqnarray}
where $\uPsi(x\pm\a,\th)$ satisfies
\begin{eqnarray}
\auD\uPsi(x-\a,\th)
+\uPsi(x+\a,\th)\auD&=&0,\\
\overrightarrow{\tilde{\uD}}\uPsi(x-\at,\th)
+\uPsi(x+\at,\th)\overrightarrow{\tilde{\uD}}&=&0.
\end{eqnarray}
The component-wise expansions for 
$\uPsi(x,\th)$ and $\overline{\uPsi}(x,\th)$ 
can be expressed using the operator $U$
defined in (\ref{Uop}) as
\begin{eqnarray}
\uPsi(x+\a,\th)=U^{-1}\uPsi^{\prime}(x+\a,\th)U,\hspace{15pt}
\overline{\uPsi}(x+\a,\th)=U\overline{\uPsi}^{\prime}(x+\a,\th)U^{-1},
\end{eqnarray}
with
\begin{eqnarray}
&&\hspace*{-30pt}
\uPsi^{\prime}(x+\a,\th)=i\uc(x+\a)
+\th_{\mu}\uomega_{\mu}(x+\m)
+i\th_1\th_2\ulambda(x+\hat{n}_1+\a_2) , \\
&&\hspace*{-30pt}\overline{\uPsi}^{\prime}(x+\a,\th)
=i\bar{\uc}(x+\a)+\th \ub(x+2\a)
+\tilde{\th}\uphi(x+\a+\at)
-i\th\tilde{\th}\urho(x+2\a+\at).
\end{eqnarray}
The locations of the component fields appeared in the above 
superfields are depicted in Fig.\ref{fig:NABFsym01} for
the symmetric choice of the shift parameters A)
while for the asymmetric choice B) one can show that the 
configurations coincide with the Abelian case (Fig.\ref{fig:superfield-asym}).
It is important to note here that in the symmetric choice A),
the fermionic fields are located on the half-integer sites namely
dual sites while the bosonic fields are on the integer sites or 
equivalently the original sites, which is in contrast
with the situation in the Abelian BF model.
It is also interesting to point out that $\uomega_{\mu}$ 
now lives on the site $x+\m$ and carries the shift $2\m$ 
and thus has a nature of link gauge variable, which is in contrast with the 
Abelian case where $\uomega_{\mu}$ lives on the site $x+\a_\mu$ and 
carries the shift $2\a_\mu$. This difference stems from the fact 
that the superfields themselves now carry the shift for the non-Abelian 
case.

The SUSY transformation can now be obtained
in a straightforward way from
\begin{eqnarray}
  \us_A \uPhi (x+\a,\th) &=& \auQ_A ~\uPhi (x+\a,\th)+ \uPhi (x+\a+ 2 \a_A,\th)~
  \auQ_A, \\
  \us_A \overline{\uPsi} (x+\a,\th) &=& 
\auQ_A ~\overline{\uPsi} (x+\a,\th)+ \overline{\uPsi} (x+\a+ 2 \a_A,\th)~  \auQ_A,
\end{eqnarray}
whose results are summarized in Table \ref{NAtrans}
with the following notations:
\begin{eqnarray}
\upar^{(s)}_{\mu}\uvarphi(x)
 &=& \uvarphi(x+\hat{n}_{\mu})-\uvarphi(x-\hat{n}_{\mu})\label{n1},\\
\left[ \uomega_{\mu}, \uc \right](x) &\equiv& \uomega_{\mu}(x+\a)\uc(x-\m)
-\uc(x+\m)\uomega_{\mu}(x-\a)\label{n2},\\
(\epsilon_{\mu\nu}\uomega_{\mu}\uomega_{\nu})(x)
&\equiv& \epsilon_{\mu\nu}\uomega_{\mu}(x+\hat{n}_{\nu})
\uomega_{\nu}(x-\hat{n}_{\mu})\label{n3},\\
\{\uc,\ulambda\}(x)&\equiv& \uc(x+\hat{n}_1+\a_2)\ulambda(x-\a)
+\ulambda(x+\a)\uc(x-\hat{n}_1-\a_2)\label{n4}.
\end{eqnarray}

\begin{table}
\begin{center}
\renewcommand{\arraystretch}{1.3}
\renewcommand{\tabcolsep}{3pt}
\begin{tabular}{c|c|c|c}
\hline
$\uphi^A$  & $\us\uphi^A$ & $\us_{\mu}\uphi^A$  & $\tilde{\us}\uphi^A$  \\
\hline
$\uc(x+\a)$ & $-\uc(x+3\a)\uc(x+\a)$ 
& $-i\uomega_{\mu}(x+\m)$  & $0$  \\
$\uomega_{\nu}(x+\n)$ & 
$\upar^{(s)}_{\nu}\uc(x+\n+\a)$ & 
$-i\epsilon_{\mu\nu}\ulambda(x+\n+\a_{\mu})$ &
  $-\epsilon_{\nu\rho}\upar^{(s)}_{\rho}\uc(x+\n+\at)$ \\[-3pt]
&  $+[\uomega_{\nu},\uc](x+\n+\a)$  & & \\
$\ulambda(x+\hat{n}_1+a_2)$ &
 $\epsilon_{\rho\sigma}\upar^{(s)}_{\rho}
\uomega_{\sigma}(x+\hat{n}_1+\hat{n}_2)$
 & $0$ &
 $-\upar^{(s)}_{\rho}\uomega_{\rho}(x)$ \\[-3pt]
&  $+(\epsilon_{\rho\sigma}
\uomega_{\rho}\uomega_{\sigma})(x+\hat{n}_1+\hat{n}_2)
$  & & \\[-5pt]
& $-\{ \uc,\ulambda\}(x+\hat{n}_1+\hat{n}_2)$ & & \\ 
\hline
$\bar{\uc}(x+\a)$ & $-i\ub(x+2\a)$  & $0$ & $-i\uphi(x+\a+\at)$ \\
$\ub(x+2\a)$ & 0 & $\upar^{(s)}_{\mu}\bar{\uc}(x+\a+\m)$ 
& $-i\urho(x+2\a+\at)$ \\
$\uphi(x+\a+\at)$ & $i\urho(x+2\a+\at)$ &
 $-\epsilon_{\mu\rho}\upar^{(s)}_{\rho}\bar{c}(x+\a-\hat{n}_{\rho})$ & 0 \\
$\urho(x+2\a+\at)$ & 0 
& $-\upar^{(s)}_{\mu}\uphi(x+2\a-|\epsilon_{\mu\rho}|\hat{n}_{\rho})$
 & 0 \\[-3pt]
& & $-\epsilon_{\mu\rho}\upar^{(s)}_{\rho}\ub (x+2\a-\hat{n}_{\rho})$ & \\
\hline
\end{tabular}
\caption{$N=2$ twisted SUSY non-Abelian transformation for the component fields.}
\label{NAtrans}
\end{center}
\end{table}

\noindent
Since we use the symmetric difference operator, the coordinate dependence 
in the SUSY transformation table is cooperated to the SUSY operation 
as in the Table \ref{T-laws_F_lat}.  

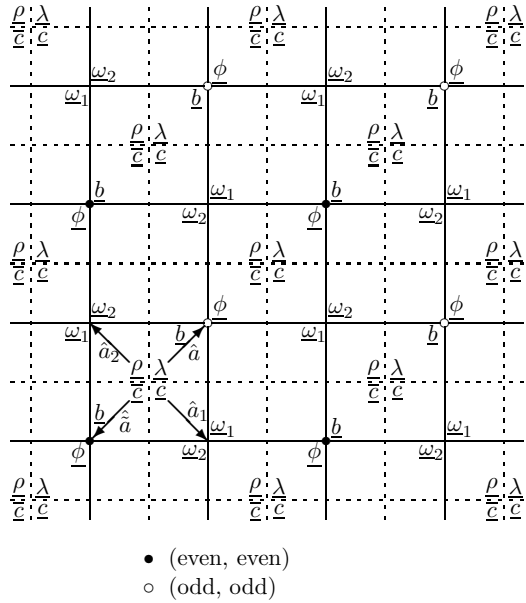
\begin{figure}
  \begin{center}
   \scalebox{.8}{
   \input{NABFfig_sym02.tex}
   }
 \end{center}
 \caption{Component fields of non-Abelian super BF action on the
 lattice for the symmetric choice of the shift parameters A).}
 \label{fig:NABFsym02}
\end{figure}

We can construct a super invariant action in the similar way as in 
the Abelian model, 
\begin{eqnarray}
S_{NABF}^{lat}
&=& \int d\theta^4\ \sum_{x}\mathrm{\tr}\ i\ \overline{\Psi}(x,\theta)
\Phi(x,\theta) \label{1stline}\\
 &=& \int d\th^4\ \sum_{x} \mathrm{\tr}\ i\ \overline{\uPsi}(x+\a,\th)
\uPhi(x+\a,\th) \label{2ndline}\\
&=& \sum_{x}  \mathrm{\tr}\ \us\tilde{\us}\us_1\us_2 
\bigl(-i\bar{\uc}(x+\a)\uc(x+\a)\bigr) \\
&=& \sum_{x} \mathrm{tr} \bigg[
 \uphi(x+\a-\at)\bigl[\epsilon_{\mu\nu}
\upar^{(s)}_{\mu}\uomega_{\nu}(x+\a-\at) \nonumber \\* 
&& +(\epsilon_{\mu\nu}\uomega_{\mu}\uomega_{\nu})(x+\a-\at)
-\{\uc,\ulambda\}(x+\a-\at)\bigr] \nonumber \\*[5pt]
&& + \ub(x)\upar^{(s)}_{\mu}\uomega_{\mu}(x)
-i\bar{\uc}(x+\a)\upar^{(s)}_{\mu}\uD^{(s)}_{\mu}\uc(x+\a)
\nonumber \\*
&& + i\urho(x-\at)\ulambda(x-\at) 
\bigg]\label{action1_1} \\
&=& \sum_{x}\mathrm{tr} \bigg[
 \uphi(x+\a-\at)\bigl[\epsilon_{\mu\nu}
\upar_{+\mu}\uomega_{\nu}(x-\a_{\mu}-\at) \nonumber \\* 
&& +(\epsilon_{\mu\nu}\uomega_{\mu}\uomega_{\nu})(x+\a-\at)
-\{\uc,\ulambda\}(x+\a-\at)\bigr] \nonumber \\*[5pt]
&& + \ub(x)\upar_{-\mu}\uomega_{\mu}(x+\m)
-i\bar{\uc}(x+\a)\upar_{-\mu}\uD_{+\mu}\uc(x+\a)
\nonumber \\*
&& + i\urho(x-\at)\ulambda(x-\at) 
\bigg]
,\label{action1_2}
\end{eqnarray}
where the second line (\ref{2ndline}) is 
obtained from the first line by inserting Eqs (\ref{Phi-shift}) and 
(\ref{T(-a)}) and moving the shift operator $T(\pm \a)$ to cancel 
each other and thus no redundant shift operator appears in the action.
The covariant difference operators 
in (\ref{action1_1}) and (\ref{action1_2}) are introduced as, 
\begin{eqnarray}
\uD^{(s)}_{\mu}\uc(x)
&\equiv& \upar^{(s)}_{\mu}\uc(x)+[\uomega_{\mu},\uc](x),\\
\uD_{+\mu}\uc(x)
&\equiv& \upar_{+\mu}\uc(x)+[\uomega_{\mu},\uc](x+\m),
\end{eqnarray}
respectively,
and again the notation (\ref{n1})$\sim$(\ref{n4}) are used.
Even with the interaction terms, the exactness of the action with respect to 
the all twisted supercharges ensures the $N=2$ twisted SUSY 
invariance by construction. 
The field configurations appeared in the action on the lattice 
are depicted in Fig.~\ref{fig:NABFsym02} for the symmetric choice of the 
shift parameters A),
while for the asymmetric choice B), one can again show that the locations 
of the component fields coincides with the Abelian case
(Fig.\ref{fig:lattaction-asym}).
For the symmetric choice A) the summation of the lattice coordinates for the 
action should cover the same coordinate sites as the Abelian case:
\be
\sum_x=\sum_{\hbox{(even,even)}} + \sum_{\hbox{(odd,odd)}},
\label{even-odd-sum-nonab}
\eeq
while it covers (even, even) sites for the asymmetric choice B) similar 
to the Abelian case again.

\section{Summary and Discussions}
\label{summary}

We have proposed a new formulation of a twisted superspace on a lattice 
based on the continuum formulation. 
The most important difference from the previous trials of formulating 
SUSY on the lattice is that we have introduced mild noncommutativity 
between the difference operators and fields to preserve the Leibniz rule on 
the lattice. As a result an exact extended SUSY for all the twisted 
supercharges is realized on the lattice.  

The origin of the noncommutative nature of the difference operator 
stems from the fact that the shift operator itself plays a role of difference 
operator in the representation space of a commutator. 
Since the twisted super algebra includes the difference operator, the twisted 
supercharges carry the noncommutative shifts correspondingly. 
We found the consistency conditions of the shift parameters explicitly 
for $N=2$ twisted SUSY algebra in two dimensions. 
Those consistency conditions naturally led to define the supercharge 
differential operators in a consistent way, where the twisted super parameters 
need to carry the opposite shifts to the corresponding supercharges. 

Parallel to the continuum twisted superspace formulation, we have introduced 
a twisted superspace on the lattice. The crucial difference from the continuum 
formulation is that each of the component fields of the superspace carries the 
shifting nature since the supercharges and parameters both carry the shifts. 
We have found a new type of unexpected three-dimensional lattice structure 
which is generated by the consistency condition of the noncommutativity 
between a $N=D=2$ twisted superfield and super parameters. 
This three-dimensional lattice has a cell structure where unit cell 
contains all the component fields of a superfield with chiral pairs per 
site. 
When the three-dimensional lattice is reduced into two dimension, the lattice 
super field still has an interesting geometrical interpretation  
about how the component fields of the superfield are scattered semilocally 
within a double size lattice in a systematic way. The reason why the component 
fields are confined within a double size lattice is due to the Grassmann odd 
nature of the twisted super parameters which carry the noncommutative shifts.  

Based on the formulation of twisted superspace on the lattice we have constructed 
explicit examples of super BF models and Wess-Zumino models in two dimensions. 
Those actions have exact $N=2$ twisted SUSY invariance on the lattice for 
all the twisted supercharges. Each term in the action has a geometrical 
correspondence with the location of the fields. 
The lattice counterpart of Wess-Zumino model which has the standard $N=2$ 
SUSY invariance on the lattice has a second derivative Dirac 
fermion term which is necessary to preserve exact SUSY and 
originally appeared from the Dirac-K\"ahler fermion mechanism on the 
lattice\cite{G,KMN,KK}. 
In the non-Abelian extension of super BF model the chiral condition required 
the same shifting property between the super covariant derivative and the 
superfield which led to enforce a nontrivial shift to the superfield itself 
in contrast with the Abelian case. Exact $N=2$ twisted SUSY invariance is 
assured even with the non-Abelian interaction terms.  

As we show in Section 5, the component fields in the action are cyclically 
permutable due to the vanishing property of total sum of the shift parameters. 
It is then effectively similar as if the underlined fields do not carry the 
noncommutative shifts and have the similar feature as the standard 
shift-less fields. It is, however, an open question if this feature remains 
true in the quantum level and in the continuum limit of the lattice.  

In this paper we have given a general formulation of twisted superspace 
formalism on the lattice. As a specific example, however, we have chosen 
particular models: super BF and Wess-Zumino models which are constructed by the 
bilinear form of chiral and anti-chiral superfields. 
It is natural to ask if we can formulate twisted super Yang-Mills action 
on the lattice which was already formulated in the continuum twisted 
superspace in two dimensions\cite{KT,KKU}. 
The answer is affirmative and it will be given elsewhere\cite{DKKN2}. 
The next immediate question could be if we can extend $N=D=2$ twisted superspace 
formalism on the lattice into four dimensions. 
We first need continuum $N=D=4$ twisted superspace formalism which will be 
published soon\cite{KKM}.  

Another important question could be to understand the twisting mechanism from 
algebraic point of view on the lattice. As it was pointed out in \cite{KT, KKU} 
that the twisting mechanism is essentially the Dirac-K\"ahler fermion mechanism 
and the $R$-rotation of the twisted super algebra is equivalent to the rotation 
of the flavor suffixes of the Dirac-K\"ahler fermion. 
We need to clear up how the ghost related fermionic fields having an integer spin 
leads to construct Dirac or Majorana half integer spin fermion on the lattice. 
This problem is essentially related to the chiral fermion problem as well.


\section*{Acknowledgments}

A. D'Adda thanks kind hospitality for his stay in Sapporo where this 
collaboration started while N. Kawamoto thanks warm hospitality 
for his stay in Torino where intensive discussions of this collaboration 
continued. This work is supported in part by Japanese 
Ministry of Education, Science, Sports and Culture under the grant number 
13640250 and 13135201.

\bigskip

\appendix
\begin{center}
{\Large \textbf{Appendix}}
\end{center}
\section{Notations}
\label{app:notation}
\renewcommand{\theequation}{\Alph {section}.\arabic{equation}}

We use the following notation for $\gamma$ matrices:
\begin{equation}
 \gamma_1=\sigma_3 \qquad \gamma_2=\sigma_1
   \qquad \gamma_5=\gamma_1 \gamma_2.
\end{equation}
$\gamma_\mu^T=\gamma_\mu(\mu=1,\ 2)$ and $\gamma_5^T = -\gamma_5$.
They satisfy
\begin{eqnarray}
 \{\gamma_\mu, \gamma_\nu\}&=& 2\delta_{\mu\nu}\\
 C\gamma_\mu C^{-1}&=&\gamma_\mu^T,
\end{eqnarray}
where $C$ is a charge conjugation matrix and can be taken as $C=1$ in
Euclidean two dimensions.

The following relations are useful in the formulation of 
Dirac-K\"ahler fermion mechanism:
\begin{eqnarray}
  \mathbf{1}_{ij} \mathbf{1}_{kl}
  + (\gamma_\mu)_{ij} (\gamma_\mu)_{kl}
  + (\gamma_5)_{ij} (\gamma_5)_{kl}
  &=& 2\delta_{ik}\delta_{jl},\\
  \mathbf{1}_{ij} \mathbf{1}_{kl}
  - (\gamma_\mu)_{ij} (\gamma_\mu)_{kl}
  + (\gamma_5)_{ij} (\gamma_5)_{kl}
  &=& 2(\gamma_5)_{ik}(\gamma_5)_{jl}.
\end{eqnarray}

\section{$N=2$ SUSY transformation for the component 
fields of Wess-Zumino action}
\label{app:transformation}
Here we explicitly give the $N=2$ SUSY transformation of fields for 
the Wess-Zumino model in two dimensions. 
These SUSY transformations can be derived by the twisted 
SUSY transformation of 
the original component fields from superfields. Dirac-K\"ahler fermion 
mechanism is essentially used to transform from the tensor suffixes to the 
spinor suffixes. 

For the scalar field $N=2$ SUSY transformation is given by
\begin{eqnarray}
 \us_{\alpha i}\uphi_1(x)
  &=& (\gamma_5)_{\alpha\beta}\uxi_{\beta j}(x) (\gamma_5)_{ji}\nonumber\\
  && {}-\frac{1}{2}\delta_{\alpha i}\delta_{j\beta}
        \bigl( \uxi_{\beta j}(x+2\a) - \uxi_{\beta j}(x)\bigr)
      +\frac{1}{2}(\gamma_5)_{\alpha i}(\gamma_5)_{j\beta}
        \bigl( \uxi_{\beta j}(x+2\at) - \uxi_{\beta j}(x)\bigr)\nonumber\\
 \us_{\alpha i}\uphi_2(x)
  &=& i\xi_{\alpha i}(x)\nonumber\\
  && {}+\frac{i}{2}\delta_{\alpha i}\delta_{j\beta}
      \bigl(\uxi_{\beta j}(x+2\a)-\uxi_{\beta j}(x)\bigr)
     -\frac{i}{2}(\gamma_5)_{\alpha i}(\gamma_5)_{j\beta}
     \bigl(\uxi_{\beta j}(x+2\at)-\uxi_{\beta j}(x)\bigr).\nonumber\\
\end{eqnarray}
For Dirac-K\"ahler fermion fields the $N=2$ SUSY transformation is given by 
\begin{eqnarray}
 \lefteqn{\us_{\alpha i}\uxi_{\beta j}(x)}\nonumber\\
 &=& -(\gamma_5)_{\alpha\beta}\delta_{ij}\uF_1(x+\a_1+\a_2)
    +i \delta_{\alpha\beta}(\gamma_5)_{ij}\uF_2(x+\a_1+\a_2)\nonumber\\
 && {}+i(\gamma_\mu\gamma_5)_{\alpha\beta}(\gamma_5)_{ij}
    \frac{\upar_{+\mu} + \upar_{-\mu}}{2} \uphi_1(x)
      -(\gamma_\mu)_{\alpha\beta}\delta_{ij}
    \frac{\upar_{+\mu} + \upar_{-\mu}}{2} \uphi_2(x)\nonumber\\
 && {}-\frac{1}{2}(\gamma_5)_{\alpha i} \delta_{\beta j}
    \bigl(\uF_1(x+\a_1+\a_2+2\at)-\uF_1(x+\a_1+\a_2)\bigr)\nonumber\\
 && {}+\frac{1}{2}\delta_{\alpha i} (\gamma_5)_{\beta j}
    \bigl(\uF_1(x+\a_1+\a_2+2\a)-\uF_1(x+\a_1+\a_2)\bigr)\nonumber\\
 && {}-\frac{i}{2}(\gamma_5)_{\alpha i} \delta_{\beta j}
    \bigl(\uF_2(x+\a_1+\a_2+2\at)-\uF_2(x+\a_1+\a_2)\bigr)\nonumber\\
 && {}+\frac{i}{2}\delta_{\alpha i} (\gamma_5)_{\beta j}
    \bigl(\uF_2(x+\a_1+\a_2+2\a)-\uF_2(x+\a_1+\a_2)\bigr)\nonumber\\
 && {}+i(\gamma_5)_{\alpha\beta}(\gamma_\mu\gamma_5)_{ij}
    \frac{\upar_{+\mu} - \upar_{-\mu}}{2} \uphi_1(x)\nonumber\\
 && {} -\frac{i}{2}(\gamma_\mu)_{\alpha i}\delta_{\beta j}\upar_{+\mu}
     \bigl(\uphi_1(x)-\uphi_1(x-2\a)\bigr)
    -\frac{i}{2}(\gamma_\mu\gamma_5)_{\alpha i}(\gamma_5)_{\beta j}
   \upar_{-\mu}\bigl(\uphi_1(x)-\uphi_1(x-2\at)\bigr)\nonumber\\
 && {}-\delta_{\alpha\beta}(\gamma_\mu)_{ij}
    \frac{\upar_{+\mu}-\upar_{-\mu}}{2}\uphi_2(x)\nonumber\\
 && {}+\frac{1}{2}(\gamma_\mu)_{\alpha i}\delta_{\beta j}
     \upar_{+\mu}\bigl( \uphi_2(x)-\uphi_2(x-2\a)\bigr)
    +\frac{1}{2}(\gamma_\mu\gamma_5)_{\alpha i}(\gamma_5)_{\beta j}
   \upar_{-\mu}\bigl( \uphi_2(x)-\uphi_2(x-2\at)\bigr).\nonumber\\
\end{eqnarray}
For the auxiliary fields we obtain the following SUSY transformation:
\begin{eqnarray}
 \us_{\alpha i}\uF_1(x+\a_1+\a_2)
 &=& i\frac{\upar_{+\mu}+\upar_{-\mu}}{2}(\gamma_\mu\gamma_5)_{\alpha\beta}
     \uxi_{\beta i}(x) \nonumber\\
 && {}+i\frac{\upar_{+\mu}-\upar_{-\mu}}{2}(\gamma_5)_{\alpha\beta}
    \uxi_{\beta j}(x) (\gamma_\mu)_{ji} \nonumber\\
 && {}-\frac{i}{2}(\gamma_5\gamma_\mu)_{\alpha i}\delta_{j\beta}
     \upar_{-\mu}\bigl(\uxi_{\beta j}(x-2\at)-\uxi_{\beta j}(x)\bigr)\nonumber\\
 && {}+\frac{i}{2}(\gamma_\mu)_{\alpha i}(\gamma_5)_{j \beta}
     \upar_{+\mu}\bigl(\uxi_{\beta j}(x-2\a)-\uxi_{\beta j}(x)\bigr)\\
 \us_{\alpha i}\uF_2(x+\a_1+\a_2)
 &=& \frac{\upar_{+\mu} + \upar_{-\mu}}{2}(\gamma_\mu)_{\alpha\beta}
     \uxi_{\beta j}(x)(\gamma_5)_{ji} \nonumber\\
 && {}+\frac{\upar_{+\mu}-\upar_{-\mu}}{2}
      \uxi_{\alpha j}(x)(\gamma_5\gamma_\mu)_{ji} \nonumber\\
 && {}-\frac{1}{2}(\gamma_5\gamma_\mu)_{\alpha i}\delta_{j\beta}\upar_{-\mu}
  \bigl( \uxi_{\beta j}(x-2\at) - \uxi_{\beta j}(x)\bigr) \nonumber\\
 && {}+\frac{1}{2}(\gamma_\mu)_{\alpha i}(\gamma_5)_{j\beta}\upar_{+\mu}
  \bigl( \uxi_{\beta j}(x-2\a) - \uxi_{\beta j}(x)\bigr).
\end{eqnarray}


\end{document}

%% file: three-dim.pstex_t
\begin{picture}(0,0)%
\includegraphics{three-dim.pstex}%
\end{picture}%
\setlength{\unitlength}{3947sp}%
\begingroup\makeatletter\ifx\SetFigFont\undefined%
\gdef\SetFigFont#1#2#3#4#5{%
  \reset@font\fontsize{#1}{#2pt}%
  \fontfamily{#3}\fontseries{#4}\fontshape{#5}%
  \selectfont}%
\fi\endgroup%
\begin{picture}(4977,2617)(4889,-1716)
\put(5074,-299){\makebox(0,0)[lb]{\smash{\SetFigFont{14}{16.8}{\familydefault}{\mddefault}{\updefault}$\a$}}}
\put(5671,-627){\makebox(0,0)[lb]{\smash{\SetFigFont{14}{16.8}{\familydefault}{\mddefault}{\updefault}$\a_2$}}}
\put(5671,-1716){\makebox(0,0)[lb]{\smash{\SetFigFont{14}{16.8}{\familydefault}{\mddefault}{\updefault}$\hat{n}_1$}}}
\put(5819,-1307){\makebox(0,0)[lb]{\smash{\SetFigFont{14}{16.8}{\familydefault}{\mddefault}{\updefault}$\a_1$}}}
\put(5236,-1184){\makebox(0,0)[lb]{\smash{\SetFigFont{14}{16.8}{\familydefault}{\mddefault}{\updefault}$\hat{n}_2$}}}
\end{picture}

%% file: three-dim-cell.pstex_t
\begin{picture}(0,0)%
\includegraphics{three-dim-cell.pstex}%
\end{picture}%
\setlength{\unitlength}{4736sp}%
\begingroup\makeatletter\ifx\SetFigFont\undefined%
\gdef\SetFigFont#1#2#3#4#5{%
  \reset@font\fontsize{#1}{#2pt}%
  \fontfamily{#3}\fontseries{#4}\fontshape{#5}%
  \selectfont}%
\fi\endgroup%
\begin{picture}(3092,1782)(1151,-2131)
\put(2653,-1596){\makebox(0,0)[lb]{\smash{\SetFigFont{12}{14.4}{\familydefault}{\mddefault}{\updefault}$\uvarphi_{12}$}}}
\put(1276,-793){\makebox(0,0)[lb]{\smash{\SetFigFont{12}{14.4}{\familydefault}{\mddefault}{\updefault}$\uvarphi_S$}}}
\put(1151,-1577){\makebox(0,0)[lb]{\smash{\SetFigFont{12}{14.4}{\familydefault}{\mddefault}{\updefault}$\a$}}}
\put(1240,-1166){\makebox(0,0)[lb]{\smash{\SetFigFont{12}{14.4}{\familydefault}{\mddefault}{\updefault}$\uvarphi$}}}
\put(2121,-1718){\makebox(0,0)[lb]{\smash{\SetFigFont{12}{14.4}{\familydefault}{\mddefault}{\updefault}$\uvarphi_1$}}}
\put(1837,-1797){\makebox(0,0)[lb]{\smash{\SetFigFont{12}{14.4}{\familydefault}{\mddefault}{\updefault}$\hat{n}_2$}}}
\put(1711,-2131){\makebox(0,0)[lb]{\smash{\SetFigFont{12}{14.4}{\familydefault}{\mddefault}{\updefault}$\hat{n}_1$}}}
\put(1632,-1888){\makebox(0,0)[lb]{\smash{\SetFigFont{12}{14.4}{\familydefault}{\mddefault}{\updefault}$\a_1$}}}
\put(1793,-675){\makebox(0,0)[lb]{\smash{\SetFigFont{12}{14.4}{\familydefault}{\mddefault}{\updefault}$\uvarphi_{S2}$}}}
\put(1976,-1366){\makebox(0,0)[lb]{\smash{\SetFigFont{12}{14.4}{\familydefault}{\mddefault}{\updefault}$\uvarphi_2$}}}
\put(2114,-1034){\makebox(0,0)[lb]{\smash{\SetFigFont{12}{14.4}{\familydefault}{\mddefault}{\updefault}$\uvarphi_{S1}$}}}
\put(2624,-1055){\makebox(0,0)[lb]{\smash{\SetFigFont{12}{14.4}{\familydefault}{\mddefault}{\updefault}$\uvarphi_{S12}$}}}
\put(1524,-1500){\makebox(0,0)[lb]{\smash{\SetFigFont{12}{14.4}{\familydefault}{\mddefault}{\updefault}$\a_2$}}}
\end{picture}

%% file: gen_sf_sym_a.tex


{
\fontsize{11pt}{15pt}
\selectfont
\setlength{\unitlength}{1pt}
\begin{picture}(300,270)(-130,-135)
 \multiput(-130,-120)(0,120){3}{\line(1,0){260}}
 \multiput(-120,-130)(120,0){3}{\line(0,1){260}}

{
 \thicklines
 \put(20,20){\vector(1,1){40}}
 \put(-20,20){\vector(-1,1){40}}
 \put(20,-20){\vector(1,-1){40}}
 \put(-20,-20){\vector(-1,-1){40}}
 \put(25,35){\makebox(0,0){$\a$}}
 \put(-25,35){\makebox(0,0){$\a_2$}}
 \put(25,-35){\makebox(0,0){$\a_1$}}
 \put(-25,-35){\makebox(0,0){$\at$}}

 \put(120,0){
  \put(-60,60){\vector(1,-1){60}}
  \put(-60,-60){\vector(1,1){60}}
 }
 \put(-120,0){
  \put(60,60){\vector(-1,-1){60}}
  \put(60,-60){\line(-1,1){40}}
  \put(5,-5){\vector(-1,1){5}}
 }
 \put(0,120){
  \put(-25,-25){\vector(1,1){25}}
  \put(50,-50){\vector(-1,1){50}}
 }
 \put(0,-120){
  \put(-65,65){\line(1,-1){10}}
  \put(-25,25){\vector(1,-1){25}}
  \put(50,50){\vector(-1,-1){50}}
 }
}

\matrixput(-60,-60)(120,0){2}(0,120){2}
{
 \put(0,0){\makebox(0,0){\line(1,0){6}}}
 \put(0,0){\makebox(0,0){\line(0,1){6}}}
}
\put(0,-120){
  \put(0,0){\makebox(0,0){$\times$}}
}
\put(-120,0){
  \put(0,0){\makebox(0,0){$\times$}}
}
\put(120,0){
  \put(0,0){\makebox(0,0){$\times$}}
}
\put(0,120){
  \put(0,0){\makebox(0,0){$\times$}}
}

\put(0,0){
 \put(0,0){\circle*4}
 \put(-3,-3){\makebox(0,0)[rt]{$\figvarphi(x)$}}
 \put(3,-3){\makebox(0,0)[lt]{$\figvarphi_{S12P}(x)$}}
 \put(-3,3){\makebox(0,0)[rb]{$\figvarphi_{SP}(x)$}}
 \put(3,3){\makebox(0,0)[lb]{$\figvarphi_{12}(x)$}}
}

\put(60,60){
 \put(-8,3){\makebox(0,0)[lb]{$\figvarphi_{S12}(x+\a)$}}
 \put(-8,18){\makebox(0,0)[lb]{$\figvarphi_S(x+\a)$}}
}

\put(-60,-60){
 \put(-8,-3){\makebox(0,0)[lt]{$\figvarphi_{12P}(x+\at)$}}
 \put(-8,-18){\makebox(0,0)[lt]{$\figvarphi_P(x+\at)$}}
}

\put(60,-60){
 \put(-8,-3){\makebox(0,0)[lt]{$\figvarphi_{S1P}(x+\a_1)$}}
 \put(-8,-18){\makebox(0,0)[lt]{$\figvarphi_1(x+\a_1)$}}
}

\put(-60,60){
 \put(-8,3){\makebox(0,0)[lb]{$\figvarphi_{S2P}(x+\a_2)$}}
 \put(-8,18){\makebox(0,0)[lb]{$\figvarphi_2(x+\a_2)$}}
}

\put(120,0){
 \put(3,-3){\makebox(0,0)[lt]{$\figvarphi_{S1}(x+\hat{n}_1)$}}
}

\put(-120,0){
 \put(3,-3){\makebox(0,0)[lt]{$\figvarphi_{2P}(x-\hat{n}_1)$}}
}

\put(0,120){
 \put(3,3){\makebox(0,0)[lb]{$\figvarphi_{S2}(x+\hat{n}_2)$}}
}

\put(0,-120){
 \put(3,-3){\makebox(0,0)[lt]{$\figvarphi_{1P}(x-\hat{n}_2)$}}
}

\end{picture}

}

%% file: gen_sf_sym_b.tex


{
\fontsize{11pt}{15pt}
\selectfont
\setlength{\unitlength}{1pt}
\begin{picture}(300,270)(-130,-135)
 \multiput(-130,-120)(0,120){3}{\line(1,0){260}}
 \multiput(-120,-130)(120,0){3}{\line(0,1){260}}

{
 \thicklines
 \put(20,20){\vector(1,1){40}}
 \put(-20,20){\vector(-1,1){40}}
 \put(20,-20){\vector(1,-1){40}}
 \put(-20,-20){\vector(-1,-1){40}}
 \put(25,35){\makebox(0,0){$\a$}}
 \put(-25,35){\makebox(0,0){$\a_2$}}
 \put(25,-35){\makebox(0,0){$\a_1$}}
 \put(-25,-35){\makebox(0,0){$\at$}}

 \put(120,0){
  \put(-60,60){\vector(1,-1){60}}
  \put(-60,-60){\vector(1,1){60}}
 }
 \put(-120,0){
  \put(60,60){\vector(-1,-1){60}}
  \put(60,-60){\line(-1,1){40}}
  \put(5,-5){\vector(-1,1){5}}
 }
 \put(0,120){
  \put(-25,-25){\vector(1,1){25}}
  \put(50,-50){\vector(-1,1){50}}
 }
 \put(0,-120){
  \put(-65,65){\line(1,-1){10}}
  \put(-25,25){\vector(1,-1){25}}
  \put(50,50){\vector(-1,-1){50}}
 }
}

\matrixput(-60,-60)(120,0){2}(0,120){2}
{
 \put(0,0){\makebox(0,0){\line(1,0){6}}}
 \put(0,0){\makebox(0,0){\line(0,1){6}}}
}
\put(0,-120){
  \put(0,0){\makebox(0,0){$\times$}}
}
\put(-120,0){
  \put(0,0){\makebox(0,0){$\times$}}
}
\put(120,0){
  \put(0,0){\makebox(0,0){$\times$}}
}
\put(0,120){
  \put(0,0){\makebox(0,0){$\times$}}
}

\put(0,0){
 \put(0,0){\circle*4}
 \put(-3,-3){\makebox(0,0)[rt]{$\figvarphi(x)$}}
 \put(3,-3){\makebox(0,0)[lt]{$\figvarphi_{S12P}(x)$}}
 \put(-3,3){\makebox(0,0)[rb]{$\figvarphi_{SP}(x)$}}
 \put(3,3){\makebox(0,0)[lb]{$\figvarphi_{12}(x)$}}
}

\put(-60,-60){
 \put(-8,-3){\makebox(0,0)[lt]{$\figvarphi_{S12}(x-\a)$}}
 \put(-8,-18){\makebox(0,0)[lt]{$\figvarphi_S(x-\a)$}}
}

\put(60,60){
 \put(-8,3){\makebox(0,0)[lb]{$\figvarphi_{12P}(x-\at)$}}
 \put(-8,18){\makebox(0,0)[lb]{$\figvarphi_P(x-\at)$}}
}

\put(-60,60){
 \put(-8,3){\makebox(0,0)[lb]{$\figvarphi_{S1P}(x-\a_1)$}}
 \put(-8,18){\makebox(0,0)[lb]{$\figvarphi_1(x-\a_1)$}}
}

\put(60,-60){
 \put(-8,-3){\makebox(0,0)[lt]{$\figvarphi_{S2P}(x-\a_2)$}}
 \put(-8,-18){\makebox(0,0)[lt]{$\figvarphi_2(x-\a_2)$}}
}

\put(-120,0){
 \put(3,-3){\makebox(0,0)[lt]{$\figvarphi_{S1}(x-\hat{n}_1)$}}
}

\put(120,0){
 \put(3,-3){\makebox(0,0)[lt]{$\figvarphi_{2P}(x+\hat{n}_1)$}}
}

\put(0,-120){
 \put(3,-3){\makebox(0,0)[lt]{$\figvarphi_{S2}(x-\hat{n}_2)$}}
}

\put(0,120){
 \put(3,3){\makebox(0,0)[lb]{$\figvarphi_{1P}(x+\hat{n}_2)$}}
}

\end{picture}

}

%% file: gen_sf_nat_a.tex


{
\fontsize{11pt}{15pt}
\selectfont
\setlength{\unitlength}{1pt}
\begin{picture}(330,270)(-130,-135)
 \multiput(-130,-120)(0,120){3}{\line(1,0){260}}
 \multiput(-120,-130)(120,0){3}{\line(0,1){260}}

{
 \thicklines
 \put(0,0){\vector(1,0){120}}
 \put(0,0){\vector(0,1){120}}
  \put(-15,-15){\line(-1,-1){90}}
  \put(-115,-115){\vector(-1,-1){5}}
 \put(8,60){\makebox(0,0){$\a_2$}}
 \put(60,-8){\makebox(0,0){$\a_1$}}
 \put(-55,-65){\makebox(0,0){$\at$}}

 \put(-120,0){
  \put(0,-120){\vector(0,1){120}}
  \put(120,120){\line(-1,-1){105}}
  \put(5,5){\vector(-1,-1){5}}
 }
 \put(0,-120){
  \put(-120,0){\vector(1,0){120}}
  \put(120,120){\line(-1,-1){105}}
  \put(5,5){\vector(-1,-1){5}}
 }
 \put(120,120){
  \put(-120,0){\vector(1,0){120}}
  \put(0,-120){\vector(0,1){120}}
 }
 \put(0,0){
  \put(120,120){\line(-1,-1){105}}
  \put(5,5){\vector(-1,-1){5}}
 }

}

\put(0,0){
 \put(0,0){\circle*4}
 \put(-3,-3){\makebox(0,0)[rt]{$\figvarphi(x)$}}
 \put(3,-3){\makebox(0,0)[lt]{$\figvarphi_{S}(x)$}}
 \put(-3,3){\makebox(0,0)[rb]{$\figvarphi_{12P}(x)$}}
 \put(3,3){\makebox(0,0)[lb]{$\figvarphi_{S12P}(x)$}}
}

\put(120,0){
 \put(3,3){\makebox(0,0)[lb]{$\figvarphi_{S1}(x+\hat{n}_1)$}}
 \put(3,-3){\makebox(0,0)[lt]{$\figvarphi_{1}(x+\hat{n}_1)$}}
}

\put(-120,0){
 \put(3,3){\makebox(0,0)[lb]{$\figvarphi_{2P}(x-\hat{n}_1)$}}
 \put(3,-3){\makebox(0,0)[lt]{$\figvarphi_{S2P}(x-\hat{n}_1)$}}
}

\put(0,120){
 \put(3,3){\makebox(0,0)[lb]{$\figvarphi_{2}(x+\hat{n}_2)$}}
 \put(3,-3){\makebox(0,0)[lt]{$\figvarphi_{S2}(x+\hat{n}_2)$}}
}

\put(0,-120){
 \put(3,3){\makebox(0,0)[lb]{$\figvarphi_{1P}(x-\hat{n}_2)$}}
 \put(3,-3){\makebox(0,0)[lt]{$\figvarphi_{S1P}(x-\hat{n}_2)$}}
}

\put(120,120){
 \put(3,3){\makebox(0,0)[lb]{$\figvarphi_{S12}(x+\hat{n}_1+\hat{n}_2)$}}
 \put(3,-3){\makebox(0,0)[lt]{$\figvarphi_{12}(x+\hat{n}_1+\hat{n}_2)$}}
}

\put(-120,-120){
 \put(3,3){\makebox(0,0)[lb]{$\figvarphi_{SP}(x-\hat{n}_1-\hat{n}_2)$}}
 \put(3,-3){\makebox(0,0)[lt]{$\figvarphi_{P}(x-\hat{n}_1-\hat{n}_2)$}}
}

\end{picture}

}

%% file: gen_sf_nat_b.tex


{
\fontsize{11pt}{15pt}
\selectfont
\setlength{\unitlength}{1pt}
\begin{picture}(330,270)(-130,-135)
 \multiput(-130,-120)(0,120){3}{\line(1,0){260}}
 \multiput(-120,-130)(120,0){3}{\line(0,1){260}}

{
 \thicklines
 \put(0,0){\vector(1,0){120}}
 \put(0,0){\vector(0,1){120}}
  \put(-15,-15){\line(-1,-1){90}}
  \put(-115,-115){\vector(-1,-1){5}}
 \put(8,60){\makebox(0,0){$\a_2$}}
 \put(60,-8){\makebox(0,0){$\a_1$}}
 \put(-55,-65){\makebox(0,0){$\at$}}

 \put(-120,0){
  \put(0,-120){\vector(0,1){120}}
  \put(120,120){\line(-1,-1){105}}
  \put(5,5){\vector(-1,-1){5}}
 }
 \put(0,-120){
  \put(-120,0){\vector(1,0){120}}
  \put(120,120){\line(-1,-1){105}}
  \put(5,5){\vector(-1,-1){5}}
 }
 \put(120,120){
  \put(-120,0){\vector(1,0){120}}
  \put(0,-120){\vector(0,1){120}}
 }
 \put(0,0){
  \put(120,120){\line(-1,-1){105}}
  \put(5,5){\vector(-1,-1){5}}
 }

}

\put(0,0){
 \put(0,0){\circle*4}
 \put(-3,-3){\makebox(0,0)[rt]{$\figvarphi(x)$}}
 \put(3,-3){\makebox(0,0)[lt]{$\figvarphi_{S}(x)$}}
 \put(-3,3){\makebox(0,0)[rb]{$\figvarphi_{12P}(x)$}}
 \put(3,3){\makebox(0,0)[lb]{$\figvarphi_{S12P}(x)$}}
}

\put(-120,0){
 \put(3,3){\makebox(0,0)[lb]{$\figvarphi_{S1}(x-\hat{n}_1)$}}
 \put(3,-3){\makebox(0,0)[lt]{$\figvarphi_{1}(x-\hat{n}_1)$}}
}

\put(120,0){
 \put(3,3){\makebox(0,0)[lb]{$\figvarphi_{2P}(x+\hat{n}_1)$}}
 \put(3,-3){\makebox(0,0)[lt]{$\figvarphi_{S2P}(x+\hat{n}_1)$}}
}

\put(0,-120){
 \put(3,3){\makebox(0,0)[lb]{$\figvarphi_{2}(x-\hat{n}_2)$}}
 \put(3,-3){\makebox(0,0)[lt]{$\figvarphi_{S2}(x-\hat{n}_2)$}}
}

\put(0,120){
 \put(3,3){\makebox(0,0)[lb]{$\figvarphi_{1P}(x+\hat{n}_2)$}}
 \put(3,-3){\makebox(0,0)[lt]{$\figvarphi_{S1P}(x+\hat{n}_2)$}}
}

\put(-120,-120){
 \put(3,3){\makebox(0,0)[lb]{$\figvarphi_{S12}(x-\hat{n}_1-\hat{n}_2)$}}
 \put(3,-3){\makebox(0,0)[lt]{$\figvarphi_{12}(x-\hat{n}_1-\hat{n}_2)$}}
}

\put(120,120){
 \put(3,3){\makebox(0,0)[lb]{$\figvarphi_{SP}(x+\hat{n}_1+\hat{n}_2)$}}
 \put(3,-3){\makebox(0,0)[lt]{$\figvarphi_{P}(x+\hat{n}_1+\hat{n}_2)$}}
}

\end{picture}

}

%% file: transformation.tex
{
\fontsize{11pt}{15pt}
\selectfont
\setlength{\unitlength}{1pt}
\begin{picture}(130,80)(-5,-15)
 \put(0,0){\circle*{4}}
 \put(0,0){\vector(1,1){60}}
 \put(30,30){\makebox(0,0){\line(1,0){5}}}
 \put(30,30){\makebox(0,0){\line(0,1){5}}}

 \put(0,0){\makebox(6,-1)[lt]{$\uvarphi(x)$}}
 \put(30,30){\makebox(-2,-3)[lt]{$\uvarphi_A(x+\a_A)=\us_A\uvarphi(x)$}}
 \put(50,52){\makebox(0,0)[br]{$2\a_A$}}
\end{picture}
}

%% file: superspace_sym.tex
{
\fontsize{11pt}{15pt}
\selectfont
\setlength{\unitlength}{1pt}
\begin{picture}(200,230)(-10,-10)
 \multiput(-10,0)(0,60){4}{\line(1,0){230}}
 \multiput(0,-10)(60,0){4}{\line(0,1){200}}
 \multiput(-10,30)(0,60){3}{\dashline{5}(0,0)(230,0)}
 \multiput(30,-10)(60,0){4}{\dashline{5}(0,0)(0,200)}
 
 {
 \thicklines
 \put(70,70){\vector(1,1){20}}
 \put(50,70){\vector(-1,1){20}}
 \put(70,50){\vector(1,-1){20}}
 \put(50,50){\vector(-1,-1){20}}
 \put(83,75){\makebox(0,0){$\a$}}
 \put(40,75){\makebox(0,0){$\a_2$}}
 \put(85,45){\makebox(0,0){$\a_1$}}
 \put(36,45){\makebox(0,0){$\at$}}
 }
 \matrixput(60,60)(120,0){2}(0,120){2}{
 \put(0,0){\circle*{4}}
 \put(6,6){\makebox(0,0){$\figlambda$}}
 \put(-6,6){\makebox(0,0){$\figrho$}}
 \put(-6,-6){\makebox(0,0){$\figcb$}}
 \put(6,-6){\makebox(0,0){$\figc$}}
 }

 \matrixput(60,60)(120,0){2}(0,120){2}{
 \put(-36,-36){\makebox(0,-2){$\figphi$}}
 \put(36,-36){\makebox(3,0){$\figomega_1$}}
 }
 \multiput(60,60)(120,0){2}{
 \put(36,36){\makebox(0,0){$\figb$}}
 \put(-36,36){\makebox(-3,0){$\figomega_2$}}
 }

 {
 \thicklines
 \put(0,0){\vector(1,0){60}}
 \put(0,0){\vector(0,1){60}}
 \put(40,-2){\makebox(0,0)[t]{$\hat{n}_1$}}
 \put(-2,40){\makebox(0,0)[r]{$\hat{n}_2$}}
}

\end{picture}

}

%% file: superspace_nat.tex
{
\fontsize{11pt}{15pt}
\selectfont
\setlength{\unitlength}{1pt}
\begin{picture}(200,230)(-10,-10)
 \multiput(-10,0)(0,60){4}{\line(1,0){230}}
 \multiput(0,-10)(60,0){4}{\line(0,1){200}}
 \multiput(-10,30)(0,60){3}{\dashline{5}(0,0)(230,0)}
 \multiput(30,-10)(60,0){4}{\dashline{5}(0,0)(0,200)}
 
 {
 \thicklines
 \put(60,60){\vector(1,0){60}}
 \put(60,60){\vector(0,1){60}}
 \put(50,50){\vector(-1,-1){50}}
 \put(100,67){\makebox(0,0){$\a_1$}}
 \put(67,100){\makebox(0,0){$\a_2$}}
 \put(13,24){\makebox(0,0){$\at$}}
 }

\matrixput(0,0)(120,0){2}(0,120){2}{
 \put(-6,-6){\makebox(0,0){$\figlambda$}}
 \put(6,-6){\makebox(0,0){$\figrho$}}
 \put(-6,6){\makebox(0,1){$\figphi$}}
}

\matrixput(60,60)(120,0){2}(0,120){2}{
 \put(-6,6){\makebox(0,0){$\figb$}}
 \put(-6,-6){\makebox(0,0){$\figcb$}}
 \put(6,-6){\makebox(0,0){$\figc$}}
 \put(-54,6){\makebox(3,0){$\figomega_1$}}
 \put(6,-54){\makebox(3,0){$\figomega_2$}}
}

 \put(100,58){\makebox(0,0)[t]{$\hat{n}_1$}}
 \put(58,100){\makebox(0,0)[r]{$\hat{n}_2$}}

\end{picture}

}

%% file: superspace_sym2.tex
{
\fontsize{11pt}{15pt}
\selectfont
\setlength{\unitlength}{1pt}
\begin{picture}(260,315)(-10,-55)
 \matrixput(2,0)(0,120){3}(120,0){2}{\line(1,0){116}}
 \multiput(-10,0)(0,120){3}{\line(1,0){8}}
 \multiput(242,0)(0,120){3}{\line(1,0){8}}
 \matrixput(0,2)(0,120){2}(120,0){3}{\line(0,1){116}}
 \multiput(0,-10)(120,0){3}{\line(0,1){8}}
 \multiput(0,242)(120,0){3}{\line(0,1){8}}
 \multiput(-10,60)(0,120){2}{\line(1,0){260}}
 \multiput(60,-10)(120,0){2}{\line(0,1){260}}
 \multiput(-10,30)(0,60){4}{\dashline{5}(0,0)(260,0)}
 \multiput(30,-10)(60,0){4}{\dashline{5}(0,0)(0,260)}
 
 {
 \thicklines
 \put(70,70){\vector(1,1){20}}
 \put(50,70){\vector(-1,1){20}}
 \put(70,50){\vector(1,-1){20}}
 \put(50,50){\vector(-1,-1){20}}
 \put(83,75){\makebox(0,0){$\a$}}
 \put(40,75){\makebox(0,0){$\a_2$}}
 \put(85,45){\makebox(0,0){$\a_1$}}
 \put(48,40){\makebox(0,0){$\at$}}
 }

 \matrixput(60,60)(120,0){2}(0,120){2}{
 \put(0,0){\circle*{4}}
 \put(6,6){\makebox(0,0){$\figlambda$}}
 \put(-6,6){\makebox(0,0){$\figrho$}}
 \put(-6,-6){\makebox(0,0){$\figcb$}}
 \put(6,-6){\makebox(0,0){$\figc$}}
 }

 \matrixput(60,60)(120,0){2}(0,120){2}{
 \put(-36,-36){\makebox(0,-2){$\figphi$}}
 \put(36,-24){\makebox(4,0){$\figomega_1$}}
 \put(-24,36){\makebox(3,0){$\figomega_2$}}
 }
 \multiput(60,180)(120,0){2}{ \put(24,24){\makebox(0,-1){$\figb$}}}
 \put(204,84){\makebox(0,-1){$\figb$}}
 \put(76,84){\makebox(0,-1){$\figb$}}

 \matrixput(0,0)(120,0){3}(0,120){3}{
 \put(0,0){\circle{4}}
 \put(6,6){\makebox(0,0){$\figlambda$}}
 \put(-6,6){\makebox(0,0){$\figrho$}}
 \put(-6,-6){\makebox(0,0){$\figcb$}}
 \put(6,-6){\makebox(0,0){$\figc$}}
 }

 \matrixput(60,60)(120,0){2}(0,120){2}{
 \put(36,36){\makebox(0,3){$\figphi$}}
 \put(-36,24){\makebox(0,0){$\figomega_1$}}
 \put(24,-36){\makebox(-1,0){$\figomega_2$}}
 }
 \multiput(60,180)(120,0){2}{\put(-24,-24){\makebox(-2,1){$\figb$}}}
 \put(116,36){\makebox(-2,1){$\figb$}}
 \put(36,42){\makebox(-2,1){$\figb$}}

 \put(60,-30){\circle*{4}}
 \put(70,-30){\makebox(0,0)[l]{(even, even)}}
 \put(60,-45){\circle{4}}
 \put(70,-45){\makebox(0,0)[l]{(odd, odd)}}

\end{picture}

}

%% file: superspace_nat2.tex
{
\fontsize{11pt}{15pt}
\selectfont
\setlength{\unitlength}{1pt}
\begin{picture}(260,295)(-10,-35)
 \multiput(-10,0)(0,60){5}{\line(1,0){260}}
 \multiput(0,-10)(60,0){5}{\line(0,1){260}}
 
 {
 \thicklines
 \put(60,60){\vector(1,0){60}}
 \put(60,60){\vector(0,1){60}}
 \put(50,50){\vector(-1,-1){50}}
 \put(100,67){\makebox(0,0){$\a_1$}}
 \put(67,100){\makebox(0,0){$\a_2$}}
 \put(13,24){\makebox(0,0){$\at$}}
 }

\matrixput(0,0)(120,0){3}(0,120){3}{
 \put(-6,-6){\makebox(0,0){$\figlambda$}}
 \put(6,-6){\makebox(0,0){$\figrho$}}
 \put(-6,6){\makebox(0,1){$\figphi$}}
}

\matrixput(60,60)(120,0){2}(0,120){2}{
 \put(0,0){\circle*{4}}
 \put(-6,6){\makebox(0,0){$\figb$}}
 \put(-6,-6){\makebox(0,0){$\figcb$}}
 \put(6,-6){\makebox(0,0){$\figc$}}
 \put(-54,6){\makebox(3,0){$\figomega_1$}}
 \put(6,-54){\makebox(3,0){$\figomega_2$}}
}
\multiput(300,60)(0,120){2}{
 \put(-54,6){\makebox(3,0){$\figomega_1$}}
}
\multiput(60,300)(120,0){2}{
 \put(6,-54){\makebox(3,0){$\figomega_2$}}
}

 \put(60,-30){\circle*{4}}
 \put(70,-30){\makebox(0,0)[l]{(even, even)}}

\end{picture}

}

%% file: b_superspace_sym.tex
{
\fontsize{11pt}{15pt}
\selectfont
\setlength{\unitlength}{1pt}
\begin{picture}(200,230)(-10,-10)
 \multiput(-10,0)(0,60){4}{\line(1,0){230}}
 \multiput(0,-10)(60,0){4}{\line(0,1){200}}
 \multiput(-10,30)(0,60){3}{\dashline{5}(0,0)(230,0)}
 \multiput(30,-10)(60,0){4}{\dashline{5}(0,0)(0,200)}
 
 {
 \thicklines
 \put(70,70){\vector(1,1){20}}
 \put(50,70){\vector(-1,1){20}}
 \put(70,50){\vector(1,-1){20}}
 \put(50,50){\vector(-1,-1){20}}
 \put(83,75){\makebox(0,0){$\a$}}
 \put(40,75){\makebox(0,0){$\a_2$}}
 \put(85,45){\makebox(0,0){$\a_1$}}
 \put(36,45){\makebox(0,0){$\at$}}
 }
 \matrixput(60,60)(120,0){2}(0,120){2}{
 \put(0,0){\circle*{4}}
 \put(6,6){\makebox(0,4){$\figphit$}}
 \put(-6,6){\makebox(0,4){$\figvarphit$}}
 \put(-6,-6){\makebox(0,0){$\figvarphi$}}
 \put(6,-6){\makebox(0,0){$\figphi$}}
 }

 \matrixput(60,60)(120,0){2}(0,120){2}{
 \put(-36,-36){\makebox(0,-2){$\uchit$}}
 \put(36,-36){\makebox(3,0){$\figpsi_1$}}
 }
 \multiput(60,60)(120,0){2}{
 \put(36,36){\makebox(0,0){$\figchi$}}
 \put(-36,36){\makebox(-3,2){$\figpsi_2$}}
 }

\end{picture}

}

%% file: b_superspace_nat.tex
{
\fontsize{11pt}{15pt}
\selectfont
\setlength{\unitlength}{1pt}
\begin{picture}(200,230)(-10,-10)
 \multiput(-10,0)(0,60){4}{\line(1,0){230}}
 \multiput(0,-10)(60,0){4}{\line(0,1){200}}
 \multiput(-10,30)(0,60){3}{\dashline{5}(0,0)(230,0)}
 \multiput(30,-10)(60,0){4}{\dashline{5}(0,0)(0,200)}
 
 {
 \thicklines
 \put(60,60){\vector(1,0){60}}
 \put(60,60){\vector(0,1){60}}
 \put(50,50){\vector(-1,-1){50}}
 \put(100,67){\makebox(0,0){$\a_1$}}
 \put(67,100){\makebox(0,0){$\a_2$}}
 \put(13,24){\makebox(0,0){$\at$}}
 }

\matrixput(0,0)(120,0){2}(0,120){2}{
 \put(-6,-6){\makebox(0,-4){$\figphit$}}
 \put(6,-6){\makebox(0,-4){$\figvarphit$}}
 \put(-6,6){\makebox(0,1){$\figchit$}}
}

\matrixput(60,60)(120,0){2}(0,120){2}{
 \put(-6,6){\makebox(0,0){$\figchi$}}
 \put(-6,-6){\makebox(0,0){$\figvarphi$}}
 \put(6,-6){\makebox(0,0){$\figphi$}}
 \put(-54,6){\makebox(3,2){$\figpsi_1$}}
 \put(6,-54){\makebox(3,2){$\figpsi_2$}}
}

\end{picture}

}

%% file: b_superspace_sym2.tex
{
\fontsize{11pt}{15pt}
\selectfont
\setlength{\unitlength}{1pt}
\begin{picture}(260,315)(-10,-55)
 \matrixput(2,0)(0,120){3}(120,0){2}{\line(1,0){116}}
 \multiput(-10,0)(0,120){3}{\line(1,0){8}}
 \multiput(242,0)(0,120){3}{\line(1,0){8}}
 \matrixput(0,2)(0,120){2}(120,0){3}{\line(0,1){116}}
 \multiput(0,-10)(120,0){3}{\line(0,1){8}}
 \multiput(0,242)(120,0){3}{\line(0,1){8}}
 \multiput(-10,60)(0,120){2}{\line(1,0){260}}
 \multiput(60,-10)(120,0){2}{\line(0,1){260}}
 \multiput(-10,30)(0,60){4}{\dashline{5}(0,0)(260,0)}
 \multiput(30,-10)(60,0){4}{\dashline{5}(0,0)(0,260)}
 
 {
 \thicklines
 \put(70,70){\vector(1,1){20}}
 \put(50,70){\vector(-1,1){20}}
 \put(70,50){\vector(1,-1){20}}
 \put(50,50){\vector(-1,-1){20}}
 \put(83,75){\makebox(0,0){$\a$}}
 \put(40,75){\makebox(0,0){$\a_2$}}
 \put(85,45){\makebox(0,0){$\a_1$}}
 \put(48,40){\makebox(0,0){$\at$}}
 }

 \matrixput(60,60)(120,0){2}(0,120){2}{
 \put(0,0){\circle*{4}}
 \put(6,6){\makebox(0,4){$\figphit$}}
 \put(-6,6){\makebox(0,2){$\figvarphit$}}
 \put(-6,-6){\makebox(0,0){$\figvarphi$}}
 \put(6,-6){\makebox(0,0){$\figphi$}}
 }

 \matrixput(60,60)(120,0){2}(0,120){2}{
 \put(-36,-36){\makebox(0,-2){$\figchit$}}
 \put(36,-24){\makebox(4,2){$\figpsi_1$}}
 \put(-24,36){\makebox(3,0){$\figpsi_2$}}
 }
 \multiput(60,180)(120,0){2}{ \put(24,24){\makebox(0,-1){$\figchi$}}}
 \put(204,84){\makebox(0,-1){$\figchi$}}
 \put(76,84){\makebox(0,-1){$\figchi$}}

 \matrixput(0,0)(120,0){3}(0,120){3}{
 \put(0,0){\circle{4}}
 \put(6,6){\makebox(0,4){$\figphit$}}
 \put(-6,6){\makebox(0,2){$\figvarphit$}}
 \put(-6,-6){\makebox(0,0){$\figvarphi$}}
 \put(6,-6){\makebox(0,0){$\figphi$}}
 }

 \matrixput(60,60)(120,0){2}(0,120){2}{
 \put(36,36){\makebox(0,3){$\figchit$}}
 \put(-36,24){\makebox(0,2){$\figpsi_1$}}
 \put(24,-36){\makebox(-1,0){$\figpsi_2$}}
 }
 \multiput(60,180)(120,0){2}{\put(-24,-24){\makebox(-2,1){$\figchi$}}}
 \put(116,36){\makebox(-2,1){$\figchi$}}
 \put(36,42){\makebox(-2,1){$\figchi$}}

 \put(60,-30){\circle*{4}}
 \put(70,-30){\makebox(0,0)[l]{(even, even)}}
 \put(60,-45){\circle{4}}
 \put(70,-45){\makebox(0,0)[l]{(odd, odd)}}

\end{picture}

}

%% file: b_superspace_nat2.tex
{
\fontsize{11pt}{15pt}
\selectfont
\setlength{\unitlength}{1pt}
\begin{picture}(260,295)(-10,-35)
 \multiput(-10,0)(0,60){5}{\line(1,0){260}}
 \multiput(0,-10)(60,0){5}{\line(0,1){260}}
 
 {
 \thicklines
 \put(60,60){\vector(1,0){60}}
 \put(60,60){\vector(0,1){60}}
 \put(50,50){\vector(-1,-1){50}}
 \put(100,67){\makebox(0,0){$\a_1$}}
 \put(67,100){\makebox(0,0){$\a_2$}}
 \put(13,24){\makebox(0,0){$\at$}}
 }

\matrixput(0,0)(120,0){3}(0,120){3}{
 \put(-6,-6){\makebox(0,-4){$\figphit$}}
 \put(6,-6){\makebox(0,-4){$\figvarphit$}}
 \put(-6,6){\makebox(0,1){$\figchit$}}
}

\matrixput(60,60)(120,0){2}(0,120){2}{
 \put(0,0){\circle*{4}}
 \put(-6,6){\makebox(0,0){$\figchi$}}
 \put(-6,-6){\makebox(0,0){$\figvarphi$}}
 \put(6,-6){\makebox(0,0){$\figphi$}}
 \put(-54,6){\makebox(3,2){$\figpsi_1$}}
 \put(6,-54){\makebox(3,2){$\figpsi_2$}}
}
\multiput(300,60)(0,120){2}{
 \put(-54,6){\makebox(3,2){$\figpsi_1$}}
}
\multiput(60,300)(120,0){2}{
 \put(6,-54){\makebox(3,2){$\figpsi_2$}}
}

 \put(60,-30){\circle*{4}}
 \put(70,-30){\makebox(0,0)[l]{(even, even)}}

\end{picture}

}

%% file: NABFfig_sym01.tex
{
\fontsize{11pt}{15pt}
\selectfont
\setlength{\unitlength}{1pt}
\begin{picture}(200,230)(-10,-10)


 \multiput(-10,30)(0,60){3}{\line(1,0){230}}
 \multiput(30,-10)(60,0){4}{\line(0,1){200}}
 \multiput(-10,0)(0,60){4}{\dashline{2}(0,0)(230,0)}
 \multiput(0,-10)(60,0){4}{\dashline{2}(0,0)(0,200)}
 
 {
 \thicklines
 \put(70,70){\vector(1,1){20}}
 \put(50,70){\vector(-1,1){20}}
 \put(70,50){\vector(1,-1){20}}
 \put(50,50){\vector(-1,-1){18.5}}
 \put(83,75){\makebox(0,0){$\a$}}
 \put(40,75){\makebox(0,0){$\a_2$}}
 \put(85,45){\makebox(0,0){$\a_1$}}
 \put(36,45){\makebox(0,0){$\at$}}
 }
 \matrixput(60,60)(120,0){2}(0,120){2}{
 \put(-30,-30){\circle*{4}}
 \put(6,6){\makebox(0,0){$\figlambda$}}
 \put(-6,6){\makebox(0,0){$\figrho$}}
 \put(-6,-6){\makebox(0,0){$\figcb$}}
 \put(6,-6){\makebox(0,0){$\figc$}}
 }

 \matrixput(60,60)(120,0){2}(0,120){2}{
 \put(-36,-36){\makebox(0,-2){$\figphi$}}
 \put(36,-36){\makebox(3,0){$\figomega_1$}}
 }
 \multiput(60,60)(120,0){2}{
 \put(36,36){\makebox(0,0){$\figb$}}
 \put(-36,36){\makebox(-3,0){$\figomega_2$}}
 }

\end{picture}

}

%% file: NABFfig_sym02.tex
{
\fontsize{11pt}{15pt}
\selectfont
\setlength{\unitlength}{1pt}
\setlength{\unitlength}{.0851em}
\begin{picture}(260,315)(-10,-55)


 \multiput(-10,0)(0,60){5}{\dashline{2}(0,0)(260,0)}
 \multiput(0,-10)(60,0){5}{\dashline{2}(0,0)(0,260)}

 \multiput(-10,90)(0,120){2}{\line(1,0){98}}
 \multiput(92,90)(0,120){2}{\line(1,0){116}}
 \multiput(212,90)(0,120){2}{\line(1,0){38}}

 \multiput(-10,30)(0,120){2}{\line(1,0){38}}
 \multiput(32,30)(0,120){2}{\line(1,0){116}}
 \multiput(152,30)(0,120){2}{\line(1,0){98}}

 \multiput(90,-10)(120,0){2}{\line(0,1){98}}
 \multiput(90,92)(120,0){2}{\line(0,1){116}}
 \multiput(90,212)(120,0){2}{\line(0,1){38}}

 \multiput(30,-10)(120,0){2}{\line(0,1){38}}
 \multiput(30,32)(120,0){2}{\line(0,1){116}}
 \multiput(30,152)(120,0){2}{\line(0,1){98}}

 {
 \thicklines
 \put(70,70){\vector(1,1){18.3}}
 \put(50,70){\vector(-1,1){20}}
 \put(70,50){\vector(1,-1){20}}
 \put(50,50){\vector(-1,-1){18.5}}
 \put(83,75){\makebox(0,0){$\a$}}
 \put(40,75){\makebox(0,0){$\a_2$}}
 \put(85,45){\makebox(0,0){$\a_1$}}
 \put(48,40){\makebox(0,0){$\at$}}
 }

 \matrixput(60,60)(120,0){2}(0,120){2}{
 \put(-30,-30){\circle*{4}}
 \put(6,6){\makebox(0,0){$\figlambda$}}
 \put(-6,6){\makebox(0,0){$\figrho$}}
 \put(-6,-6){\makebox(0,0){$\figcb$}}
 \put(6,-6){\makebox(0,0){$\figc$}}
 }

 \matrixput(60,60)(120,0){2}(0,120){2}{
 \put(-36,-36){\makebox(0,-2){$\figphi$}}
 \put(36,-24){\makebox(4,0){$\figomega_1$}}
 \put(-24,36){\makebox(3,0){$\figomega_2$}}
 }
 \multiput(60,180)(120,0){2}{ 
\put(30,30){\circle{4}}
\put(24,24){\makebox(0,-1){$\figb$}}}

 \put(204,84){\makebox(0,-1){$\figb$}}
 \put(76,84){\makebox(0,-1){$\figb$}}
\put(90,90){\circle{4}}
\put(210,90){\circle{4}}

 \matrixput(0,0)(120,0){3}(0,120){3}{
 \put(6,6){\makebox(0,0){$\figlambda$}}
 \put(-6,6){\makebox(0,0){$\figrho$}}
 \put(-6,-6){\makebox(0,0){$\figcb$}}
 \put(6,-6){\makebox(0,0){$\figc$}}
 }

 \matrixput(60,60)(120,0){2}(0,120){2}{
 \put(36,36){\makebox(0,3){$\figphi$}}
 \put(-36,24){\makebox(0,0){$\figomega_1$}}
 \put(24,-36){\makebox(-1,0){$\figomega_2$}}
 }
 \multiput(60,180)(120,0){2}{\put(-24,-24){\makebox(-2,1){$\figb$}}}
 \put(156,36){\makebox(-2,1){$\figb$}}
 \put(36,43){\makebox(-2,1){$\figb$}}

 \put(60,-30){\circle*{4}}
 \put(70,-30){\makebox(0,0)[l]{(even, even)}}
 \put(60,-45){\circle{4}}
 \put(70,-45){\makebox(0,0)[l]{(odd, odd)}}

\end{picture}

}